\documentclass[11pt]{article}

\usepackage[final]{acl}

\usepackage{times}
\usepackage{latexsym}

\usepackage[T1]{fontenc}

\usepackage[utf8]{inputenc}

\usepackage{microtype}

\usepackage{inconsolata}

\usepackage{graphicx}

\usepackage{amsmath}    %
\usepackage{amssymb}    %
\usepackage{subcaption}
\usepackage{booktabs}   %
\usepackage{colortbl}   %
\usepackage{xcolor}     %
\usepackage{multirow}   %
\usepackage{enumitem}

\usepackage[capitalise]{cleveref}

\usepackage{fontawesome5}

\usepackage{xcolor}
\newif\ifhl
\hltrue   %

\title{When VLMs `Fix' Students: Identifying and Penalizing Over-Correction in the Evaluation of Multi-line Handwritten Math OCR}

\author{
 \textbf{Jin Seong\textsuperscript{1}}\thanks{\ Equal contribution.},
 \textbf{Wencke Liermann\textsuperscript{1}}\footnotemark[\value{footnote}],
 \textbf{Minho Kim\textsuperscript{1}}\thanks{Work done while interning at ETRI.},
 \textbf{Jong-hun Shin\textsuperscript{1}},
 \textbf{Soojong Lim\textsuperscript{1}}
\\
 \textsuperscript{1}Electronics and Telecommunications Research Institute, Republic of Korea
\\
 \small{
   \textbf{Correspondence:} \href{mailto:real_castle@etri.re.kr}{real\_castle@etri.re.kr}
 }
}

\begin{document}
\maketitle
\begin{abstract}
Accurate transcription of handwritten mathematics is crucial for educational AI systems, yet current benchmarks fail to evaluate this capability properly. Most prior studies focus on single-line expressions and rely on lexical metrics such as BLEU, which fail to account for the severity of differences.
In this paper, we present the first systematic study of multi-line handwritten math Optical Character Recognition (OCR), revealing a critical failure mode of Vision-Language Models (VLMs): over-correction. Instead of faithfully transcribing a student's work, these models often "fix" errors, thereby hiding the very mistakes that educational assessment aims to detect.
To quantify this phenomenon, we propose PINK (Penalized INK-based score), a semantic evaluation metric that leverages a Large Language Model (LLM) for rubric-based grading and explicitly penalizes differences that increase the grade.
Our comprehensive evaluation of 15 state-of-the-art VLMs on the FERMAT dataset reveals substantial ranking reversals compared to BLEU: models like GPT-4o are heavily penalized for aggressive over-correction, whereas Gemini 2.5 Flash emerges as the most faithful transcriber.
Furthermore, a human expert study shows that PINK aligns significantly better with human judgment (55.0\% preference over BLEU's 39.5\%), providing a more reliable evaluation framework for handwritten math OCR in educational settings.

\smallskip
\noindent
\faGithub\ \textbf{Code:}\ 
\href{https://github.com/9115jin/PINK}
{\texttt{github.com/9115jin/PINK}}

\end{abstract}

\section{Introduction}
\label{sec:intro}

\begin{figure}[t]
  \centering
  \includegraphics[width=\linewidth]{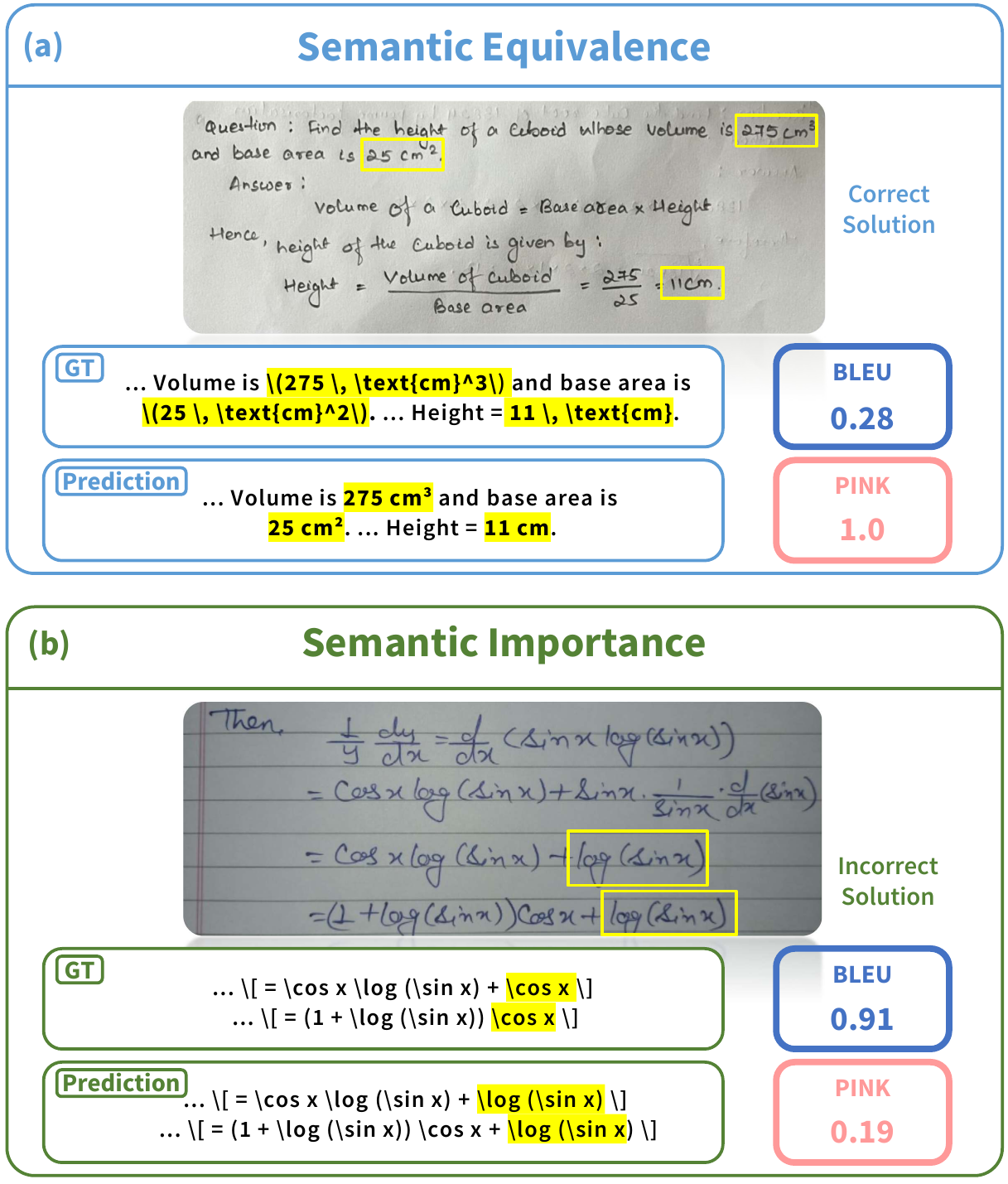}
  \caption{\textbf{BLEU’s Blind Spots in Multi-line Handwritten Math OCR.} (a) BLEU penalizes semantically equivalent expressions. (b) BLEU assigns a high score even when the OCR misses a key part of the solution, treating a critical omission as insignificant.}
  \label{fig:intro_example}
\end{figure}

The deployment of Vision-Language Models (VLMs) in educational AI systems promises personalized learning at scale; however, it requires one fundamental capability: the faithful interpretation of students' handwritten work. Here, a key challenge is that, while existing Optical Character Recognition (OCR) research has matured around single-line expressions~\cite{zanibbi2016crohme, deng2017im2markup, yuan2022syntax}, real student solutions span multiple lines, interweaving mathematical notation with explanatory text throughout the problem-solving process. As modern VLMs, equipped with remarkable capabilities in mathematical reasoning, autoregressively process this longer context, they build a strong, logically consistent internal narrative, which usually helps to solve ambiguities in the transcription process. However, when the model encounters an erroneous visual token (e.g., ”5” in ”2 × 3 = 5”), this token could create a conflict with the powerful internal context that implies the token in that spot must be ”6”. Consequently, a model might dismiss the visual evidence and generate a token that aligns with its logical reasoning \citep{lihidden}, erasing evidence of the student's misconception in the process, a phenomenon we call Over-Correction. Over-Correction has been observed across all modalities, including the fields of automatic speech \citep{udagawa2024robust} and grammar error recognition \citep{park2025leveraging}.

Using the FERMAT \cite{fermat2025} dataset, we quantify the phenomenon for handwritten multi-line math OCR, delivering evidence that Over-Correction is not limited to isolated sample instances but a pervasive failure mode of modern VLMs. The implications for educational AI are severe: an automated tutor receiving such "corrected" transcriptions would miss critical learning opportunities, potentially reinforcing misconceptions by failing to address them.

Our analysis employs PINK (Penalized INK-based score), a novel metric specifically designed to evaluate educational faithfulness. It leverages an LLM-based auto-grader with a carefully designed rubric system that assesses semantic mathematical correctness across five dimensions and allows for nuanced partial credit, i.e., 0-20 points, mirroring human grading. In a two-stage evaluation pipeline, PINK explicitly detects and penalizes over-corrections. It first computes an oracle baseline score by grading the ground truth transcription, before comparing it to the score assigned to the transcription generated by the OCR model. When a model's output scores higher than the oracle (indicating it "improved" the student's work), PINK applies graduated penalties, ensuring models are held accountable for transcription faithfulness. PINK also addresses another common deficit of traditional OCR evaluation metrics, e.g. BLEU~\cite{papineni2002bleu}. It can recognize semantic equivalence and reflect semantic importance: a critical unit mistake (cm→m) receives more weight than minor formatting variations. As \cref{fig:intro_example} illustrates, mathematically identical expressions like '\textbackslash frac\{1\}\{2\}' and '0.5' are penalized by BLEU due to LaTeX formatting differences, while PINK correctly identifies them as expressions of equal meaning. To quantify this advantage, we analyzed 2,244 FERMAT~\cite{fermat2025} samples transcribed by Ovis2 8B VLM and found that 60.9\% of discrepancies flagged by BLEU were limited to (8.2\%) or included (52.7\%) formatting differences.  

We evaluated 15 SoTA VLMs (ranging from 7B to 124B parameters) using our framework, and the results revealed a complete departure from conventional BLEU-based rankings. Under BLEU, GPT-4o achieves 3rd place with 0.735; under PINK, it drops to 6th with 0.907, heavily penalized for aggressive over-correction. Conversely, Gemini 2.5 Flash rises from 10th (BLEU: 0.614) to 1st (PINK: 0.935) by maintaining exceptional transcription faithfulness. A human expert evaluation demonstrates that the ranking under PINK offers a more human-aligned perspective: given a choice between BLEU and PINK scores for the same OCR output, experts preferred PINK 55.0\% of the time versus BLEU's 39.5\%. Beyond rankings, our results also reveal a strong positive correlation between model scale and over-correction frequency, that is, larger models within the same family consistently exhibit higher over-correction rates, suggesting this behavior may be an emergent property of advanced reasoning capabilities~\cite{wei2022emergent}.

In summary, our contributions are as follows.
\begin{itemize}[nosep, leftmargin=*] 
\item \textbf{First multi-line math OCR evaluation framework with semantic evaluation}: We adapt the FERMAT dataset to the task of OCR performance assessment and compare the abilities of 15 state-of-the-art VLMs.
\item \textbf{Discovery and quantification of Over-Correction}: We identify and systematically analyze a critical VLM failure mode affecting 42.1\%-66.2\% of transcriptions, where models spontaneously "fix" student errors rather than faithfully transcribing them.
\item \textbf{PINK metric validated by human preference}: We propose a faithfulness-aware metric that combines semantic accuracy with transcription fidelity, achieving 55\% human preference over BLEU's 39.5\% and revealing dramatic ranking reversals among state-of-the-art VLMs.
\end{itemize}

\section{Related Work}
\label{sec:related_works}

\noindent \textbf{Datasets for Math OCR.}  

\noindent Most handwritten math OCR work has matured around the \textbf{CROHME} challenges, which evaluate symbol recognition and structure matching via Symbol Layout Trees (SLT) and ExpRate, but these benchmarks focus on \textit{single expressions} only~\cite{zanibbi2016crohme}. Large-scale datasets such as \textbf{Im2LaTeX-100k}~\cite{deng2017im2markup}, \textbf{HME100K}~\cite{yuan2022syntax}, and \textbf{MathWriting}~\cite{MathWriting} extend the scale, yet they remain largely isolated-expression oriented. A first step toward \textbf{multi-line evaluation} was introduced in the ICDAR 2023 \textbf{MLHMER} competition~\cite{wu2023icdar}, but standardized document-level settings remain nascent.

\noindent\textbf{Evaluation Metrics for Math OCR.}

\noindent\textit{Text-based.} String-level metrics like BLEU and edit distance are widely adopted in image-to-LaTeX tasks ~\cite{deng2017im2markup}, but are brittle to format variations and fail to capture semantic equivalence.

\noindent\textit{Embedding-based.} BERTScore~\cite{bertscore} and COMET~\cite{comet} fail in mathematical contexts because they measure local textual similarity but do not verify step-by-step global reasoning. An output may look similar while containing an invalid transformation that breaks the derivation.

\noindent\textit{Rule/Structure-based.} Symbol Layout Trees~\cite{zanibbi2016crohme} represent symbols as labeled groups and spatial relationships as labeled directed edges. They can be used to determine equivalence in symbol-layout structure, but not on its own determine mathematical equivalence. In contrast, Computer Algebra Systems~\cite{meurer2017sympy} can check whether two terms are equivalent by transforming them into a common form. However, both fail to account for derivation fidelity, and cross-line consistency.

\noindent\textit{Image-based.} Rendering-based comparisons (PSNR, SSIM, LPIPS) mitigate markup variance, and the recent Character Detection Matching (CDM)~\cite{wang2025image} evaluates recognition at the rendered character level. However, these approaches assume renderable outputs and remain limited to single-line formulas.

\noindent\textit{LLM Evaluators.} Recent work demonstrates LLMs' effectiveness as evaluators for reasoning tasks, achieving strong human alignment across a range of tasks \cite{zheng2023judging, fu2023gptscore,chiang2024assignment,geval2023}, including educational grading~\cite{chiang2024assignment,choi2026exploring}. 

However, no existing metric jointly addresses semantic correctness, multi-line coherence, and
  transcription faithfulness—the gap our work targets.

\section{The PINK Score}
\label{sec:methods}
\begin{figure*}[t]
    \centering
    \includegraphics[width=\linewidth]{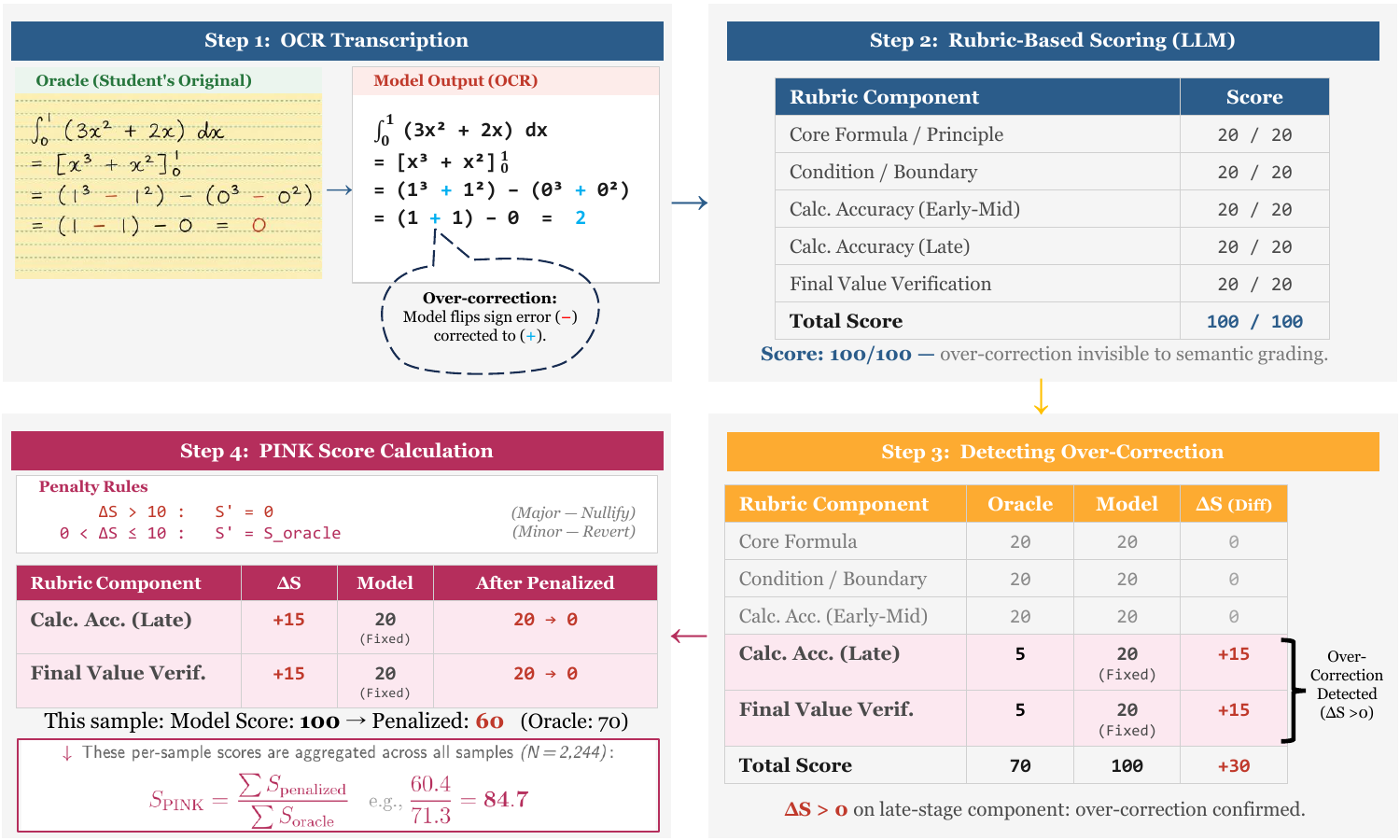}
    \caption{\textbf{Overview of the proposed PINK scoring system.} 
  Step~1: a VLM OCR output over-corrects a student error. 
  Step~2: the auto-grading system assigns a seemingly high score. 
  Step~3: comparison with the oracle score reveals an over-correction event. 
  Step~4: penalties are applied to compute the final PINK score.}
    \label{fig:pipeline}
\end{figure*}

While conventional metrics such as BLEU fail to capture semantic equivalence and the varying importance of differences, we first use an LLM-based auto-grader to assign semantic scores to both the oracle transcription of the solution path and the OCR transcription. These scores reflect the educational impact of OCR outputs more directly than lexical matching, since they evaluate whether the transcribed solution preserves the reasoning needed for downstream assessment.
However, semantic auto-grading alone cannot account for over-correction, where an OCR output receives a higher score by implicitly fixing a student's mistake. 
To address this limitation, we propose the PINK (Penalized INK-based) Score, a two-stage framework illustrated in \cref{fig:pipeline}: the first stage obtains semantic scores via LLM-based auto-grading, and the second explicitly detects and penalizes over-corrections according to their severity.

\subsection{Semantic Auto-Grading System}
The first stage of PINK obtains a semantic evaluation of the VLM’s OCR output. For this purpose, we employ GPT-5 (`gpt-5-2025-08-07') as our auto-grader system, leveraging the GPT family's capabilities in both evaluation tasks \cite{geval2023,chiang2024assignment} and mathematical reasoning \cite{AIME2025Leaderboard, gpt5}.

This system assesses mathematical correctness and logical flow irrespective of surface text forms. 
It employs a process-oriented rubric grounded in educational assessment principles. Following the philosophy that evaluating only final answers overlooks crucial problem-solving aspects~\cite{polya1945solve, schoenfeld1985mathematical}, we adopt partial-credit scoring similar to standardized tests like AP Calculus~\cite{collegeboard2023ap, cambridge2022alevel}. 
Our rubric decomposes solutions into five components assessing the entire problem-solving arc, each worth 20 points for a total of 100, as detailed in \cref{fig:pipeline} (Step 2). The full 
pedagogical basis and rubric specification are provided in 
~\cref{subsec:rubric_details}; detailed prompts appear in ~\cref{supp:prompts}.

The auto-grader applies the same rubric to two solution texts, producing comparable scores:

\begin{itemize} [leftmargin=*, nosep]
    \item \textbf{Model Score} ($S_{\text{model}}$): Score for the VLM's OCR output
    \item \textbf{Oracle Score} ($S_{\text{oracle}}$): Score for the ground-truth transcription (including any student errors)
\end{itemize}

As illustrated in \cref{fig:pipeline}, when a student makes a sign error yielding $0$ but the VLM "corrects" it to $2$, our grader initially assigns a perfect score ($S_{\text{model}}=100$), whereas the true student solution achieves a lower score $S_{\text{oracle}}=70$. To validate that such values are robust to grader choice and inherent LLM variance, we later replicated the full pipeline with the open-source Qwen3-80B as an alternative judge, achieving near-identical VLM-level PINK scores ($r{=}0.99$, $\tau{=}0.90$). Further analyses confirming stability across repeated runs and prompt variants appear in \cref{app:reproducibility}.

\subsection{Detecting Over-Correction}
The second stage is designed to detect over-correction events that remain invisible to the auto-grader alone. Such situations are particularly problematic in educational contexts, where the goal is to preserve---not overwrite---the student’s reasoning. In this context over-corrections are just as detrimental as wrongfully modified correct parts.

To capture this, we define over-correction events as cases where the model score exceeds the oracle score. For rubric component $r \in \{1,\dots,R\}$:
\begin{equation}
    \Delta S_r = S_{\text{model},r} - S_{\text{oracle},r}
\end{equation}

An event occurs whenever $\Delta S_r > 0$. The example in Step 3 of \cref{fig:pipeline} shows a $+15$ difference on the final answer rubric, flagging an over-correction. The magnitude of $\Delta S_r$ determines the penalty applied in the next stage.

\subsection{Penalty Mechanism}
Having identified over-correction, the next step is to decide how strongly it should be penalized. We adopt a two-tiered penalty system that separates minor variations from severe distortions, with the threshold set at 10 points—half the maximum score for any single rubric component. 

\noindent \textbf{Minor Over-Correction (Reversion):} A score inflation of 1--10 points is considered a minor discrepancy. We attribute such differences to the probabilistic nature of LLM grading or trivial semantic variations that do not fundamentally alter the student's solution. The penalty is therefore a simple \textit{reversion}: the component's score is replaced with the original oracle score $S_{\text{oracle},r}$ .

\noindent \textbf{Major Over-Correction (Nullification):} A score inflation of 11 points or more exceeds the halfway mark for a rubric component (20 points). This indicates that the model has fundamentally altered or ``corrected'' a clear and significant error in the student’s work. Such behavior is a serious distortion of the student's reasoning and is penalized with \textit{nullification}, reducing the component's score to 0.

This logic is formally defined in the calculation of the penalized score for each rubric component:
\begin{equation} 
S'_{r} = \begin{cases}
S_{\text{model},r}, & \Delta S_r \leq 0, \\
S_{\text{oracle},r}, & 1 \leq \Delta S_r \leq 10 \quad \text{(Minor)}, \\
0, & \Delta S_r > 10 \quad \text{(Major)}.
\end{cases} 
\end{equation}

For a given test item, let $R$ be the number of rubric components; the overall penalized score is:
\begin{equation}
S_{\text{penalized}} = \sum_{r=1}^{R} S'_{r}
\end{equation}

In Step~4 of \cref{fig:pipeline}, two rubric scores are nullified from $20/20$ to $0/20$, lowering the overall score from $100$ to $60$. This ensures that inflated correctness is not misinterpreted as faithfulness.

Empirically, setting the threshold at $T{=}10$, which corresponds to half of the maximum rubric score (20 points), 
provides a robust balance between sensitivity and stability. 
Approximately 80\% of all over-correction events fall below this boundary, while model ranking stability peaks near this value (Kendall's $\tau{>}0.9$~\cite{kendalls_tau}). These findings confirm that $T{=}10$ is both pedagogically meaningful and empirically optimal (see \cref{supp:penalty_threshold} for detailed ablation studies).

\subsection{Final PINK Score Calculation}
A raw penalized score alone can be misleading, since the oracle transcription in our perturbation dataset does not reach 100 (averaging 71.3). To obtain a comparable and interpretable metric, we normalize the penalized score by the oracle score.

Let $N$ be the number of test items. For each test item $n$, let $S_{\text{penalized},n}$ and $S_{\text{oracle},n}$ denote its penalized and oracle scores, respectively. The final \textbf{PINK score} ($S_{\text{PINK}}$) is defined as

\begin{equation}
S_{\text{PINK}} =
\frac{\sum_{n=1}^{N} S_{\text{penalized},n}}
     {\sum_{n=1}^{N} S_{\text{oracle},n}}
\end{equation}

Under this definition, a model whose penalized scores equal the oracle scores achieves 1.0, and deviations from 1.0 quantify departures from faithful transcription.

\section{Experimental Setup}
\label{sec:experiments}

\paragraph{Dataset}
We employ FERMAT~\cite{fermat2025}, a dataset designed to assess VLMs’ ability to detect, localize and correct mistakes in handwritten mathematical content. FERMAT is derived from 609 manually curated math problems. Based on a manually designed, comprehensive taxonomy of common student mistakes (computational, conceptual, notational, presentation, superficial), the authors adopted a human-in-the-loop approach to introduce targeted perturbations into correct solution paths. The resulting perturbed solution paths were transcribed by 43 human annotators, yielding 2,244 authentic handwritten samples. Notably, the original authors did not use FERMAT to explicitly assess OCR performance, and we are the first to adapt it for this purpose.

\paragraph{Baseline Models}
We selected 15 state-of-the-art VLMs with diverse architectures and scales (ranging from 7B to 124B) for a comprehensive performance comparison:
Ovis2~\cite{ovis}, InternVL3~\cite{internvl3}, Qwen2.5-VL~\cite{qwen2_5}, LLaMA3.2~\cite{llama3_2}, Pixtral~\cite{pixtral}, GPT-4o~\cite{gpt4, openai2024gpt4ocard}, and Gemini 2.5 Flash~\cite{gemini2_5_flash}. Full results for all models are reported in \cref{tab:heatmap-open-closed}. We focus exclusively on VLMs, as established non-VLM OCR models are inapplicable to FERMAT: TrOCR lacks math-symbol coverage, Nougat cannot process handwriting, and Pix2tex is limited to single-line expressions. We also standardize the OCR prompt across all models, and prompt-sensitivity analyses show stable rankings under grading- and OCR-prompt paraphrases~\cref{sec:autograder_prompt_sensitivity,sec:ocr_prompt_sensitivity}.

\paragraph{Evaluation Metrics}
The primary metric used in our evaluation is the proposed \textbf{PINK score}. PINK measures semantic correctness while penalizing over-correction, providing a faithfulness-oriented alternative to conventional text-based metrics. For comparison, we also report \textbf{BLEU}, the most widely used baseline for math OCR, as well as \textbf{Edit Distance}, a standard string-level measure included for completeness. We omit Expression Rate (ExpRate), the standard metric for isolated formula recognition, as it is inapplicable to multi-line solutions mixing prose and equations and collapses to near 0\% for all VLMs on FERMAT.

To support PINK computation, we also report the \textbf{Oracle Score} ($S_{\text{oracle}}$), the auto-grader’s score on the ground-truth transcription. This baseline verifies that the perturbations reflect genuine mathematical errors and provides the normalization anchor used by PINK. The perturbed solutions average 71.3, while 500 clean samples score 95.1.

\section{Results}

\subsection{Over-Correction is Pervasive and Systematic}
\label{sec:main_results}
Over-correction is pervasive: across all 15 evaluated VLMs, \textbf{42--66\%} of transcriptions contain at least one instance where the model overwrites the student's original work. To understand this pervasive failure, we characterize the distribution and mechanism of this behavior below.

\begin{figure}[t]
    \centering
    \includegraphics[width=\linewidth]{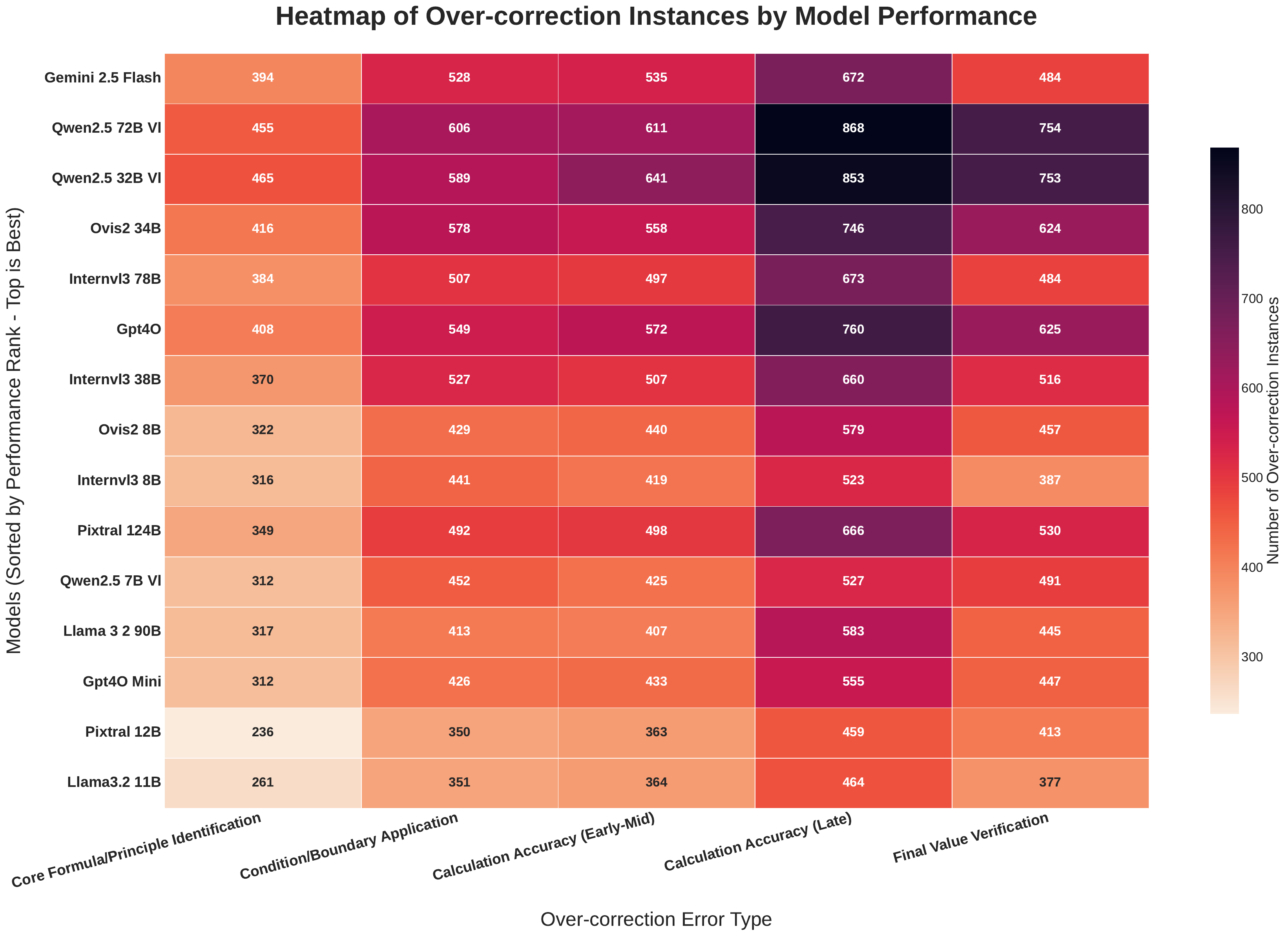}
    \caption{\textbf{Distribution of over-correction instances across rubric items for 15 models}, sorted by Penalized Score (Top is Best). Over-corrections are heavily concentrated in the final two stages, particularly in higher-performing models, consistent with autoregressive context accumulation overriding visual evidence.}
    \label{fig:heatmap}
\end{figure}

\noindent \textbf{Where and When Does Over-Correction Occur?}
Our analysis reveals that over-correction is not a random perceptual glitch but a systematic behavior tied to the model’s reasoning process. The heatmap in \cref{fig:heatmap} exposes two prominent trends. First, over-corrections cluster heavily in the final stages of problem-solving---specifically in \textit{\textbf{Calculation Accuracy (Late)}} and \textit{\textbf{Final Value Verification}}. Second, this concentration is markedly more pronounced in higher-performing models. This suggests that over-correction is an emergent property of sophisticated reasoning rather than a simple perceptual failure. As the model autoregressively processes a solution, it builds a strong, logically consistent internal narrative. When encountering visually inconsistent tokens (e.g., ``5’’ in ``$2 \times 3 = 5$’’), these contradictions are often dismissed as noise. Larger and more capable models, which rely more heavily on their internal context, are therefore more likely to override visual evidence and produce logically inferred tokens instead.

\begin{figure}[t]
    \centering
    \includegraphics[width=0.9\linewidth]{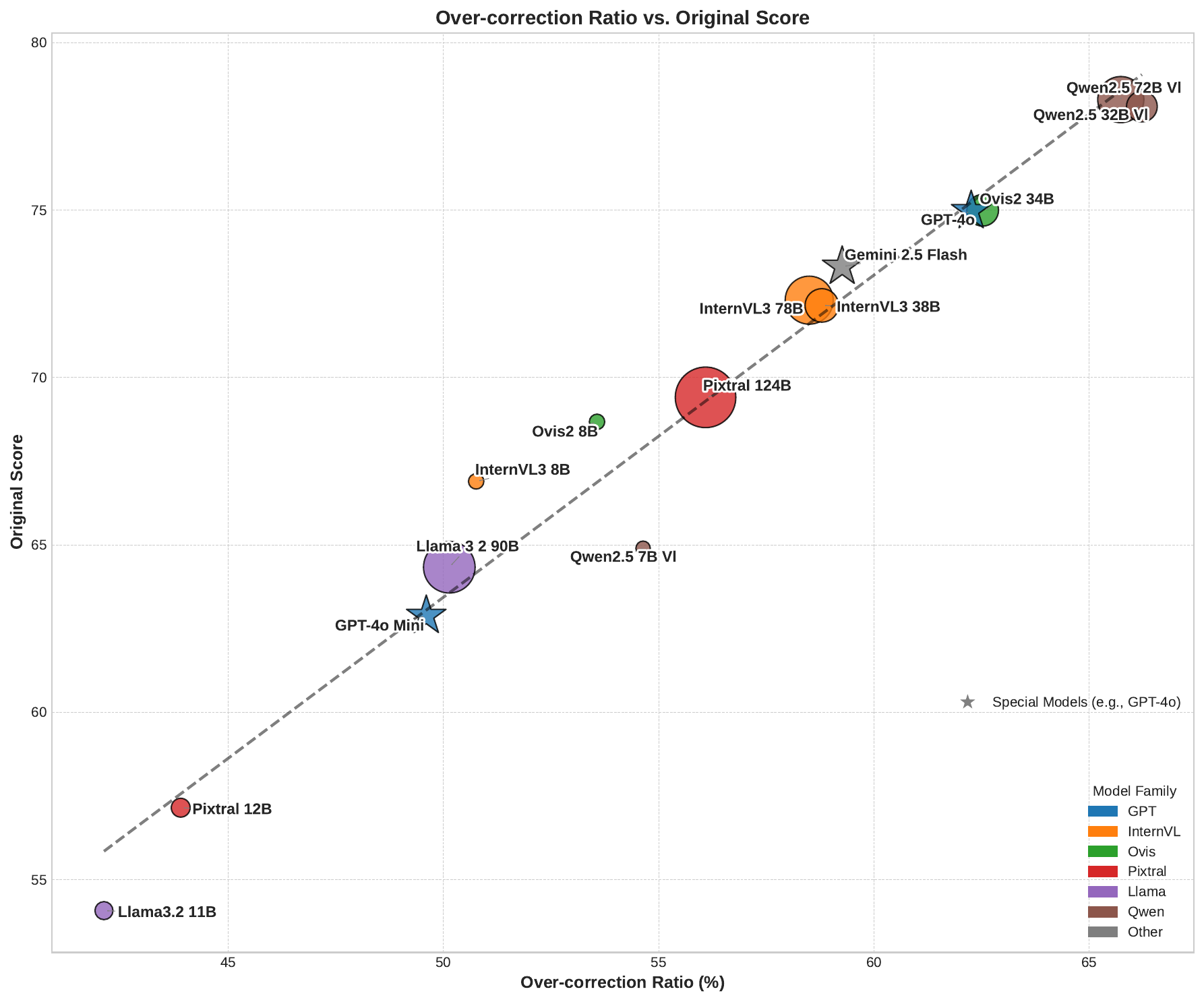}
    \caption{\textbf{Correlation between Original Score and Over-correction Ratio.} Bubble size represents the scale of the model.}
    \label{fig:bubble_chart}
\end{figure}

\noindent \textbf{Correlation with Model Scale and Properties.}
To investigate how model characteristics relate to over-correction, we analyze the bubble chart in \cref{fig:bubble_chart}. The chart plots each model’s over-correction ratio against its rubric-based score, with bubble size indicating parameter count (see \cref{supp:oc_rates} for full per-model frequencies).

A clear pattern emerges: \textbf{stronger models over-correct more}. Models achieving higher rubric-based scores also exhibit higher over-correction ratios. Over-correction is pervasive yet variable, ranging from \textbf{66.2\%} (Qwen2.5 32B VL) to \textbf{42.1\%} (Llama3.2 11B). High-performing models such as \textbf{Gemini 2.5 Flash} and \textbf{Qwen2.5 72B} cluster in the top-right region, reflecting both strong mathematical ability and a greater tendency to overwrite the student’s work. The trend is consistent with a greater reliance on internal context, which can resemble the context-driven  generation biases discussed in prior work~\cite{Ji_2023}.

\subsection{Visual Evidence of Over-Correction}
\label{sec:visual_evidence}

\begin{figure*}[t]
  \centering
  \includegraphics[width=\textwidth]{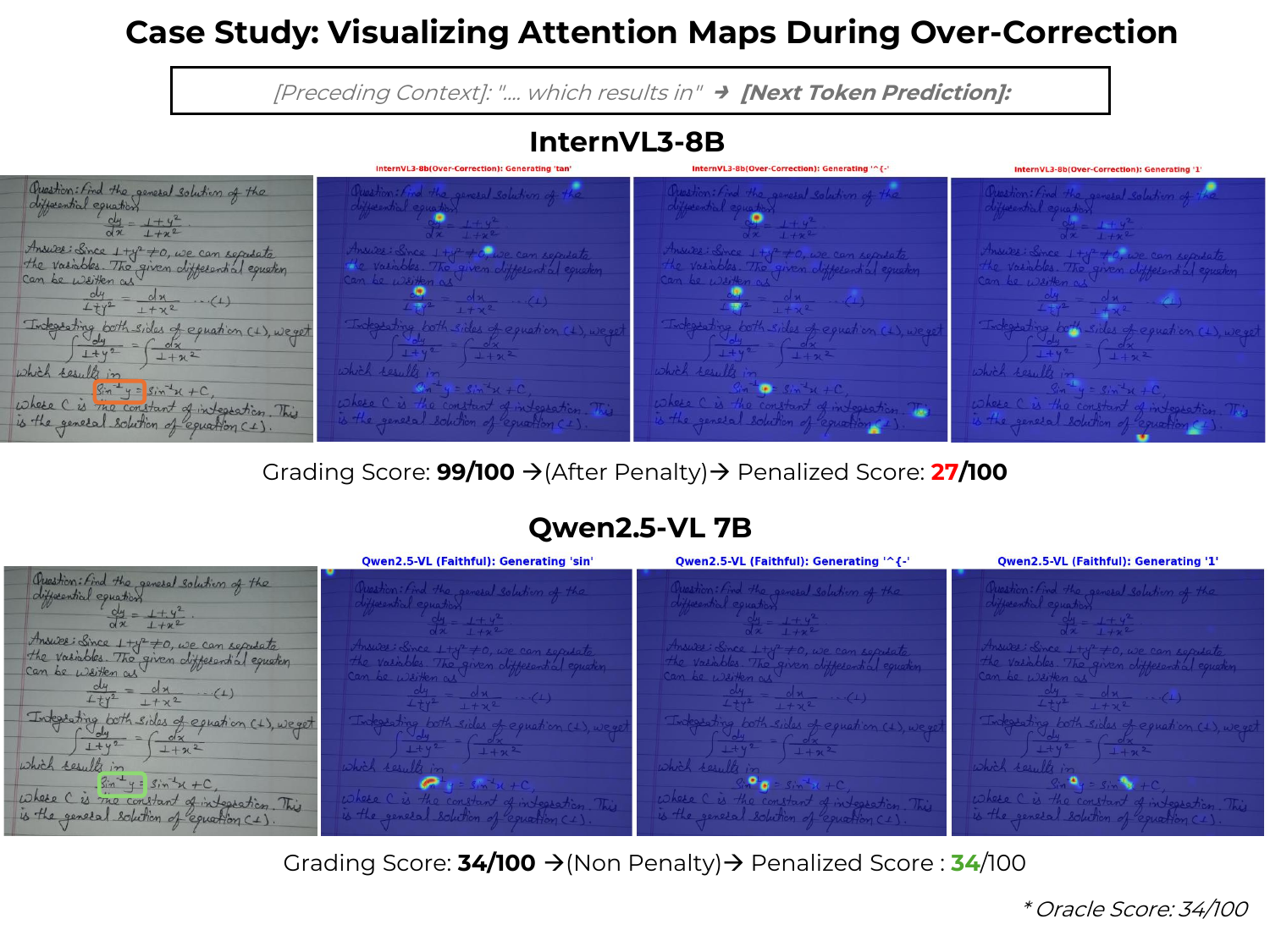} 
  \caption{
  \textbf{Visualizing Attention Maps and Penalty Application.} 
  We visualize the attention mechanism at the moment of generating the target token, conditioned on the preceding integral context.
  \textbf{(Top) InternVL3-8B} produces the mathematically consistent but unfaithful token ``tan'', while showing little attention to the handwritten ``sin''. Its high raw score of \textbf{99} is reduced to \textbf{27} after applying the PINK penalty.
  \textbf{(Bottom) Qwen2.5-VL-7B} faithfully preserves ``sin'' while maintaining attention on the relevant handwritten strokes. Its score remains \textbf{34}, matching the oracle score, with no penalty. This contrast is consistent with accumulated mathematical context attenuating visual grounding during over-correction.
  }
  \label{fig:attention_viz}
\end{figure*}

We analyze attention maps during over-corrected token generation (\cref{fig:attention_viz}). When InternVL3-8B corrects a student’s ``sin’’ to ``tan’’, attention on the handwritten error region drops sharply; a similarly sized faithful transcriber, Qwen2.5-VL 7B, maintains attention on the strokes. We observe similar patterns in additional cases, consistent with context accumulation attenuating visual grounding and motivating PINK’s explicit penalty for score improvements over the oracle. Further details on attention extraction and case-level analysis are provided in \cref{subsec:attention_analysis}.

\subsection{Standard Text Metrics Miss Over-Correction}
\label{sec:metric_failure}

Having established that over-correction is both 
pervasive and systematic, we now ask: do standard 
evaluation metrics reflect these differences in 
model faithfulness? We compare rankings under 
BLEU and PINK to test if a faithfulness-aware 
metric yields meaningfully different assessments.

\begin{figure}[t]
    \centering
    \includegraphics[width=\linewidth]{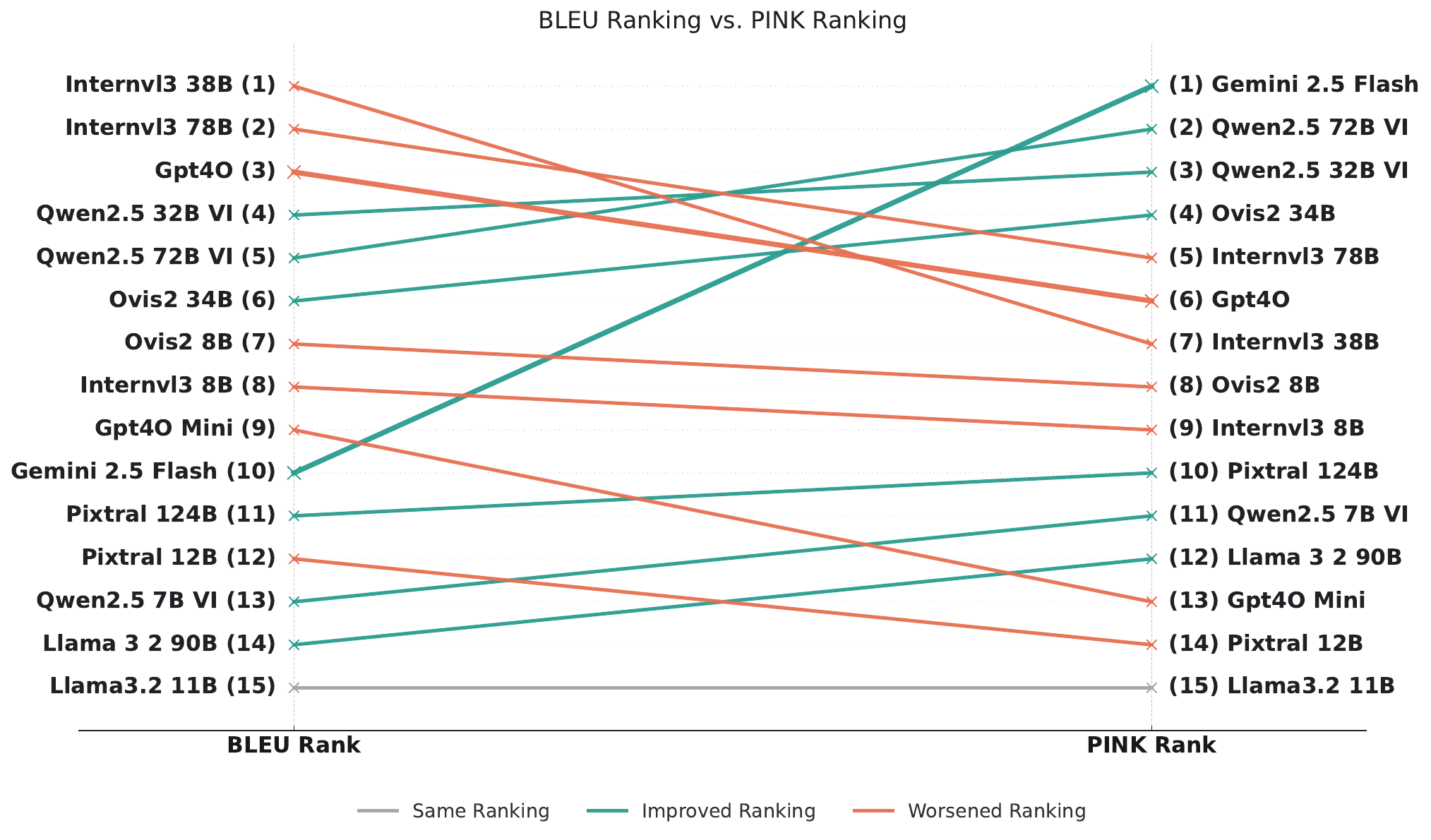}
    \caption{\textbf{Ranking Reversal of VLMs.} The leaderboard based on conventional BLEU scores (left) is dramatically overturned when re-evaluated with our PINK metric (right), which penalizes over-correction. Models like \textit{GPT-4o} drop significantly, while \textit{Gemini 2.5 Flash} ascends nine places to the top rank, showing that existing metrics miss the crucial dimension of faithfulness.}
    \label{fig:ranking_reversal}
\end{figure}

We compare the rankings of 15 open-source and proprietary VLMs on the FERMAT dataset~\cite{fermat2025} under two metrics: the conventional text-based BLEU score~\cite{papineni2002bleu} and our proposed PINK metric, which penalizes over-correction. The results in \cref{fig:ranking_reversal} are striking. An OCR leaderboard based on BLEU is dramatically overturned when re-evaluated with PINK. Models that achieve top ranks under BLEU, such as \textit{GPT-4o} and \textit{InternVL}, experience a substantial drop in rank. Notably, \textit{Gemini 2.5 Flash}, previously a mid-tier model under BLEU, ascends nine places to claim the top position under PINK. This reversal provides strong evidence that over-correction is not a rare edge case, but a frequent failure mode in SoTA VLMs. Crucially, this divergence is not explained by rubric grading alone: under raw rubric scores the two models rank similarly (Gemini 5th, GPT-4o 4th). The gap emerges only when transcription faithfulness is accounted for (see \cref{app:ranking_reshuffle} for the full transition across all 15 VLMs).

\begin{figure}[t]
    \centering
    \begin{subfigure}[b]{0.48\linewidth}
        \includegraphics[width=\linewidth]{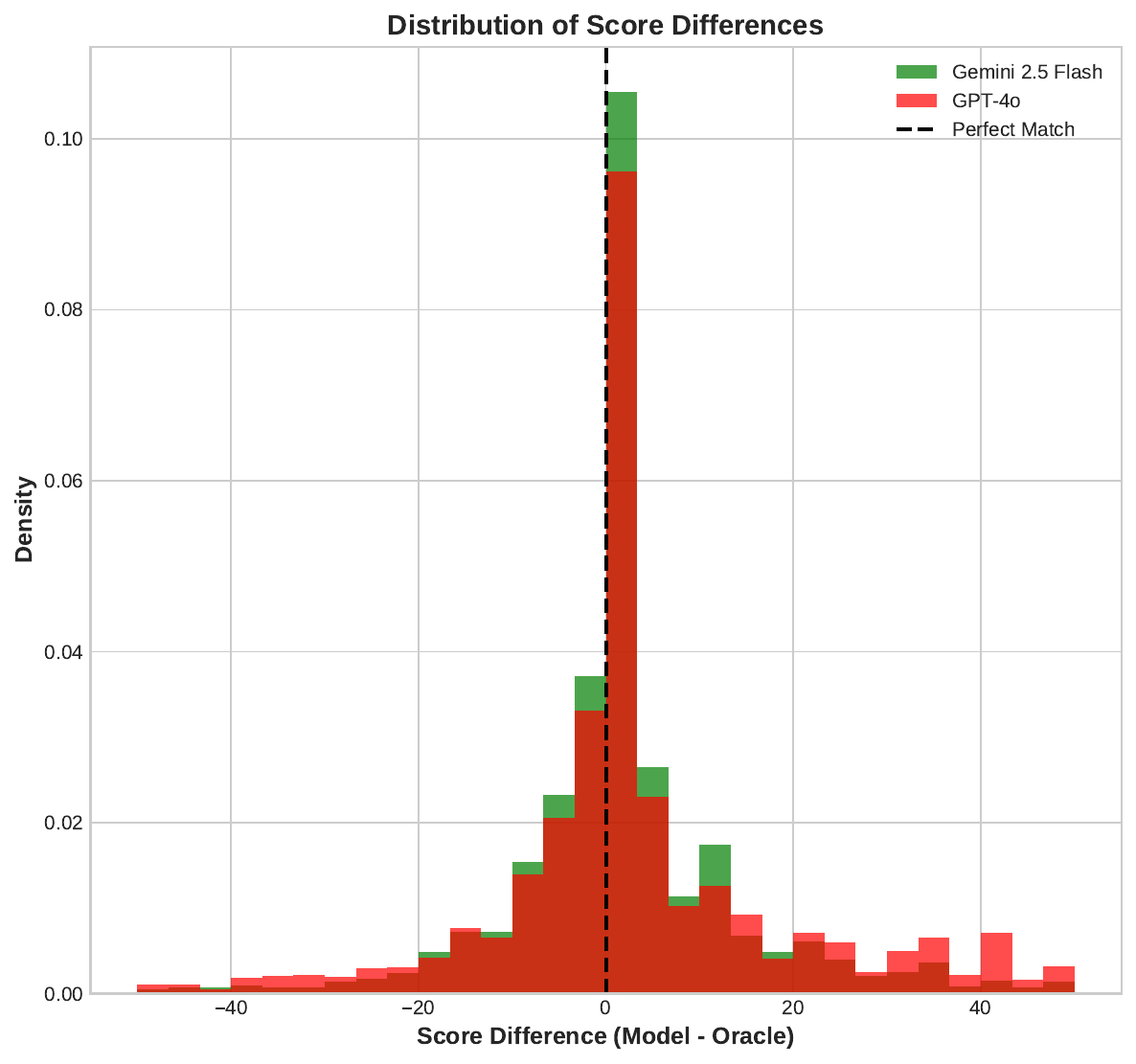}
        \caption{Score Difference Distribution}
        \label{fig:distribution_plot}
    \end{subfigure}
    \hfill
    \begin{subfigure}[b]{0.48\linewidth}
        \includegraphics[width=\linewidth]{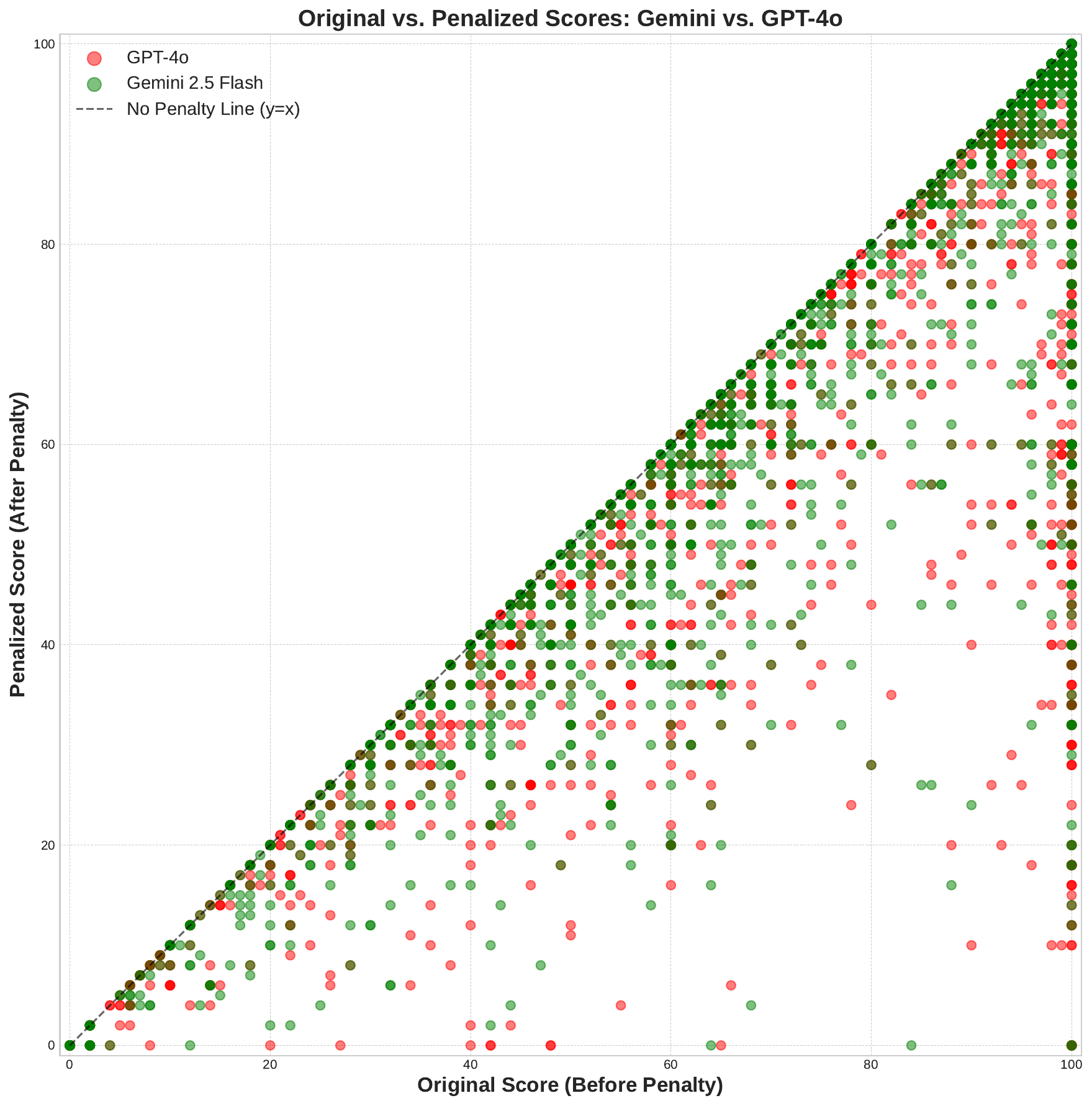}
        \caption{Penalty Analysis Scatter Plot}
        \label{fig:scatter_plot}
    \end{subfigure}
    \caption{\textbf{Over-Correction Rate}: Comparison between \textit{GPT-4o} (red) and \textit{Gemini 2.5 Flash} (green).}
    \label{fig:analysis_plots}
\end{figure}

To better understand this ranking reversal, we closely examine the two models with the largest discrepancy: \textit{GPT-4o} and \textit{Gemini 2.5 Flash}. The distribution plot in \cref{fig:distribution_plot} shows the difference between the rubric-based score assigned to a model’s OCR transcription and that assigned to the ground-truth transcription, where a positive value indicates over-correction. \textit{GPT-4o} exhibits a \textbf{longer and heavier right tail} than \textit{Gemini 2.5 Flash}, indicating both more frequent and more severe over-corrections. The scatter plot in \cref{fig:scatter_plot} further supports this trend by comparing rubric-based scores before and after applying the over-correction penalty: \textit{Gemini 2.5 Flash} remains tightly clustered around the $y = x$ diagonal, while many outputs from \textit{GPT-4o} fall far below it. Some \textit{GPT-4o} samples drop from near 100 to near zero after penalization, showing that the model can produce logically consistent transcriptions that nevertheless severely distort the student’s original work. For completeness, \cref{tab:heatmap-open-closed} reports the full quantitative results for all models across PINK, BLEU, and Edit Distance~\cite{editdistance1965}, supporting the leaderboard reversal described above.

\definecolor{LightGray}{gray}{0.95}
\definecolor{LightBlue}{rgb}{0.9,0.95,1.0}
\definecolor{LightOrange}{rgb}{1.0,0.92,0.85}
\definecolor{skyblue}{rgb}{0.53, 0.81, 0.92}

\begin{table}[t]
\centering
\caption{\textbf{Performance comparison of open-source and closed-source models.} The table evaluates models using the PINK, BLEU, and Edit Distance scores. The background color intensity indicates the relative performance for each metric.}
\label{tab:heatmap-open-closed}
\resizebox{\linewidth}{!}{
\begin{tabular}{lccc}
\toprule
\textbf{Model} & \textbf{\textcolor{pink}P\textcolor{pink}I\textcolor{pink}N\textcolor{pink}K Score} $\uparrow$ & \textbf{\textcolor{skyblue}B\textcolor{skyblue}L\textcolor{skyblue}E\textcolor{skyblue}U Score} $\uparrow$ & \textbf{Edit Distance} $\downarrow$ \\
\midrule
\rowcolor{gray!10}
\multicolumn{4}{c}{\textbf{Open-source Models}} \\
\midrule
Qwen2.5 72B VL      & \cellcolor{pink!66}0.920 & \cellcolor{skyblue!68}0.733 & \cellcolor{orange!42}0.1978 \\
Qwen2.5 32B VL      & \cellcolor{pink!65}0.919 & \cellcolor{skyblue!68}0.734 & \cellcolor{orange!42}0.1975 \\
Ovis2 34B           & \cellcolor{pink!64}0.916 & \cellcolor{skyblue!66}0.723 & \cellcolor{orange!41}0.1997 \\
InternVL3 78B       & \cellcolor{pink!62}0.911 & \cellcolor{skyblue!70}0.749 & \cellcolor{orange!43}0.1969 \\
InternVL3 38B       & \cellcolor{pink!60}0.905 & \cellcolor{skyblue!75}\textbf{0.771} & \cellcolor{orange!50}0.1925 \\
Ovis2 8B            & \cellcolor{pink!52}0.887 & \cellcolor{skyblue!50}0.652 & \cellcolor{orange!30}0.2321 \\
InternVL3 8B        & \cellcolor{pink!49}0.873 & \cellcolor{skyblue!48}0.642 & \cellcolor{orange!29}0.2364 \\
Pixtral 124B         & \cellcolor{pink!40}0.850 & \cellcolor{skyblue!35}0.606 & \cellcolor{orange!22}0.2836 \\
Qwen2.5 7B VL       & \cellcolor{pink!36}0.832 & \cellcolor{skyblue!25}0.565 & \cellcolor{orange!18}0.3332 \\
Llama-3.2 90B VI    & \cellcolor{pink!33}0.814 & \cellcolor{skyblue!20}0.512 & \cellcolor{orange!18}0.3339 \\
Pixtral 12B       & \cellcolor{pink!18}0.722 & \cellcolor{skyblue!35}0.606 & \cellcolor{orange!27}0.2495 \\
Llama-3.2 11B VI    & \cellcolor{pink!10}0.676 & \cellcolor{skyblue!10}0.407 & \cellcolor{orange!10}0.4649 \\
\midrule
\rowcolor{gray!10}
\multicolumn{4}{c}{\textbf{Closed-source Models}} \\
\midrule
Gemini 2.5 Flash    & \cellcolor{pink!70}\textbf{0.935} & \cellcolor{skyblue!35}0.614 & \cellcolor{orange!40}0.2007 \\
GPT-4o              & \cellcolor{pink!61}0.907 & \cellcolor{skyblue!69}0.735 & \cellcolor{orange!70}\textbf{0.1676} \\
GPT-4o Mini         & \cellcolor{pink!24}0.787 & \cellcolor{skyblue!35}0.618 & \cellcolor{orange!25}0.2698 \\
\bottomrule
\end{tabular}}
\end{table}

\begin{figure*}[t]
    \centering %

    \begin{subfigure}[b]{0.45\textwidth}
        \centering
        \includegraphics[width=\linewidth]{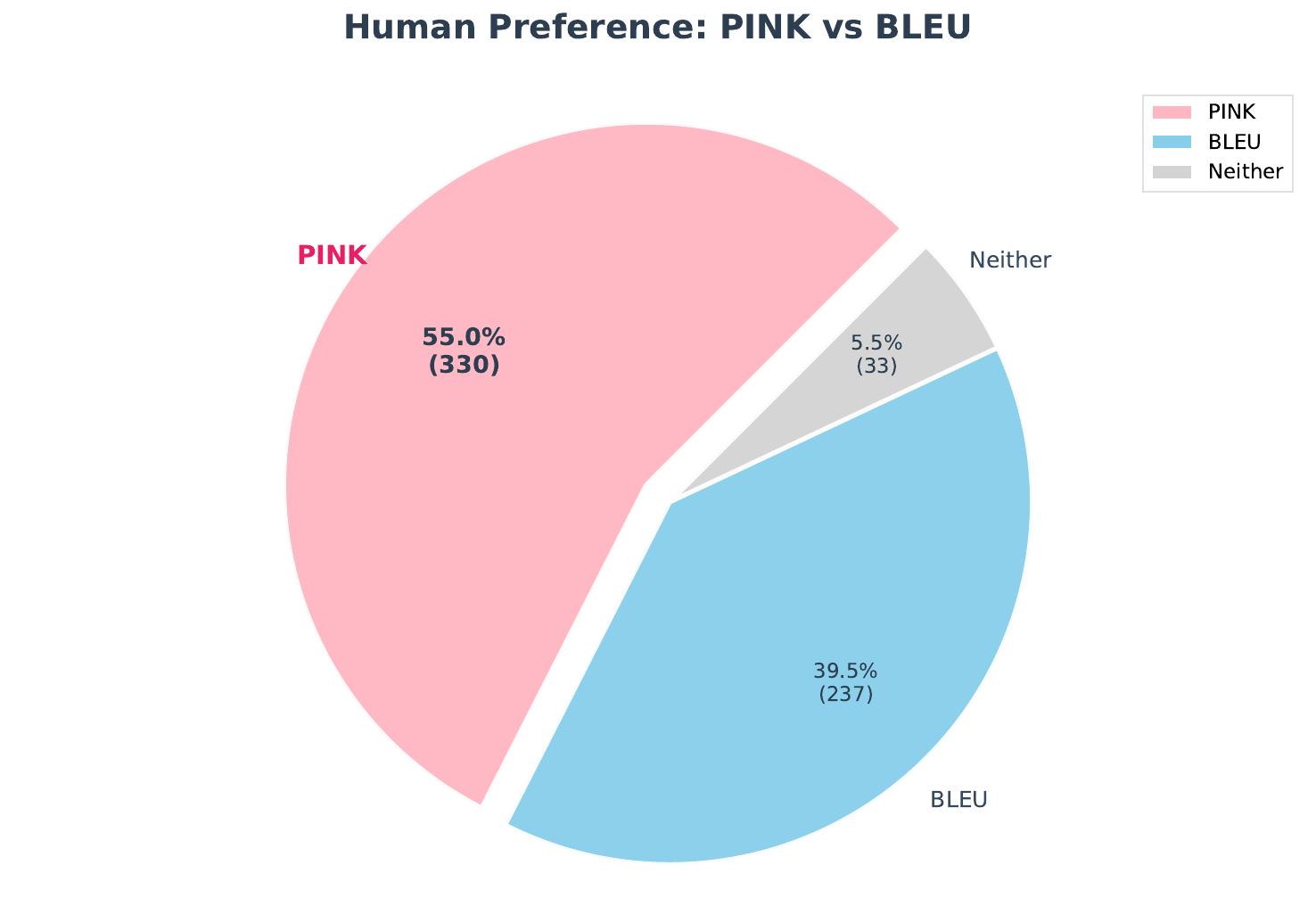}
        \caption{Human Preference: PINK vs BLEU.}
        \label{fig:human_pref_pie}
    \end{subfigure}
    \hfill %
    \begin{subfigure}[b]{0.45\textwidth}
        \centering
        \includegraphics[width=\linewidth]{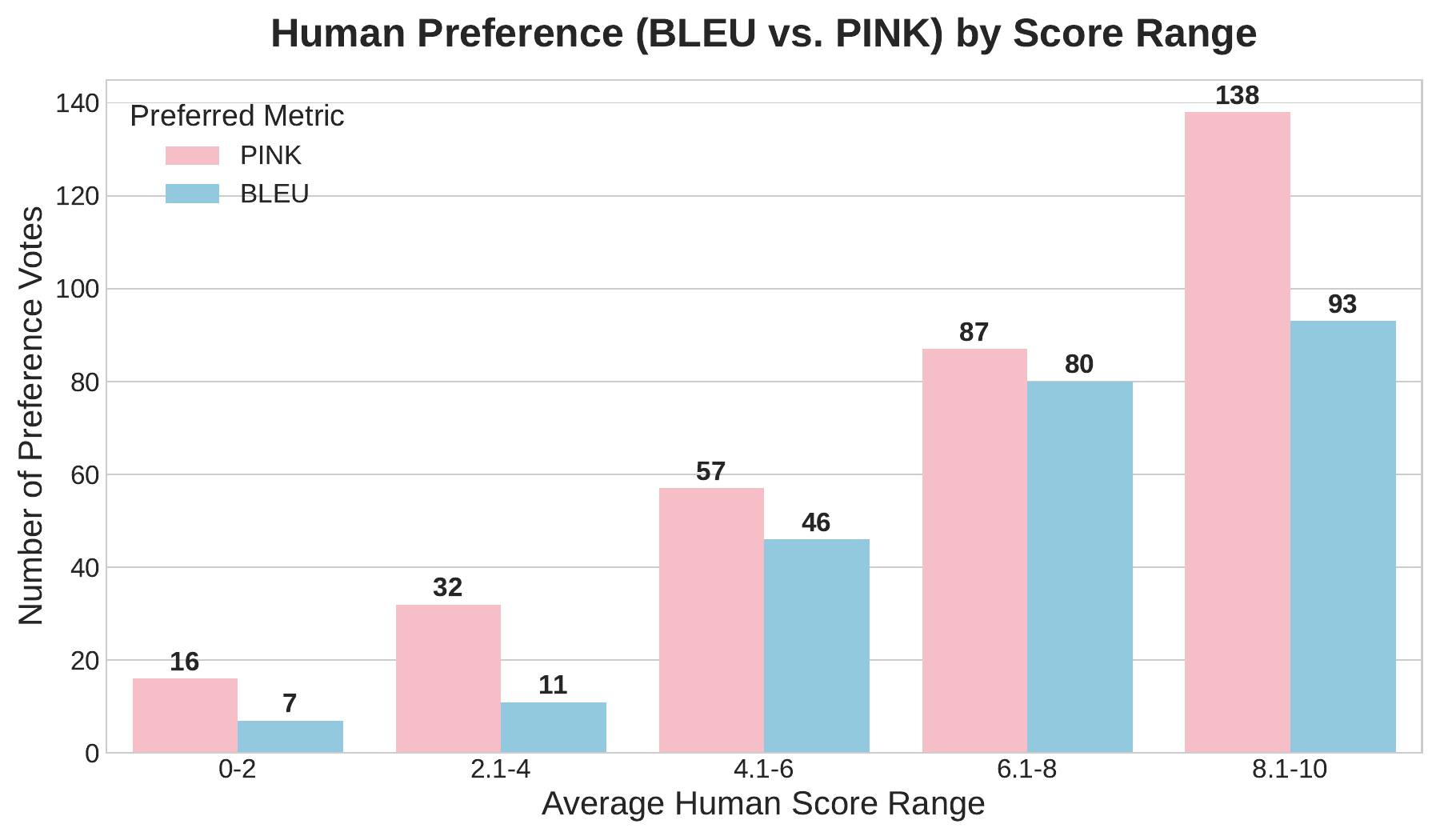}
        \caption{Preference analysis by score range.}
        \label{fig:human_pref_bar}
    \end{subfigure}

    \caption{\textbf{Human evaluation results.} (a) Experts showed a clear overall preference for PINK over BLEU. (b) PINK was consistently preferred across all score ranges, demonstrating its robustness.}
    \label{fig:human_pref} 
\end{figure*}

\subsection{Human Evaluation}
\label{sec:human_eval}
To validate whether PINK aligns better with human judgment than BLEU, we designed an evaluation around two research questions:

\begin{itemize} [leftmargin=*, nosep]
    \item \textbf{RQ1 (Reliability):} How consistent is the LLM-based automated scoring system, which underpins the PINK score, with evaluations from human experts?
    \item \textbf{RQ2 (Preference):} Do human experts prefer the PINK score over BLEU for assessing the quality of OCR results?
\end{itemize}

We recruited three expert annotators with university-level mathematical knowledge and an average of over two years of experience in grading mathematics assignments. The evaluation was performed on 200 samples extracted from the FERMAT dataset, transcribed by Ovis2-8B VLM. To ensure this sample was an unbiased representation, we confirmed its BLEU score distribution statistically mirrors the full dataset (mean 0.650 vs. 0.652). The annotators were assigned two tasks:

\textbf{Direct Grading (for RQ1):} For each of the 200 OCR outputs, annotators assigned an overall score on a 0-10 scale, guided by the same five evaluation categories used in our automated scoring system.

\textbf{Preference Selection (for RQ2):} This task directly measures which metric aligns better with human judgment. Annotators were first presented with a student's original handwritten solution and the correct solution. Then, the PINK and BLEU scores, calculated for a single OCR output of the student's work, were presented anonymously as `Score A' and `Score B'. The assignment of metrics to `Score A' or `Score B' was randomized for each sample to eliminate positional bias. Annotators then selected which score they believed better reflected the faithfulness of the OCR result.

The results of our evaluation are as follows.

\textbf{ RQ1: Auto-Grading System Correlates Strongly with Experts.}
To assess the reliability of our human evaluation, we measured inter-rater reliability, with a Cohen's weighted kappa of 0.69, indicating a substantial level of agreement among annotators. We then compared the average expert scores with our LLM-based automated scores, obtaining a Pearson correlation coefficient of 0.78. These results support the use of automated scoring within PINK as a proxy for expert judgment.

\textbf{RQ2: Experts Show a Clear and Consistent Preference for PINK over BLEU.}
As shown in \cref{fig:human_pref_pie}, experts more often judged PINK to better reflect OCR quality than BLEU. Overall, PINK was preferred in 55.0\% of cases, compared to BLEU's 39.5\% ($p < 0.001$, two-sided binomial test on item-level majority votes). The trend is not BLEU-specific: comparisons against MathBERTa~\cite{mathberta} and Edit Distance~\cite{editdistance1965} show consistent results (\cref{sec:extended_baselines}). Experts also favored PINK across all score brackets (\cref{fig:human_pref_bar}), suggesting better alignment with perceived OCR quality. These findings support that penalizing over-correction improves alignment with expert judgment.

\section{Conclusion and Future Work}
\label{sec:Conclusion}

Over-correction is not a rare edge case but a systematic failure mode affecting 42--66\% of transcriptions across all 15 evaluated models, with stronger models over-correcting more frequently. Standard metrics like BLEU fail to detect this behavior; PINK, which explicitly penalizes over-correction, reverses model rankings and aligns better with human judgment (55\% preference vs.\ BLEU's 39.5\%). Our attention map analysis further suggests that autoregressive context accumulation attenuates visual grounding, indicating that stronger reasoning may come at the cost of transcription faithfulness---a trade-off that warrants further investigation. As VLMs are increasingly deployed in educational settings, ensuring that these models preserve rather than erase student reasoning must become a first-class evaluation criterion.

Our preliminary mitigation experiments reveal that over-correction is not easily resolved through prompting alone (\cref{app:mitigation}). While explicit instructions marginally reduced over-correction, they simultaneously degraded overall transcription accuracy, resulting in no net improvement. This suggests that over-correction is deeply entangled with the models' core reasoning capabilities, likely requiring architecture- or training-level interventions, such as targeted model editing approaches for LVLMs \citep{seong2025memeic}

\section*{Limitations}
\label{sec:limitation}
Despite our extensive experiments and analyses, this work still has several limitations.

First, a high PINK score does not imply a universally superior VLM or OCR model. PINK specifically measures transcription faithfulness with respect to over-correction and should be interpreted as one evaluation axis rather than a holistic quality indicator. Second, our evaluation pipeline relies on an LLM-based auto-grader, with GPT-5 serving as the primary judge. This introduces a dependency on proprietary infrastructure that limits both reproducibility and accessibility. To mitigate this, we conducted cross-grader validation using the open-source Qwen3-80B, achieving strong agreement. We plan to release the complete evaluation codebase, including the open-source grading pipeline, to support independent replication.
Third, all experiments are conducted on a single dataset, FERMAT. This choice was driven by the current absence of alternative publicly available datasets that simultaneously feature multi-line, handwritten mathematical solutions with intentionally introduced student errors. Extending this evaluation across diverse data sources will be an important direction for future work.
Finally, we acknowledge the limited scale of our human evaluation due to the difficulty of recruiting annotators with combined expertise in \LaTeX{}, mathematics, and pedagogy. Nonetheless, the preference trends and score correlations strongly support PINK's alignment with expert judgment.

Our critique of over-correction should not be read as a negative assessment of VLM progress. The challenge lies in enabling models to distinguish when to reason and when to faithfully transcribe---we hope PINK can serve as a useful tool for navigating this trade-off.

\section*{Acknowledgements}
This work was supported by Electronics and Telecommunications Research Institute (ETRI) grant funded by the Korean government (26CB1200, Development and Application of Science-Specialized Multimodal Foundation Models). It was also supported by the Institute of Information \& Communications Technology Planning \& Evaluation (IITP) grant funded by the Korea Government (MSIT) (No. RS-2023-00216011, Development of Artificial Complex Intelligence for Conceptually Understanding and Inferring like Human) and Institute of Information and Communications Technology Planning and Evaluation (IITP) grant funded by the Korea government (MSIT) (RS-2019-II190004, Development of semi-supervised learning language intelligence technology and Korean tutoring service for foreigners).

\bibliography{main}

\begin{thebibliography}{40}
\providecommand{\natexlab}[1]{#1}

\bibitem[{Agrawal et~al.(2024)Agrawal, Antoniak, Hanna, Bout, Chaplot,
  Chudnovsky, Costa, Monicault, Garg, Gervet, Ghosh, Héliou, Jacob, Jiang,
  Khandelwal, Lacroix, Lample, Casas, Lavril, Scao, Lo, Marshall, Martin,
  Mensch, Muddireddy, Nemychnikova, Pellat, Platen, Raghuraman, Rozière,
  Sablayrolles, Saulnier, Sauvestre, Shang, Soletskyi, Stewart, Stock, Studnia,
  Subramanian, Vaze, Wang, and Yang}]{pixtral}
Pravesh Agrawal, Szymon Antoniak, Emma~Bou Hanna, Baptiste Bout, Devendra
  Chaplot, Jessica Chudnovsky, Diogo Costa, Baudouin~De Monicault, Saurabh
  Garg, Theophile Gervet, Soham Ghosh, Amélie Héliou, Paul Jacob, Albert~Q.
  Jiang, Kartik Khandelwal, Timothée Lacroix, Guillaume Lample, Diego~Las
  Casas, Thibaut Lavril, and 23 others. 2024.
\newblock \href {https://arxiv.org/abs/2410.07073} {Pixtral 12b}.
\newblock \emph{Preprint}, arXiv:2410.07073.

\bibitem[{AIME(2025)}]{AIME2025Leaderboard}
AIME. 2025.
\newblock Aime 2025 leaderboard.
\newblock \url{https://www.kaggle.com/benchmarks/open-benchmarks/aime-2025}.
\newblock Accessed: 2025-11-13.

\bibitem[{Bai et~al.(2025)Bai, Chen, Liu, Wang, Ge, Song, Dang, Wang, Wang,
  Tang, Zhong, Zhu, Yang, Li, Wan, Wang, Ding, Fu, Xu, Ye, Zhang, Xie, Cheng,
  Zhang, Yang, Xu, and Lin}]{qwen2_5}
Shuai Bai, Keqin Chen, Xuejing Liu, Jialin Wang, Wenbin Ge, Sibo Song, Kai
  Dang, Peng Wang, Shijie Wang, Jun Tang, Humen Zhong, Yuanzhi Zhu, Mingkun
  Yang, Zhaohai Li, Jianqiang Wan, Pengfei Wang, Wei Ding, Zheren Fu, Yiheng
  Xu, and 8 others. 2025.
\newblock \href {https://arxiv.org/abs/2502.13923} {Qwen2.5-vl technical
  report}.
\newblock \emph{Preprint}, arXiv:2502.13923.

\bibitem[{{Cambridge International Education}(2024)}]{cambridge2022alevel}
{Cambridge International Education}. 2024.
\newblock {Cambridge International AS \& A Level Mathematics 9709/12, Paper 1
  Pure Mathematics 1, Mark Scheme, May/June 2024}.
\newblock {Published mark scheme; document code: 9709\_s24\_ms\_12}.

\bibitem[{Chiang et~al.(2024)Chiang, Chen, Kuan, Yang, and
  Lee}]{chiang2024assignment}
Cheng-Han Chiang, Wei-Chih Chen, Chun-Yi Kuan, Chienchou Yang, and Hung-yi Lee.
  2024.
\newblock \href {https://doi.org/10.18653/v1/2024.emnlp-main.146} {Large
  language model as an assignment evaluator: Insights, feedback, and challenges
  in a 1000+ student course}.
\newblock In \emph{Proceedings of the 2024 Conference on Empirical Methods in
  Natural Language Processing}, pages 2489--2513, Miami, Florida, USA.
  Association for Computational Linguistics.

\bibitem[{Choi et~al.(2026)Choi, Kang, Seong, and Huang}]{choi2026exploring}
Hongseok Choi, Myeong-Cheol Kang, Jin Seong, and Jin-Xia Huang. 2026.
\newblock \href {https://doi.org/10.1007/s10618-026-01193-z} {Exploring
  zero-shot essay scoring: from feature-based to llm-based approaches}.
\newblock \emph{Data Mining and Knowledge Discovery}, 40:35.

\bibitem[{{College Board}(2023)}]{collegeboard2023ap}
{College Board}. 2023.
\newblock \href
  {https://apcentral.collegeboard.org/media/pdf/ap23-sg-calculus-bc.pdf} {{AP
  Calculus BC 2023 Scoring Guidelines}}.
\newblock {Free-response scoring guidelines}.

\bibitem[{Comanici et~al.(2025)Comanici, Bieber, Schaekermann, Pasupat,
  Sachdeva, Dhillon, Blistein, Ram, Zhang, Rosen, Marris, Petulla, Gaffney,
  Aharoni, Lintz, Pais, Jacobsson, Szpektor, Jiang, Haridasan, Omran, Saunshi,
  Bahri, Mishra, Chu, Boyd, Hekman, Parisi, Zhang, Kawintiranon, Bedrax-Weiss,
  Wang, Xu, Purkiss, Mendlovic, Deutel, Nguyen, Langley, Korn, Rossazza, Ramé,
  Waghmare, Miller, Byrd, Sheshan, Hadsell, Bhardwaj, Janus, Rissa, Horgan,
  Abdagic, Belenki, Allingham, Singh, Guidroz, Srinivasan, Schmit, Chiafullo,
  Elisseeff, Jha, Kolhar, Berrada, Ding, Si, Mallick, Och, Erell, Ni, Latkar,
  Yang, Sirkovic, Feng, Leland, Hornung, Wu, Blundell, Alvari, Huang, Yip,
  Deur, Liu, Surita, Duque, Damen, Jia, Guez, Mircea, Sinha, Magni, Stradomski,
  Marian, Galić, Chen, Husain, Singhal, Grewe, Aubet, Song, Blanco, Rechis,
  Ho, Munoz, Zheng, Hamrick, Mather, Taitelbaum, Rutherford, Lei, Chen, Shukla,
  Moreira, Doi, Isik, Shabat, Rogozińska, Kolipaka, Chang, Vušak,
  Venkatachary, Noghabi, Bharti, Jun, Zaks, Green, Challagundla, Wong,
  Mohammad, Hirsch, Cheng, Naim, Proleev, Vincent, Singh, Krikun, Krishnan,
  Ghahramani, Atias, Aggarwal, Kirov, Vytiniotis, Koh, Chronopoulou, Dogra,
  Ion, Tyen, Lee, Weissenberger, Strohman, Balakrishna, Rae, Velic,
  de~Liedekerke, Elyada, Yuan, Liu, Shani, Kishchenko, Alessio, Li, Song, Kwei,
  Jankowski, Pappu, Namiki, Ma, Tripuraneni, Cherry, Ikonomidis, Ling, Ji,
  Westberg, Wright, Yu, Parkinson, Ramaswamy, Connor, Yeganeh, Grover,
  Kenwright, Litchev, Apps, Tomala, Halim, Castro-Ros, Li, Boral, Sho, Yarom,
  Malmi, Klinghoffer, Lin, Ansell, S, Zhao, Zuo, Santoro, Cheng, Demmessie,
  Liu, Brichtova, Culp, Braun, Graur, Ng, Mehta, Phillips, Sundberg, Godbole,
  Liu, Katariya, Rim, Seyedhosseini, Ammirati, Valfridsson, Malihi, Knight,
  Toor, Lampe, Ittycheriah, Chiang, Yeung, Fréchette, Rao, Wang, Srivastava,
  Zhang, Rhodes, Brand, Weesner, Figotin, Gimeno, Fellinger, Marcenac, Leal,
  Marcus, Cotruta, Cabrera, Luo, Garrette, Axelrod, Baltateanu, Barker, Chen,
  Toma, Ingram, Riesa, Kulkarni, Zhang, Liu, Wang, Polacek, Wu, Hui, Reyes, Su,
  Barnes, Malhi, Siddiqui, Feng, Damaschin, Pighin, Steiner, Yang, Boppana,
  Ivanov, Kandoor, Shah, Mujika, Huang, Choquette-Choo, Patel, Yu, Creswell,
  Jerry, Liu, Barros, Razeghi, Roy, Culliton, Xiong, Pan, Strohmann, Powell,
  Seal, DeCarlo, Shyam, Katircioglu, Wang, Hardin, Odisho, Broder, Chang, Nair,
  Shtefan, O'Brien, Agarwal, Potluri, Goyal, Jhindal, Thakur, Stuken, Lyon,
  Toutanova, Feng, Wu, Horn, Wang, Cullum, Taubman, Shrivastava, Shi,
  Tomlinson, Patel, Tu, Oflazer, Pongetti, Yang, Taïga, Perot, Pierse, Han,
  Drori, Iturrate, Chakrabarti, Yeung, Dopson, ting Chen, Kulshreshtha, Guo,
  Pham, Schuster, Chen, Polozov, Xing, Zhou, Kacham, Kukliansky, Miech,
  Yaroshenko, Chi, Douglas, Fei, Blondel, Myla, Madmoni, Wu, Keysers, Kjems,
  Albuquerque, Yu, D'sa, Plantan, Ionescu, Elias, Gupta, Vuyyuru, Alcober,
  Zhou, Ji, Hartmann, Puttagunta, Song, Amid, Stefanoiu, Lee, Pucciarelli,
  Wang, Raul, Petrov, Tian, Anklin, Nti, Gomes, Schumacher, Vesom,
  Panagopoulos, Bousmalis, Andor, Jacob, Zhang, Rosgen, Kecman, Tung, Belias,
  Goodman, Covington, Wieder, Saxena, Davoodi, Huang, Maddineni, Roulet,
  Campbell-Ajala, Sessa, Xintian, Wu, Lai, Collins, Haig, Sakenas, Xu,
  Giustina, Shafey, Charoenpanit, Garg, Ainslie, Severson, Arenas, Pathak,
  Rajayogam, Feng, Bakker, Li, Wichers, Rogers, Geng, Li, Jagerman, Jia,
  Olmert, Sharon, Mauger, Mariserla, Ma, Mohabey, Kim, Andreev, Pollom, Love,
  Jain, Agrawal, Schroecker, Fortin, Warmuth, Liu, Leach, Blok, Girirajan,
  Aharoni, Uria, Sozanschi, Goldberg, Ionita, Ribeiro, Zlocha, Birodkar,
  Lachgar, Yuan, Choudhury, Ginsberg, Zheng, Dibb, Graves, Lokhande, Rasskin,
  Muraru, Quick, Tata, Sermanet, Chawla, Karo, Wang, Zhang, Keller, Dragan, Su,
  Chou, Liu, Tao, Prabhakara, Wilson, Liu, Wang, Evans, Du, Castaño, Prasad,
  Mahdy, Gerlach, Reid, Kahn, Zait, Pillai, Ulrich, Wang, Wassenberg, Farkash,
  Yalasangi, Wang, Bauza, Bucher, Liu, Yan, Leung, Sindhwani, Barnes, Singh,
  Jurin, Chang, Bhumihar, Eiger, Citovsky, Withbroe, Li, Xue, Santo, Stoyanov,
  Raimond, Zheng, Gao, Listík, Kwasiborski, Saputro, Ozturel, Mallya,
  Majmundar, West, Caron, Wei, Castrejon, Vikram, Ramachandran, Dhawan, Park,
  Smoot, van~den Driessche, Blau, Malik, Liang, Hirsch, dos Santos, Weinstein,
  van~den Oord, Lall, FitzGerald, Jiang, Yang, Webster, Elqursh, Pope, Rotival,
  Raposo, Zhu, Dean, Alabed, Tran, Gupta, Gleicher, Austin, Rosseel, Umekar,
  Das, Sun, Chen, Misiunas, Zhou, Di, Loo, Newlan, Li, Ramasesh, Xu, Chen,
  Gandhe, Soricut, Gupta, Hu, El-Sayed, Garcia, Brusilovsky, Chen, Bolt, Huang,
  Gurney, Zhang, Pritzel, Wilkiewicz, Seybold, Shamanna, Fischer, Dean, Gill,
  Mcilroy, Bhowmick, Selier, Yang, Cheng, Magay, Tan, Varma, Walder, Kocisky,
  Nakashima, Natsev, Kwong, Gog, Zhang, Dieleman, Jimma, Ryabtsev, Brahma,
  Steiner, Du, Žužul, Žanić, Raghavachari, Gierke, Zheng, Petrova, Dauphin,
  Liu, Kessler, Hand, Duvarney, Kim, Lee, Hussenot, Hui, Smith, Jain, Xia,
  Tomar, Amiri, Phan, Fuchs, Weyand, Tomasev, Cordell, Liu, Mallinson, Joshi,
  Crawford, Suggala, Chien, Fernando, Sanchez-Vargas, Williams, Crone, Luo,
  Karpov, Shan, Thurk, Strudel, Voigtlaender, Patil, Dozat, Khodaei, Singla,
  Ambroszczyk, Wu, Chang, Roark, Hegde, Ding, Filos, Wu, Pinto, Liu, Khanna,
  Pandey, Mcloughlin, Li, Haves, Zhou, Buchatskaya, Leal, de~Boursac, Akazawa,
  Anderson, Chen, Somandepalli, Liang, Goenka, Winkler, Grushetsky, Ding,
  Smith, Ye, Pont-Tuset, Li, Li, Golany, Wegner, Jiang, Barak, Shangguan,
  Vértes, Wong, Bornschein, Tudor, Bevilacqua, Schaul, Rawat, Zhao, Axiotis,
  Meng, McLean, Lai, Beattie, Kushman, Liu, Kutzman, Lang, Ye, Netrapalli,
  Mishra, Khan, Goel, Willoughby, Tian, Zhuang, Chen, Tsai, Kementsietsidis,
  Khare, Keeling, Xu, Waters, Altché, Popat, Mittal, Saxton, Badawy, Mathieu,
  Zheng, Zhou, Ranka, Shin, Duan, Salimans, Mihailescu, Shaham, Chang, Assael,
  Dikkala, Izzard, Cohen-Addad, Graves, Feinberg, Chung, Strouse, Karmon,
  Sharifzadeh, Ashwood, Pham, Blanton, Vasiloff, Barber, Geller, Zhou, Zubach,
  Huang, Zhang, Gupta, Young, Proskurnia, Votel, Gabeur, Barcik, Tripathi, Yu,
  Yan, Changpinyo, Pavetić, Coyle, Fujii, Mendez, Zhou, Rajamani, Hechtman,
  Cao, Juan, Tan, Dalibard, Du, Clay, Yao, Jia, Vijaykumar, Zhou, Bai, Hung,
  Pecht, Todorov, Khadke, Gupta, Lahoti, Autef, Duddu, Lee-Thorp, Bykovsky,
  Misiunas, Flennerhag, Thangaraj, McGiffin, Nado, Kunesch, Noever, Hertz,
  Liang, Stone, Palmer, Daruki, Pramanik, Põder, Kyker, Khan, Sluzhaev,
  Ritter, Ruderman, Zhou, Nagpal, Vodrahalli, Necula, Barham, Pavlick,
  Hartford, Shafran, Zhao, Mikuła, Eccles, Shimokawa, Garg, Vilnis, Chen,
  Shumailov, Lee, Abdelhamed, Xie, Cohen, Hlavnova, Malkin, Sitawarin, Lottes,
  Coquinot, Yu, Kumar, Zhang, Mahendru, Ahmed, Martens, Chen, Boag, Peng,
  Devin, Klimovskiy, Phuong, Vainstein, Xie, Ramabhadran, Howard, Yu, Goswami,
  Cui, Shleifer, Pinto, Yeh, Yang, Javanmardi, Ethier, Lee, Orbay, Kotecha,
  Bromberg, Shaw, Thornton, Rosenthal, Gu, Thomas, Gemp, Ayyar, Ushio, Selvan,
  Wee, Liu, Majzoubi, Yu, Abernethy, Liechty, Pan, Nguyen, Qiong, Hu, Perrin,
  Arora, Pitler, Wang, Shivakumar, Prost, Limonchik, Wang, Gao, Cour, Buch,
  Gui, Ivanova, Neubeck, Chan, Kim, Chen, Goyal, Chung, Liu, Su, Petrushkina,
  Shen, Joulin, Xu, Lin, Kulizhskaya, Chelba, Vasudevan, Collins, Bashlovkina,
  Lu, Fritz, Park, Zhou, Su, Tanburn, Sushkov, Rasquinha, Li, Prendki, Li, LV,
  Sharma, Fitoussi, Huang, Dai, Dao, Burrows, Prior, Qin, Pundak, Sjoesund,
  Khurshudov, Zhu, Webson, Kemp, Tan, Agrawal, Sargsyan, Cheng, Stephan,
  Kwiatkowski, Reid, Byravan, Michaely, Heess, Zhou, Goenka, Carpenter,
  Levskaya, Wang, Roberts, Leblond, Chikkerur, Ginzburg, Chang, Riachi,
  Chuqiao, Xu, Borsos, Pliskin, Pawar, Lustman, Kirkwood, Anand, Chaudhary,
  Kalb, Milan, Augenstein, Goldie, Prince, Raman, Sun, Xia, Cohen, Huo, Camp,
  Ellis, Zilka, Torres, Patel, Arora, Chan, Adler, Ayoub, Liang, Jamil, Jiang,
  Baumgartner, Sun, Karov, Akulov, Zheng, Cai, Fantacci, Rubin, Acha, Wang,
  D'Souza, Sathyanarayana, Dai, Rowe, Simanovsky, Goldman, Kuang, Pan,
  Rosenberg, Rojas-Esponda, Dutta, Zeng, Jurenka, Farquhar, Bansal, Iqbal,
  Roelofs, Joung, Beak, Ryu, Poplin, Wu, Alayrac, Buthpitiya, Ronneberger,
  Habtegebriel, Li, Cavallaro, Wei, Bensky, Denk, Ganapathy, Stanway, Joshi,
  Bertolini, Lo, Ma, Charles, Sampemane, Sahni, Chen, Askham, Gaddy, Young,
  Tan, Eyal, Bražinskas, Zhong, Wu, Epstein, Bailey, Hard, Lee, Goldshtein,
  Ruiz, Badawi, Lochbrunner, Kearns, Brown, Pardo, Weber, Yang, Jiang, Akin,
  Fu, Wainwright, Zou, Gaba, Manzagol, Kan, Song, Zainullina, Lin, Ko,
  Deshmukh, Jindal, Svensson, Tyam, Zhao, Kaeser-Chen, Baird, Moradi, Hall,
  Guo, Tsang, Liang, Pereira, Ganesh, Korotkov, Adamek, Thiagarajan, Tran,
  Chen, Tar, Jain, Dasgupta, Bilal, Reitter, Zhao, Vezzani, Gehman, Mehta,
  Beltrone, Dotiwalla, Guadarrama, Abbas, Karp, Georgiev, Ferng, Brockschmidt,
  Peng, Hirnschall, Verma, Bi, Xiao, Dabush, Xu, Wallis, Parker, Wang, Xu,
  Safarli, Tewari, Zhang, Kim, Gesmundo, Thomas, Levi, Chowdhury, Rao, Garst,
  Conway-Rahman, Ran, McKinney, Xiao, Yu, Agrawal, Stjerngren, Ionescu, Chen,
  Sharma, Chiu, Liu, Franko, Sanford, Cai, Michel, Ganapathy, Labanowski,
  Garrett, Vargas, Sun, Gale, Buschmann, Desjardins, Ghelani, Jain, Verma,
  Asawaroengchai, Eisenschlos, Harlalka, Kazawa, Metzler, Howland, Jian, Ades,
  Shah, Gangwani, Lee, Ring, Hernandez, Reich, Sinha, Sathe, Kovac, Gill,
  Kannan, D'olimpio, Sevenich, Whang, Kim, Sim, Chen, Zhang, Lall, Matias, Jia,
  Friesen, Nasso, Thapliyal, Perozzi, Yu, Shekhawat, Huda, Grabowski, Wang,
  Sreevatsa, Dib, Hassen, Schuh, Milutinovic, Welty, Quinn, Shah, Wang,
  Barth-Maron, Frye, Axelsson, Zhu, Ma, Giannoumis, Sedghi, Ye, Luan, Aydin,
  Chandra, Sampathkumar, Huang, Lavrenko, Eleryan, Hong, Hansen, Carthy,
  Samanta, Ćevid, Wang, Li, Voznesensky, Hoffman, Terzis, Sehwag, Fidel, He,
  Cai, He, Feng, Nikoltchev, Phatale, Chase, Lawton, Zhang, Ouyang, Tragut,
  Manshadi, Narayanan, Shen, Gao, Bolukbasi, Roy, Li, Golovin, Panait, Qin,
  Han, Anthony, Kudugunta, Patraucean, Ray, Chen, Yang, Bhatia, Talluri,
  Morris, Ražnatović, Brownfield, An, Peng, Kane, Zheng, Duduta, Kessinger,
  Noraky, Liu, Rong, Veličković, Rush, Goldin, Wei, Garlapati, Pantofaru,
  Kwon, Ni, Noland, Trapani, Beaufays, Roy, Chow, Turker, Cideron, Mei, Clark,
  Dou, Bošnjak, Leith, Du, Yazdanbakhsh, Nasr, Kwak, Sheth, Kaskasoli, Anand,
  Lakshminarayanan, Jerome, Bieber, Chu, Senges, Shen, Sridhar, Ndebele,
  Beyret, Mohamed, Chen, Freitag, Guo, Liu, Roit, Chen, Yan, Stone, Co-Reyes,
  Cole, Scellato, Azizi, Hashemi, Jin, Iyer, Valentine, György, Ahuja, Diaz,
  Lee, Clement, Kong, Garmon, Watts, Bhatia, Gupta, Miecnikowski, Vallet, Taly,
  Loper, Joshi, Atwood, Chick, Collier, Iliopoulos, Trostle, Gunel,
  Leal-Cavazos, Hrafnkelsson, Guzman, Ju, Forbes, Emond, Chauhan, Caine, Xiao,
  Zeng, Moufarek, Murphy, Meng, Gupta, Riedel, Das, Lawal, Narayan, Sosea,
  Swirhun, Friso, Neyshabur, Lu, Girgin, Wunder, Yvinec, Pyne, Carbune,
  Rijhwani, Guo, Doshi, Briukhov, Bain, Hitron, Wang, Gupta, Chen, Du, Zhang,
  Shah, Akula, Dylla, Kachra, Kuo, Zou, Wang, Xu, Zhu, Snyder, Menon, Firat,
  Mordatch, Yuan, Ponomareva, Blevins, Moore, Wang, Chen, Scholz, Dwornik, Lin,
  Li, Antognini, I, Song, Miller, Kalra, Raveret, Akerlund, Wu, Nystrom,
  Godbole, Liu, DeBalsi, Zhao, Liu, Caciularu, Lax, Khandelwal, Langston,
  Bailey, Lattanzi, Wang, Kovelamudi, Mondal, Guruganesh, Hua, Roval,
  Wesołowski, Ingale, Halcrow, Sohn, Angermueller, Raad, Stickgold, Lu, Kosik,
  Xie, Lillicrap, Huang, Zhang, Paulus, Farabet, Wertheim, Wang, Joshi, ling
  Ko, Wu, Agrawal, Lin, Sheng, Sung, Breland-King, Butterfield, Gawde, Singh,
  Zhang, Apte, Shetty, Hutter, Li, Salesky, Lebron, Kanerva, Paganini, Nguyen,
  Vallu, Peter, Velury, Kao, Hoover, Bortsova, Bishop, Jakobovits, Agostini,
  Agarwal, Liu, Kwong, Tavakkol, Bica, Greve, GP, Marcus, Hou, Duerig,
  Moroshko, Lacey, Davis, Amelot, Wang, Kim, Strinopoulos, Wan, Lan, Krishnan,
  Tang, Humphreys, Bai, Shtacher, Machado, Pang, Burke, Liu, Aravamudhan, Song,
  Hirst, Singh, Jou, Bai, Piccinno, Fu, Alazard, Meiri, Winter, Chen, Zhang,
  Heitkaemper, Lambert, Lee, Frömmgen, Rogulenko, Nair, Niemczyk, Bulyenov,
  Xu, Shemtov, Zadimoghaddam, Toropov, Wirth, Dai, Gollapudi, Zheng, Kurakin,
  Lee, Bullard, Serrano, Balazevic, Li, Schalkwyk, Murphy, Zhang, Sequeira,
  Datta, Agrawal, Sutton, Attaluri, Chiang, Farhan, Thornton, Lin, Choma,
  Nguyen, Dasgupta, Robinson, Comşa, Riley, Pillai, Mustafa, Golan, Zandieh,
  Lespiau, Porter, Ross, Rajayogam, Agarwal, Venugopalan, Shahriari, Yan, Xu,
  Tobin, Dubov, Shi, Recasens, Kovsharov, Borgeaud, Dery, Vasanth, Gribovskaya,
  Qiu, Mahdieh, Skut, Nielsen, Zheng, Yu, Bostock, Gupta, Archer, Rawles,
  Davies, Svyatkovskiy, Tsai, Halpern, Reisswig, Wydrowski, Chang, Puigcerver,
  Taege, Li, Schnider, Li, Dena, Xu, Telang, Shi, Zen, Kastner, Ko,
  Subramaniam, Kumar, Blois, Dai, Wieting, Lu, Zeldes, Xie, Hauth, Ţifrea, Li,
  El-Husseini, Abolafia, Zhou, Ding, Ghalebikesabi, Guía, Maksai, Ágoston
  Weisz, Arik, Sukhanov, Świetlik, Jia, Yu, Wang, Brand, Bloxwich, Kirmani,
  Chen, Go, Sprechmann, Kannen, Carin, Sandhu, Edkins, Nooteboom, Gupta,
  Maggiore, Azizi, Pritch, Yin, Gupta, Tarlow, Smith, Ivanov, Babaeizadeh,
  Goel, Kambala, Chu, Kastelic, Liu, Soltau, Stone, Agrawal, Kim, Soparkar,
  Tadepalli, Bunyan, Soh, Kannan, Kim, Chen, Halumi, Roy, Wang, Sercinoglu,
  Gibson, Bhatnagar, Sano, von Dincklage, Ren, Mitrevski, Olšák, She,
  Doersch, Jilei, Wang, Liu, Tan, Yakar, Warkentin, Ramirez, Lebsack, Dillon,
  Mathews, Cobley, Wu, Chen, Simon, Nath, Sainath, Bendebury, Julian,
  Mankalale, Ćurko, Zacchello, Brown, Sodhia, Howard, Caelles, Gupta, Evans,
  Bulanova, Katzen, Goldenberg, Tsitsulin, Stanton, Schillings, Kovalev, Fry,
  Shah, Lin, Upadhyay, Li, Radpour, Maggioni, Xiong, Haas, Brennan, Kamath,
  Savinov, Nagrani, Yacovone, Kappedal, Andriopoulos, Lao, Li, Rozhdestvenskiy,
  Hashimoto, Audibert, Austin, Rodriguez, Ruoss, Honke, Karkhanis, Xiong, Wei,
  Huang, Leng, Premachandran, Bileschi, Evangelopoulos, Mensink, Pavagadhi,
  Teplyashin, Chang, Xue, Tanzer, Goldman, Patel, Li, Wiesner, Zheng,
  Stewart-Binks, Han, Li, Luo, Lenc, Lučić, Xue, Mullins, Guseynov, Chang,
  Galatzer-Levy, Zhang, Bingham, Hu, Hartman, Ma, Griffith, Irpan, Radebaugh,
  Yue, Fan, Ungureanu, Sorokin, Teufel, Li, Anil, Paparas, Wang, Lin, Peng,
  Shum, Petrovic, Brady, Nguyen, Macherey, Li, Singh, Yenugula, Iinuma, Chen,
  Kopparapu, Stern, Dave, Thekkath, Perot, Kumar, Li, Xiao, Bilotti, Bateni,
  Noble, Lee, Vázquez-Reina, Salazar, Yang, Wang, Gruzewska, Rao, Raghuram,
  Xu, Ben-David, Mei, Dalmia, Zhang, Liu, Bansal, Pankov, Schwarcz, Burns,
  Chan, Sanghai, Liang, Liang, He, Stuart, Narayanan, Zhu, Frank, Fatemi,
  Sabne, Lang, Bhattacharya, Settle, Wang, McMahan, Tacchetti, Soares, Hadian,
  Cabi, Chung, Putikhin, Li, Chen, Tarango, Michalewski, Kazemi, Masoom,
  Sheftel, Shivanna, Vadali, Comanescu, Reid, Moore, Neelakantan, Sander,
  Herzig, Rosenberg, Dehghani, Choi, Fink, Hayes, Ge, Weng, Ho, Karro, Krishna,
  Thiet, Skerry-Ryan, Eppens, Andreetto, Sarma, Bonacina, Ayan, Nawhal, Shan,
  Dusenberry, Thakoor, Gubbi, Nguyen, Tsarfaty, Albanie, Mitrović, Gandhi,
  Chen, Epasto, Stephanov, Jin, Gehman, Amini, Weber, Behbahani, Xu, Allamanis,
  Chen, Ott, Sha, Jastrzebski, Qi, Greene, Wu, Toki, Vlasic, Shapiro,
  Kotikalapudi, Shen, Saeki, Xie, Cassirer, Bharadwaj, Kiyono, Bhojanapalli,
  Rosenfeld, Ritter, Mao, Oliveira, Egyed, Bandemer, Parisotto, Kinoshita,
  Pluto, Maniatis, Li, Guo, Ghiasi, Tarbouriech, Chatterjee, Jin, Katrina, Xu,
  Palomaki, Arnold, Sewak, Piccinini, Sharma, Albrecht, Purser-haskell,
  Vaswani, Chen, Wisniewski, Cao, Aslanides, Phu, Sieb, Agubuzu, Zheng, Sohn,
  Selvi, Andreassen, Subudhi, Eruvbetine, Woodman, Mery, Krause, Ren, Ma, Luo,
  Chen, Fan, Griffiths, Schuler, Li, Zhang, Sarr, Luo, Patana, Watson,
  Naboulsi, Collins, Sidhwani, Hoogeboom, Silver, Caveness, Zhao, Rodriguez,
  Deines, Bai, Griffin, Tagliasacchi, Xue, Babbula, Pang, Ding, Shen, Peake,
  Crocker, Raghvendra, Swisher, Han, Singh, Wu, Pchelin, Munkhdalai, Alon,
  Bacon, Robles, Bulian, Johnson, Powell, Ferreira, Li, Benzing, Velimirović,
  Soyer, Kong, Tony, Nguyên, Yang, Liu, van Amersfoort, Gillick, Sun,
  Rauschmayr, Zhang, Zhan, Zhou, Frolov, Yang, Vnukov, Rouillard, Li, Mandhane,
  Fallen, Venkataraman, Hu, Brennan, Lee, Chang, Sundermeyer, Pan, Ke, Tong,
  Fabrikant, Bono, Gu, Foley, Mao, Delakis, Bhaswar, Frostig, Li, Zipori, Hope,
  Kozlova, Mishra, Djolonga, Schiff, Merey, Briakou, Morgan, Wan, Hassidim,
  Skerry-Ryan, Sengupta, Jasarevic, Kallakuri, Kunkle, Brennan, Lieber,
  Mansoor, Walker, Zhang, Xie, Žužić, Chukwuka, Druinsky, Cho, Yao, Naeem,
  Butt, Kim, Jia, Jordan, Lelkes, Kurzeja, Wang, Zhao, Over, Chakladar,
  Prasetya, Jha, Ganapathy, Cong, Shroff, Saroufim, Miryoosefi, Hammad, Nasir,
  Xi, Gao, Maeng, Hora, Cheng, Haghani, Lewenberg, Lu, Matysiak, Raisinghani,
  Wang, Baugher, Sukthankar, Giang, Schultz, Fiedel, Chen, Lee, Dey, Zheng,
  Paul, Smith, Ly, Wang, Bansal, Perz, Ricco, Blank, Keshava, Sharma, Chow,
  Lad, Jalan, Osindero, Swanson, Scott, Ilić, Li, Jonnalagadda, Soudagar,
  Xiong, Batsaikhan, Jarrett, Kumar, Shah, Lawlor, Waters, Graham, May, Ramos,
  Lefdal, Cankara, Cano, O'Donoghue, Borovik, Liu, Grimstad, Alnahlawi,
  Tsihlas, Hudson, Grigorev, Jia, Huang, Igwe, Lebedev, Tang, Krivokon, Garcia,
  Tan, Jia, Stys, Vashishth, Liang, Venkatraman, Gu, Kementsietsidis, Zhu,
  Jung, Bai, Hosseini, Ahmed, Gupta, Yuan, Ashraf, Nigam, Vasudevan, Awasthi,
  Gilady, Mariet, Eskander, Li, Hu, Garrido, Schlattner, Zhang, Saxena, Dević,
  Muralidharan, Murthy, Zhou, Choi, Wongpanich, Wang, Shah, Xu, Huang, Spencer,
  Chen, Cohan, Wang, Tompson, Wu, Haroun, Li, Huergo, Yang, Yin, Wendt,
  Bendersky, Chaabouni, Snaider, Ferret, Jindal, Thompson, Xue, Bishop, Phal,
  Sharma, Sung, Radhakrishnan, Shomrat, Ingle, Vij, Gilmer, Istin, Sobell, Lu,
  Nottage, Sadigh, Willcock, Zhang, Xu, Brown, Lee, Wang, Zhu, Tay, Kim,
  Gutierrez, Sharma, Xian, Seo, Cui, Pochernina, Baetu, Jastrzębski, Ly,
  Elhawaty, Suh, Sezener, Wang, Yuen, Tucker, Cai, Yang, Wang, Muzio, Qian,
  Yoo, Lockhart, McKee, Guo, Mehrotra, Mendonça, Mehta, Ben, Tekur, Mu, Zhu,
  Krakovna, Lee, Maschinot, Cevey, Choe, Bai, Srinivasan, Gasaway, Young,
  Siegler, Holtmann-Rice, Piratla, Baumli, Yogev, Hofer, van Hasselt, Grant,
  Chervonyi, Silver, Hogue, Agarwal, Wang, Singh, Flynn, Lipschultz, David,
  Bellot, Yang, Le, Graziano, Olszewska, Hui, Maurya, Parotsidis, Chen,
  Oguntebi, Kelley, Baddepudi, Mauerer, Shaw, Siegman, Yang, Shetty, Roy, Song,
  Stokowiec, Burnell, Savant, Busa-Fekete, Miao, Ghosh, MacDermed, Lippe,
  Dektiarev, Behrman, Mentzer, Nguyen, Wei, Verma, Knutsen, Dasari, Yan,
  Mitrichev, Wang, Shejwalkar, Austin, Sunkara, Potti, Virin, Wright, Liu,
  Riva, Pot, Kochanski, Le, Balasubramaniam, Dhar, Liao, Bloniarz, Shukla,
  Cole, Lee, Zhang, Kafle, Vashishtha, Mahmoudieh, Chen, Hoffmann, Srinivasan,
  Lago, Shalom, Wang, Elabd, Sharma, Oh, Kothawade, Le, Monteiro, Yang,
  Alarakyia, Geirhos, Mincu, Garnes, Kobayashi, Mariooryad, Krasowiak, Zhixin,
  Lai, Mourad, Wang, Bu, Aharoni, Chen, Goyal, Zubov, Bapna, Dabir, Kothari,
  Lamerigts, Cao, Shar, Yew, Kulkarni, Mahaarachchi, Joshi, Zhu, Lichtarge,
  Zhou, Muckenhirn, Selo, Vinyals, Chen, Brohan, Mehta, Cogan, Wang, Geri, Ko,
  Chen, Viola, Shivam, Wang, Elish, Popa, Pereira, Liu, Koster, Kim, Zhang,
  Ebrahimi, Talukdar, Zheng, Poklukar, Mikhalap, Johnson, Vijayakumar,
  Omernick, Dibb, Dubey, Hu, Suman, Aggarwal, Kornakov, Xia, Lowe, Kolganov,
  Xiao, Nikolaev, Hemingray, Li, Iljazi, Rybiński, Sandhu, Lu, Luong,
  Jenatton, Govindaraj, Hui, Li, Dulac-Arnold, Park, Wang, Modi, Pouget-Abadie,
  Greller, Gupta, Berry, Ramachandran, Xie, McCafferty, Wang, Gupta, Lim,
  Bratanič, Brock, Akolzin, Sproch, Karliner, Kim, Goedeckemeyer, Shazeer,
  Schmid, Calandriello, Bhatia, Choromanski, Montgomery, Dua, Ramalho, King,
  Gao, Nguyen, Lindner, Pitta, Johnson, Salama, Ardila, Han, Farnese, Odoom,
  Wang, Ding, Rink, Smith, Lehri, Cohen, Vats, He, Gopavarapu, Paszke, Patel,
  Gansbeke, Loher, Castro, Voitovich, von Glehn, George, Niklaus, Eaton-Rosen,
  Rakićević, Jue, Perel, Zhang, Bahat, Pouget, Xing, Huot, Shenoy, Bos,
  Coriou, Richter, Noy, Wang, Ontanon, Qin, Makarchuk, Hassabis, Li, Sharma,
  Venkatesan, Kemaev, Daniel, Huang, Shah, Ponce, Warren, Chen, Faruqui, Wu,
  Andačić, Payrits, McDuff, Hume, Cao, Tessler, Wang, Wang, Rendulic,
  Agustsson, Johnson, Lando, Howard, Padmanabhan, Daswani, Banino, Kilgore,
  Heek, Ji, Caceres, Li, Kassner, Vlaskin, Liu, Grills, Hou, Sukkerd, Cheon,
  Shetty, Markeeva, Stanczyk, Iyer, Gong, Gao, Gopalakrishnan, Blyth, Reynolds,
  Bhoopchand, Bilenko, Gharibian, Zayats, Faust, Singh, Ma, Jiao,
  Vijayanarasimhan, Aroyo, Yadav, Chakera, Kakarla, Meshram, Gregor, Botea,
  Senter, Jia, Kovacs, Sharma, Baur, Kang, He, Zhuo, Kostelac, Laish, Peng,
  O'Bryan, Kasenberg, Rao, Leurent, Zhang, Stevens, Salazar, Zhang, Lobov,
  Walker, Porter, Redshaw, Ke, Rao, Lee, Lam, Moffitt, Kim, Qiao, Koo, Dadashi,
  Song, Sundararajan, Xu, Kawamoto, Zhong, Barbu, Reddy, Verzetti, Li,
  Papamakarios, Klimczak-Plucińska, Cassin, Kavukcuoglu, Swavely, Vaucher,
  Zhao, Hemsley, Tschannen, Ge, Menghani, Yu, Ha, He, Wu, Song, Sterneck,
  Zinke, Calian, Marsden, Ruiz, Hessel, Gueta, Lee, Farris, Gupta, Li, Saleh,
  Misra, Xiao, Mendolicchio, Buttimore, Krayvanova, Nayakanti, Wiethoff, Pande,
  Mirhoseini, Lao, Liu, Hua, Chen, Malkov, Kalashnikov, Gupta, Audhkhasi, Zhai,
  Kopalle, Jain, Ofek, Meyer, Baatarsukh, Strejček, Qian, Freedman, Figueira,
  Sokolik, Bachem, Lin, Kharrat, Hidey, Xu, Duan, Li, Ersoy, Everett, Cen,
  Santamaria-Fernandez, Taubenfeld, Mackinnon, Deng, Zablotskaia, Viswanadha,
  Goel, Yates, Deng, Choy, Chen, Sinha, Mossin, Wang, Szlam, Hao, Rubenstein,
  Toksoz-Exley, Aperghis, Zhong, Ahn, Isard, Lacombe, Luisier, Anastasiou,
  Kalley, Prabhu, Dunleavy, Bijwadia, Mao-Jones, Chen, Pasumarthi, Wood,
  Dostmohamed, Hurley, Simsa, Parrish, Pajarskas, Harvey, Skopek, Kochinski,
  Rey, Rieser, Zhou, Lee, Acharya, Li, Jiang, Zhang, Gipson, Mahintorabi,
  Gelmi, Khajehnouri, Yeh, Lee, Matthey, Baker, Pham, Fu, Pak, Gupta,
  Vasconcelos, Sadovsky, Walker, Hsiao, Zochbauer, Marzoca, Velan, Zeng,
  Baechler, Driess, Jain, Huang, Tao, Maggs, Levine, Schneider, Gemzer, Petit,
  Han, Fisher, Zelle, Biles, Ie, Fadeeva, Liu, Franco, Collister, Zhang, Wang,
  Zhao, Kieliger, Shuster, Zhu, Gong, Chan, Sun, Basu, Zimmermann, Hayes,
  Bapna, Snoek, Yang, Datta, Abdallah, Kilgour, Li, Mah, Jun, Rivière,
  Karmarkar, Spalink, Huang, Gonzalez, Tran, Nowak, Palowitch, Chadwick,
  Talius, Mehta, Sellam, Fränken, Nicosia, He, Kini, Amos, Basu, Jobe, Shaw,
  Xu, Evans, Ikeda, Yan, Jin, Wang, Yadav, Labzovsky, Sampath, Ma, Schumann,
  Siddhant, Shah, Youssef, Agarwal, Dabney, Tonioni, Ambar, Li, Guyon, Li,
  Soergel, Fang, Karadzhov, Udrescu, Trinh, Raunak, Noury, Guo, Gupta,
  Finkelstein, Petek, Liang, Billock, Sun, Wood, Song, Yu, Matejovicova, Cohen,
  Andra, D'Ambrosio, Deng, Nallatamby, Songhori, Dangovski, Lampinen, Botadra,
  Hillier, Cao, Baddi, Kuncoro, Yoshino, Bhagatwala, Ranzato, Schaeffer, Liu,
  Ye, Sarvana, Nham, Kuang, Gao, Baek, Mittal, Wahid, Gergely, Ni, Feldman,
  Muir, Lamblin, Macherey, Dyer, Kilpatrick, Campos, Bhutani, Fort, Ahmad,
  Severyn, Chatziprimou, Ferludin, Dimarco, Kusupati, Heyward, Bahir, Villela,
  Millican, Marcus, Bahargam, Unlu, Roth, Wei, Gopal, Ghoshal, Lee, Lin, Lees,
  Lee, Hosseini, Fan, Neel, Wu, Altun, Cai, Piqueras, Woodward, Bissacco,
  Haykal, Bordbar, Sundaram, Hodkinson, Toyama, Polovets, Myers, Sinha,
  Levinboim, Krishnakumar, Chhaparia, Sholokhova, Gundavarapu, Jawahar,
  Qureshi, Hu, Momchev, Rahtz, Wu, S, Dhamdhere, Guo, Gupta, Eslami, Schain,
  Blokzijl, Welling, Orr, Bolelli, Perez-Nieves, Sirotenko, Prasad, Kar, Pigem,
  Terzi, Weisz, Ghosh, Mavalankar, Madeka, Daugaard, Adam, Shah, Berman, Tran,
  Baker, Andrejczuk, Chole, Raboshchuk, Mirzazadeh, Kagohara, Wu, Schallhart,
  Orlando, Wang, Rrustemi, Xiong, Liu, Vezer, Ramsden, yiin Chang, Mudgal, Li,
  Vieillard, Hoshen, Ahmad, Slone, Hua, Potikha, Rossini, Stritar, Prakash,
  Wang, Dong, Nazari, Nehoran, Tekelioglu, Li, Badola, Funkhouser, Li, Yerram,
  Ganeshan, Formoso, Langner, Shi, Li, Yamamori, Panda, Saade, Scarpati,
  Breaux, Carey, Zhou, Hsieh, Bridgers, Butryna, Gupta, Tulsyan, Woo, Eltyshev,
  Grathwohl, Parks, Benjamin, Panigrahy, Dodhia, Freitas, Sauer, Song, Alet,
  Tolins, Paduraru, Zhou, Albert, Zhang, Shu, Bansal, Nguyen, Globerson, Xiao,
  Manyika, Hennigan, Rong, Matak, Bakalov, Sharma, Sinopalnikov, Pierson,
  Roller, Brown, Gao, Fukuzawa, Ghafouri, Vassigh, Barr, Wang, Korsun, Jayaram,
  Ren, Zaman, Khan, Lunts, Deutsch, Uthus, Katz, Samsikova, Khalifa, Sethi,
  Sun, Tang, Alon, Luo, Yu, Nayyar, Petrini, Truong, Hellendoorn, Chinaev,
  Alberti, Wang, Hu, Mirrokni, Balashankar, Aharon, Mehta, Iscen, Kready,
  Manning, Mohananey, Chen, Tripathi, Wu, Petrovski, Hwang, Baeuml,
  Chandrakaladharan, Liu, Coaguila, Chen, Ma, Tafti, Tatineni, Spitz, Ye,
  Vicol, Rosca, Puigdomènech, Yahav, Ghemawat, Lin, Kirk, Nabulsi, Brin,
  Bohnet, Caluwaerts, Veerubhotla, Zheng, Dai, Petrov, Xu, Mehran, Xu,
  Zintgraf, Choi, Hombaiah, Thoppilan, Reddi, Lew, Li, Webster, Sawhney,
  Lamprou, Shakeri, Lunayach, Chen, Bagri, Salcianu, Chen, Donchev, Magister,
  Nørly, Rodrigues, Izo, Noga, Zou, Köppe, Zhou, Lee, Long, Eisenbud, Chen,
  Schenck, To, Zhong, Taropa, Truong, Levy, Martins, Zhang, Semturs, Zhang,
  Yakubovich, Moreno, McConnaughey, Lu, Redmond, Weerts, Bitton, Refice,
  Lacasse, Conmy, Tallec, Odell, Forbes-Pollard, Socala, Hoech, Kohli, Walton,
  Wang, Sazanovich, Zhu, Kapishnikov, Galt, Denton, Murdoch, Sikora, Mohamed,
  Wei, First, McConnell, Cobo, Qin, Avrahami, Balle, Watanabe, Louis, Kraft,
  Ariafar, Gu, Rives, Yoon, Rusu, Cobon-Kerr, Hahn, Luo, Yuvein, Zhu, Ahuja,
  Benenson, Kaufman, Yu, Hightower, Zhang, Ni, Hendricks, Wang, Yona, Jain,
  Barrio, Bhupatiraju, Velusamy, Dafoe, Riedel, Thomas, Yuan, Bellaiche,
  Panthaplackel, Kloboves, Jauhari, Akbulut, Davchev, Gladchenko, Madras,
  Chuklin, Hill, Yuan, Madhavan, Leonhard, Scandinaro, Chen, Niu, Douillard,
  Damoc, Onoe, Pedregosa, Bertsch, Leichner, Pagadora, Malmaud, Ponda, Twigg,
  Duzhyi, Shen, Wang, Garg, Chen, Evci, Lee, Liu, Kojima, Yamaguchi, Rajendran,
  Piergiovanni, Rajendran, Fornoni, Ibagon, Ragan, Khan, Blitzer, Bunner, Sun,
  Kosakai, Lundberg, Elue, Guu, Park, Park, Narayanaswamy, Wu, Mudigonda, Cohn,
  Mu, Kumar, Graesser, Zhang, Killam, Zhuang, Giménez, Jishi, Ley-Wild, Zhai,
  Osawa, Cedillo, Liu, Upadhyay, Sieniek, Sharma, Paine, Angelova, Addepalli,
  Parada, Majumder, Lamp, Kumar, Deng, Myaskovsky, Sabolić, Dudek, York,
  de~Chaumont~Quitry, Nie, Cattle, Gunjan, Piot, Khawaja, Bang, Wang,
  Khodadadeh, R, Rawlani, Powell, Lee, Griesser, Oh, Magalhaes, Li, Tokumine,
  Vogel, Hsu, BC, Jindal, Cohen, Yang, Yuan, de~Cesare, Bruguier, Xu, Roy,
  Jacovi, Belov, Arya, Meadowlark, Cohen-Ganor, Ye, Morris-Suzuki, Banzal,
  Song, Ponnuramu, Zhang, Scrivener, Zaiem, Rochman, Han, Ghazi, Lee, Drath,
  Suo, Girgis, Shenoy, Nguyen, Eck, Gupta, Yan, Carreira, Gulati, Sang,
  Mirylenka, Cooney, Chou, Ling, Fan, Coleman, Tubone, Kumar, Baldridge,
  Hernandez-Campos, Lazaridou, Besley, Yona, Bulut, Wellens, Pierigiovanni,
  George, Green, Han, Tao, Clark, You, Abdolmaleki, Fu, Chen, Chaugule,
  Chandorkar, Rahman, Thompson, Koanantakool, Bernico, Ren, Vlasov,
  Vassilvitskii, Kula, Liang, Kim, Huang, Ye, Lepikhin, and
  Helmholz}]{gemini2_5_flash}
Gheorghe Comanici, Eric Bieber, Mike Schaekermann, Ice Pasupat, Noveen
  Sachdeva, Inderjit Dhillon, Marcel Blistein, Ori Ram, Dan Zhang, Evan Rosen,
  Luke Marris, Sam Petulla, Colin Gaffney, Asaf Aharoni, Nathan Lintz,
  Tiago~Cardal Pais, Henrik Jacobsson, Idan Szpektor, Nan-Jiang Jiang, and 3416
  others. 2025.
\newblock \href {https://arxiv.org/abs/2507.06261} {Gemini 2.5: Pushing the
  frontier with advanced reasoning, multimodality, long context, and next
  generation agentic capabilities}.
\newblock \emph{Preprint}, arXiv:2507.06261.

\bibitem[{Deng et~al.(2017)Deng, Kanervisto, Ling, and
  Rush}]{deng2017im2markup}
Yuntian Deng, Anssi Kanervisto, Jeffrey Ling, and Alexander~M Rush. 2017.
\newblock Image-to-markup generation with coarse-to-fine attention.
\newblock In \emph{International Conference on Machine Learning}, pages
  980--989. PMLR.

\bibitem[{Fu et~al.(2024)Fu, Ng, Jiang, and Liu}]{fu2023gptscore}
Jinlan Fu, See-Kiong Ng, Zhengbao Jiang, and Pengfei Liu. 2024.
\newblock \href {https://doi.org/10.18653/v1/2024.naacl-long.365} {{GPTS}core:
  Evaluate as you desire}.
\newblock In \emph{Proceedings of the 2024 Conference of the North American
  Chapter of the Association for Computational Linguistics: Human Language
  Technologies (Volume 1: Long Papers)}, pages 6556--6576, Mexico City, Mexico.
  Association for Computational Linguistics.

\bibitem[{Gao et~al.(2023)Gao, Liu, Yao, Bai, Bai, Jin, and Liu}]{wu2023icdar}
Chenyang Gao, Yuliang Liu, Shiyu Yao, Jinfeng Bai, Xiang Bai, Lianwen Jin, and
  Cheng-Lin Liu. 2023.
\newblock Icdar 2023 competition on recognition of multi-line handwritten
  mathematical expressions.
\newblock In \emph{Document Analysis and Recognition - ICDAR 2023}, pages
  566--576, Cham. Springer Nature Switzerland.

\bibitem[{Gervais et~al.(2025)Gervais, Fadeeva, and Maksai}]{MathWriting}
Philippe Gervais, Anastasiia Fadeeva, and Andrii Maksai. 2025.
\newblock \href {https://doi.org/10.1145/3711896.3737436} {Mathwriting: A
  dataset for handwritten mathematical expression recognition}.
\newblock In \emph{Proceedings of the 31st ACM SIGKDD Conference on Knowledge
  Discovery and Data Mining V.2}, KDD '25, page 5459–5469, New York, NY, USA.
  Association for Computing Machinery.

\bibitem[{Grattafiori et~al.(2024)Grattafiori, Dubey, Jauhri, Pandey, Kadian,
  Al-Dahle, Letman, Mathur, Schelten, Vaughan, Yang, Fan, Goyal, Hartshorn,
  Yang, Mitra, Sravankumar, Korenev, Hinsvark, Rao, Zhang, Rodriguez,
  Gregerson, Spataru, Roziere, Biron, Tang, Chern, Caucheteux, Nayak, Bi,
  Marra, McConnell, Keller, Touret, Wu, Wong, Ferrer, Nikolaidis, Allonsius,
  Song, Pintz, Livshits, Wyatt, Esiobu, Choudhary, Mahajan, Garcia-Olano,
  Perino, Hupkes, Lakomkin, AlBadawy, Lobanova, Dinan, Smith, Radenovic,
  Guzmán, Zhang, Synnaeve, Lee, Anderson, Thattai, Nail, Mialon, Pang,
  Cucurell, Nguyen, Korevaar, Xu, Touvron, Zarov, Ibarra, Kloumann, Misra,
  Evtimov, Zhang, Copet, Lee, Geffert, Vranes, Park, Mahadeokar, Shah, van~der
  Linde, Billock, Hong, Lee, Fu, Chi, Huang, Liu, Wang, Yu, Bitton, Spisak,
  Park, Rocca, Johnstun, Saxe, Jia, Alwala, Prasad, Upasani, Plawiak, Li,
  Heafield, Stone, El-Arini, Iyer, Malik, Chiu, Bhalla, Lakhotia,
  Rantala-Yeary, van~der Maaten, Chen, Tan, Jenkins, Martin, Madaan, Malo,
  Blecher, Landzaat, de~Oliveira, Muzzi, Pasupuleti, Singh, Paluri, Kardas,
  Tsimpoukelli, Oldham, Rita, Pavlova, Kambadur, Lewis, Si, Singh, Hassan,
  Goyal, Torabi, Bashlykov, Bogoychev, Chatterji, Zhang, Duchenne, Çelebi,
  Alrassy, Zhang, Li, Vasic, Weng, Bhargava, Dubal, Krishnan, Koura, Xu, He,
  Dong, Srinivasan, Ganapathy, Calderer, Cabral, Stojnic, Raileanu, Maheswari,
  Girdhar, Patel, Sauvestre, Polidoro, Sumbaly, Taylor, Silva, Hou, Wang,
  Hosseini, Chennabasappa, Singh, Bell, Kim, Edunov, Nie, Narang, Raparthy,
  Shen, Wan, Bhosale, Zhang, Vandenhende, Batra, Whitman, Sootla, Collot,
  Gururangan, Borodinsky, Herman, Fowler, Sheasha, Georgiou, Scialom,
  Speckbacher, Mihaylov, Xiao, Karn, Goswami, Gupta, Ramanathan, Kerkez,
  Gonguet, Do, Vogeti, Albiero, Petrovic, Chu, Xiong, Fu, Meers, Martinet,
  Wang, Wang, Tan, Xia, Xie, Jia, Wang, Goldschlag, Gaur, Babaei, Wen, Song,
  Zhang, Li, Mao, Coudert, Yan, Chen, Papakipos, Singh, Srivastava, Jain,
  Kelsey, Shajnfeld, Gangidi, Victoria, Goldstand, Menon, Sharma, Boesenberg,
  Baevski, Feinstein, Kallet, Sangani, Teo, Yunus, Lupu, Alvarado, Caples, Gu,
  Ho, Poulton, Ryan, Ramchandani, Dong, Franco, Goyal, Saraf, Chowdhury,
  Gabriel, Bharambe, Eisenman, Yazdan, James, Maurer, Leonhardi, Huang, Loyd,
  Paola, Paranjape, Liu, Wu, Ni, Hancock, Wasti, Spence, Stojkovic, Gamido,
  Montalvo, Parker, Burton, Mejia, Liu, Wang, Kim, Zhou, Hu, Chu, Cai, Tindal,
  Feichtenhofer, Gao, Civin, Beaty, Kreymer, Li, Adkins, Xu, Testuggine, David,
  Parikh, Liskovich, Foss, Wang, Le, Holland, Dowling, Jamil, Montgomery,
  Presani, Hahn, Wood, Le, Brinkman, Arcaute, Dunbar, Smothers, Sun, Kreuk,
  Tian, Kokkinos, Ozgenel, Caggioni, Kanayet, Seide, Florez, Schwarz, Badeer,
  Swee, Halpern, Herman, Sizov, Guangyi, Zhang, Lakshminarayanan, Inan,
  Shojanazeri, Zou, Wang, Zha, Habeeb, Rudolph, Suk, Aspegren, Goldman, Zhan,
  Damlaj, Molybog, Tufanov, Leontiadis, Veliche, Gat, Weissman, Geboski, Kohli,
  Lam, Asher, Gaya, Marcus, Tang, Chan, Zhen, Reizenstein, Teboul, Zhong, Jin,
  Yang, Cummings, Carvill, Shepard, McPhie, Torres, Ginsburg, Wang, Wu, U,
  Saxena, Khandelwal, Zand, Matosich, Veeraraghavan, Michelena, Li, Jagadeesh,
  Huang, Chawla, Huang, Chen, Garg, A, Silva, Bell, Zhang, Guo, Yu, Moshkovich,
  Wehrstedt, Khabsa, Avalani, Bhatt, Mankus, Hasson, Lennie, Reso, Groshev,
  Naumov, Lathi, Keneally, Liu, Seltzer, Valko, Restrepo, Patel, Vyatskov,
  Samvelyan, Clark, Macey, Wang, Hermoso, Metanat, Rastegari, Bansal,
  Santhanam, Parks, White, Bawa, Singhal, Egebo, Usunier, Mehta, Laptev, Dong,
  Cheng, Chernoguz, Hart, Salpekar, Kalinli, Kent, Parekh, Saab, Balaji,
  Rittner, Bontrager, Roux, Dollar, Zvyagina, Ratanchandani, Yuvraj, Liang,
  Alao, Rodriguez, Ayub, Murthy, Nayani, Mitra, Parthasarathy, Li, Hogan,
  Battey, Wang, Howes, Rinott, Mehta, Siby, Bondu, Datta, Chugh, Hunt, Dhillon,
  Sidorov, Pan, Mahajan, Verma, Yamamoto, Ramaswamy, Lindsay, Lindsay, Feng,
  Lin, Zha, Patil, Shankar, Zhang, Zhang, Wang, Agarwal, Sajuyigbe, Chintala,
  Max, Chen, Kehoe, Satterfield, Govindaprasad, Gupta, Deng, Cho, Virk,
  Subramanian, Choudhury, Goldman, Remez, Glaser, Best, Koehler, Robinson, Li,
  Zhang, Matthews, Chou, Shaked, Vontimitta, Ajayi, Montanez, Mohan, Kumar,
  Mangla, Ionescu, Poenaru, Mihailescu, Ivanov, Li, Wang, Jiang, Bouaziz,
  Constable, Tang, Wu, Wang, Wu, Gao, Kleinman, Chen, Hu, Jia, Qi, Li, Zhang,
  Zhang, Adi, Nam, Yu, Wang, Zhao, Hao, Qian, Li, He, Rait, DeVito, Rosnbrick,
  Wen, Yang, Zhao, and Ma}]{llama3_2}
Aaron Grattafiori, Abhimanyu Dubey, Abhinav Jauhri, Abhinav Pandey, Abhishek
  Kadian, Ahmad Al-Dahle, Aiesha Letman, Akhil Mathur, Alan Schelten, Alex
  Vaughan, Amy Yang, Angela Fan, Anirudh Goyal, Anthony Hartshorn, Aobo Yang,
  Archi Mitra, Archie Sravankumar, Artem Korenev, Arthur Hinsvark, and 542
  others. 2024.
\newblock \href {https://arxiv.org/abs/2407.21783} {The llama 3 herd of
  models}.
\newblock \emph{Preprint}, arXiv:2407.21783.

\bibitem[{Ji et~al.(2023)Ji, Lee, Frieske, Yu, Su, Xu, Ishii, Bang, Madotto,
  and Fung}]{Ji_2023}
Ziwei Ji, Nayeon Lee, Rita Frieske, Tiezheng Yu, Dan Su, Yan Xu, Etsuko Ishii,
  Ye~Jin Bang, Andrea Madotto, and Pascale Fung. 2023.
\newblock \href {https://doi.org/10.1145/3571730} {Survey of hallucination in
  natural language generation}.
\newblock \emph{ACM Comput. Surv.}, 55(12).

\bibitem[{Kendall(1938)}]{kendalls_tau}
M.~G. Kendall. 1938.
\newblock \href {http://www.jstor.org/stable/2332226} {A new measure of rank
  correlation}.
\newblock \emph{Biometrika}, 30(1/2):81--93.

\bibitem[{Levenshtein(1966)}]{editdistance1965}
Vladimir~I. Levenshtein. 1966.
\newblock \href
  {http://profs.sci.univr.it/~liptak/ALBioinfo/2012_2013/2011_2012/files/levenshtein66.pdf}
  {Binary codes capable of correcting deletions, insertions, and reversals}.
\newblock \emph{Soviet physics. Doklady}, 10:707--710.

\bibitem[{Li et~al.(2025)Li, Shi, Gao, Liu, Wang, Chen, Liu, Zhao, Wang, and
  Metaxas}]{lihidden}
Zhuowei Li, Haizhou Shi, Yunhe Gao, Di~Liu, Zhenting Wang, Yuxiao Chen, Ting
  Liu, Long Zhao, Hao Wang, and Dimitris~N. Metaxas. 2025.
\newblock \href {https://proceedings.mlr.press/v267/li25ca.html} {The hidden
  life of tokens: Reducing hallucination of large vision-language models via
  visual information steering}.
\newblock In \emph{Proceedings of the 42nd International Conference on Machine
  Learning}, volume 267 of \emph{Proceedings of Machine Learning Research},
  pages 35799--35819. PMLR.

\bibitem[{Liu et~al.(2023)Liu, Iter, Xu, Wang, Xu, and Zhu}]{geval2023}
Yang Liu, Dan Iter, Yichong Xu, Shuohang Wang, Ruochen Xu, and Chenguang Zhu.
  2023.
\newblock \href {https://doi.org/10.18653/v1/2023.emnlp-main.153} {{G}-eval:
  {NLG} evaluation using gpt-4 with better human alignment}.
\newblock In \emph{Proceedings of the 2023 Conference on Empirical Methods in
  Natural Language Processing}, pages 2511--2522, Singapore. Association for
  Computational Linguistics.

\bibitem[{Lu et~al.(2024)Lu, Li, Chen, Xu, Luo, Zhang, and Ye}]{ovis}
Shiyin Lu, Yang Li, Qing-Guo Chen, Zhao Xu, Weihua Luo, Kaifu Zhang, and
  Han-Jia Ye. 2024.
\newblock \href {https://arxiv.org/abs/2405.20797} {Ovis: Structural embedding
  alignment for multimodal large language model}.
\newblock \emph{Preprint}, arXiv:2405.20797.

\bibitem[{Meurer et~al.(2017)Meurer, Smith, Paprocki, \v{C}ert\'{i}k,
  Kirpichev, Rocklin, Kumar, Ivanov, Moore, Singh, Rathnayake, Vig, Granger,
  Muller, Bonazzi, Gupta, Vats, Johansson, Pedregosa, Curry, Terrel,
  Rou\v{c}ka, Saboo, Fernando, Kulal, Cimrman, and Scopatz}]{meurer2017sympy}
Aaron Meurer, Christopher~P. Smith, Mateusz Paprocki, Ond\v{r}ej
  \v{C}ert\'{i}k, Sergey~B. Kirpichev, Matthew Rocklin, AMiT Kumar, Sergiu
  Ivanov, Jason~K. Moore, Sartaj Singh, Thilina Rathnayake, Sean Vig, Brian~E.
  Granger, Richard~P. Muller, Francesco Bonazzi, Harsh Gupta, Shivam Vats,
  Fredrik Johansson, Fabian Pedregosa, and 8 others. 2017.
\newblock \href {https://doi.org/10.7717/peerj-cs.103} {Sympy: symbolic
  computing in python}.
\newblock \emph{PeerJ Computer Science}, 3:e103.

\bibitem[{Mouchère et~al.(2016)Mouchère, Viard-Gaudin, Zanibbi, and
  Garain}]{zanibbi2016crohme}
Harold Mouchère, Christian Viard-Gaudin, Richard Zanibbi, and U.~Garain. 2016.
\newblock \href {https://doi.org/10.1109/ICFHR.2016.0116} {Icfhr2016 crohme:
  Competition on recognition of online handwritten mathematical expressions}.
\newblock In \emph{2016 15th International Conference on Frontiers in
  Handwriting Recognition (ICFHR)}, pages 607--612.

\bibitem[{Nath et~al.(2025)Nath, Bathina, Khan, and Khapra}]{fermat2025}
Oikantik Nath, Hanani Bathina, Mohammed Safi Ur~Rahman Khan, and Mitesh~M
  Khapra. 2025.
\newblock \href {https://doi.org/10.18653/v1/2025.acl-long.720} {Can
  vision-language models evaluate handwritten math?}
\newblock In \emph{Proceedings of the 63rd Annual Meeting of the Association
  for Computational Linguistics (Volume 1: Long Papers)}, pages 14784--14814,
  Vienna, Austria. Association for Computational Linguistics.

\bibitem[{{National Council of Teachers of Mathematics}(2014)}]{nctm2014}
{National Council of Teachers of Mathematics}. 2014.
\newblock \emph{Principles to Actions: Ensuring Mathematical Success for All}.
\newblock National Council of Teachers of Mathematics, {Reston, VA}.

\bibitem[{Novotn{\'y} and {\v{S}}tef{\'a}nik(2022)}]{mathberta}
V{\'i}t Novotn{\'y} and Michal {\v{S}}tef{\'a}nik. 2022.
\newblock \href {http://ceur-ws.org/Vol-3180/paper-06.pdf} {Combining sparse
  and dense information retrieval: Soft vector space model and {MathBERTa} at
  {ARQMath-3} task~1 (answer retrieval)}.
\newblock In \emph{Proceedings of the Working Notes of {CLEF} 2022}, volume
  3180 of \emph{CEUR Workshop Proceedings}, pages 104--118. {CEUR-WS}.

\bibitem[{OpenAI(2023)}]{gpt4}
OpenAI. 2023.
\newblock Gpt-4 technical report.
\newblock \emph{arXiv preprint arXiv:2303.08774}.

\bibitem[{OpenAI(2024)}]{openai2024gpt4ocard}
OpenAI. 2024.
\newblock \href {https://arxiv.org/abs/2410.21276} {Gpt-4o system card}.
\newblock \emph{Preprint}, arXiv:2410.21276.

\bibitem[{OpenAI(2025)}]{gpt5}
OpenAI. 2025.
\newblock Introducing gpt-5.
\newblock \url{https://openai.com/index/introducing-gpt-5/}.
\newblock Accessed: 2025-11-13.

\bibitem[{Papineni et~al.(2002)Papineni, Roukos, Ward, and
  Zhu}]{papineni2002bleu}
Kishore Papineni, Salim Roukos, Todd Ward, and Wei-Jing Zhu. 2002.
\newblock \href {https://doi.org/10.3115/1073083.1073135} {{B}leu: a method for
  automatic evaluation of machine translation}.
\newblock In \emph{Proceedings of the 40th Annual Meeting of the Association
  for Computational Linguistics}, pages 311--318, Philadelphia, Pennsylvania,
  USA. Association for Computational Linguistics.

\bibitem[{Park et~al.(2025)Park, Do, and Lee}]{park2025leveraging}
Taehee Park, Heejin Do, and Gary Lee. 2025.
\newblock \href {https://doi.org/10.18653/v1/2025.emnlp-main.1431} {Leveraging
  what{'}s overfixed: Post-correction via {LLM} grammatical error
  overcorrection}.
\newblock In \emph{Proceedings of the 2025 Conference on Empirical Methods in
  Natural Language Processing}, pages 28195--28207, Suzhou, China. Association
  for Computational Linguistics.

\bibitem[{Polya(1945)}]{polya1945solve}
George Polya. 1945.
\newblock \emph{How to solve it: A new aspect of mathematical method}.
\newblock Princeton university press.

\bibitem[{Rei et~al.(2020)Rei, Stewart, Farinha, and Lavie}]{comet}
Ricardo Rei, Craig Stewart, Ana~C Farinha, and Alon Lavie. 2020.
\newblock \href {https://doi.org/10.18653/v1/2020.emnlp-main.213} {{COMET}: A
  neural framework for {MT} evaluation}.
\newblock In \emph{Proceedings of the 2020 Conference on Empirical Methods in
  Natural Language Processing (EMNLP)}, pages 2685--2702, Online. Association
  for Computational Linguistics.

\bibitem[{Schoenfeld(1985)}]{schoenfeld1985mathematical}
Alan~H. Schoenfeld. 1985.
\newblock \emph{Mathematical Problem Solving}.
\newblock Academic Press.

\bibitem[{Seong et~al.(2025)Seong, Park, Liermann, Choi, Nam, Kim, Lim, and
  Lee}]{seong2025memeic}
Jin Seong, Jiyun Park, Wencke Liermann, Hongseok Choi, Yoonji Nam, Hyun Kim,
  Soojong Lim, and Namhoon Lee. 2025.
\newblock \href {https://doi.org/10.52202/085713-4304} {Memeic: A step toward
  continual and compositional knowledge editing}.
\newblock In \emph{Advances in Neural Information Processing Systems}, volume
  38, Main Conference, pages 129205--129242. Curran Associates, Inc.

\bibitem[{Udagawa et~al.(2024)Udagawa, Suzuki, Muraoka, and
  Kurata}]{udagawa2024robust}
Takuma Udagawa, Masayuki Suzuki, Masayasu Muraoka, and Gakuto Kurata. 2024.
\newblock \href {https://doi.org/10.18653/v1/2024.emnlp-industry.20} {Robust
  {ASR} error correction with conservative data filtering}.
\newblock In \emph{Proceedings of the 2024 Conference on Empirical Methods in
  Natural Language Processing: Industry Track}, pages 256--266, Miami, Florida,
  US. Association for Computational Linguistics.

\bibitem[{Wang et~al.(2025)Wang, Wu, Ouyang, Gu, Zhang, Xia, Shi, Zhang, and
  He}]{wang2025image}
Bin Wang, Fan Wu, Linke Ouyang, Zhuangcheng Gu, Rui Zhang, Renqiu Xia, Botian
  Shi, Bo~Zhang, and Conghui He. 2025.
\newblock Image over text: Transforming formula recognition evaluation with
  character detection matching.
\newblock In \emph{Proceedings of the IEEE/CVF Conference on Computer Vision
  and Pattern Recognition (CVPR)}, pages 19681--19690.

\bibitem[{Wei et~al.(2022)Wei, Tay, Bommasani, Raffel, Zoph, Borgeaud,
  Yogatama, Bosma, Zhou, Metzler, Chi, Hashimoto, Vinyals, Liang, Dean, and
  Fedus}]{wei2022emergent}
Jason Wei, Yi~Tay, Rishi Bommasani, Colin Raffel, Barret Zoph, Sebastian
  Borgeaud, Dani Yogatama, Maarten Bosma, Denny Zhou, Donald Metzler, Ed~H.
  Chi, Tatsunori Hashimoto, Oriol Vinyals, Percy Liang, Jeff Dean, and William
  Fedus. 2022.
\newblock \href {https://openreview.net/forum?id=yzkSU5zdwD} {Emergent
  abilities of large language models}.
\newblock \emph{Transactions on Machine Learning Research}.
\newblock Survey Certification.

\bibitem[{Yuan et~al.(2022)Yuan, Liu, Dikubab, Liu, Ji, Wu, and
  Bai}]{yuan2022syntax}
Ye~Yuan, Xiao Liu, Wondimu Dikubab, Hui Liu, Zhilong Ji, Zhongqin Wu, and Xiang
  Bai. 2022.
\newblock Syntax-aware network for handwritten mathematical expression
  recognition.
\newblock In \emph{Proceedings of the IEEE/CVF Conference on Computer Vision
  and Pattern Recognition (CVPR)}, pages 4553--4562.

\bibitem[{Zhang et~al.(2020)Zhang, Kishore, Wu, Weinberger, and
  Artzi}]{bertscore}
Tianyi Zhang, Varsha Kishore, Felix Wu, Kilian~Q. Weinberger, and Yoav Artzi.
  2020.
\newblock \href {https://openreview.net/forum?id=SkeHuCVFDr} {Bertscore:
  Evaluating text generation with bert}.
\newblock In \emph{International Conference on Learning Representations}.

\bibitem[{Zheng et~al.(2023)Zheng, Chiang, Sheng, Zhuang, Wu, Zhuang, Lin, Li,
  Li, Xing, Zhang, Gonzalez, and Stoica}]{zheng2023judging}
Lianmin Zheng, Wei-Lin Chiang, Ying Sheng, Siyuan Zhuang, Zhanghao Wu, Yonghao
  Zhuang, Zi~Lin, Zhuohan Li, Dacheng Li, Eric Xing, Hao Zhang, Joseph
  Gonzalez, and Ion Stoica. 2023.
\newblock \href
  {https://proceedings.neurips.cc/paper_files/paper/2023/file/91f18a1287b398d378ef22505bf41832-Paper-Datasets_and_Benchmarks.pdf}
  {Judging llm-as-a-judge with mt-bench and chatbot arena}.
\newblock In \emph{Advances in Neural Information Processing Systems},
  volume~36, pages 46595--46623. Curran Associates, Inc.

\bibitem[{Zhu et~al.(2025)Zhu, Wang, Chen, Liu, Ye, Gu, Tian, Duan, Su, Shao,
  Gao, Cui, Wang, Cao, Liu, Wei, Zhang, Wang, Xu, Li, Wang, Deng, Li, He,
  Jiang, Luo, Wang, He, Shi, Zhang, Shao, He, Xiong, Qu, Sun, Jiao, Lv, Wu,
  Zhang, Deng, Ge, Chen, Wang, Dou, Lu, Zhu, Lu, Lin, Qiao, Dai, and
  Wang}]{internvl3}
Jinguo Zhu, Weiyun Wang, Zhe Chen, Zhaoyang Liu, Shenglong Ye, Lixin Gu, Hao
  Tian, Yuchen Duan, Weijie Su, Jie Shao, Zhangwei Gao, Erfei Cui, Xuehui Wang,
  Yue Cao, Yangzhou Liu, Xingguang Wei, Hongjie Zhang, Haomin Wang, Weiye Xu,
  and 32 others. 2025.
\newblock \href {https://arxiv.org/abs/2504.10479} {Internvl3: Exploring
  advanced training and test-time recipes for open-source multimodal models}.
\newblock \emph{Preprint}, arXiv:2504.10479.

\end{thebibliography}

\clearpage
\appendix

\section{Penalty Threshold Sensitivity}
\label{supp:penalty_threshold}

\subsection{Distribution of Over-correction Magnitudes}
We analyzed over-correction scores $\Delta S = S_{\text{model}} - S_{\text{oracle}}$ across all models on the FERMAT dataset. 
As shown in Table~\ref{tab:overcorrection_dist}, 79.3\% of all over-corrections fall within the 0–10 point range, with 60.4\% concentrated below 5 points. 
Thresholds below 10 therefore over-penalize minor variations, while thresholds 
above 10 fail to capture significant over-corrections.

\begin{table}[h]
\renewcommand{\arraystretch}{0.9}
\centering
\caption{Distribution of over-correction magnitudes across all models.}
\label{tab:overcorrection_dist}
\begin{tabular}{lrr}
\toprule
Range (points) & Occurrences & Percentage \\
\midrule
0–5  & 22,483 & 60.4\% \\
5–10 & 7,038  & 18.9\% \\
10–15 & 3,548 & 9.5\% \\
15–20 & 4,131 & 11.1\% \\
\bottomrule
\end{tabular}
\end{table}

\subsection{Model-specific Behavioral Patterns}
\begin{figure}[h]
\centering
\includegraphics[width=\linewidth]{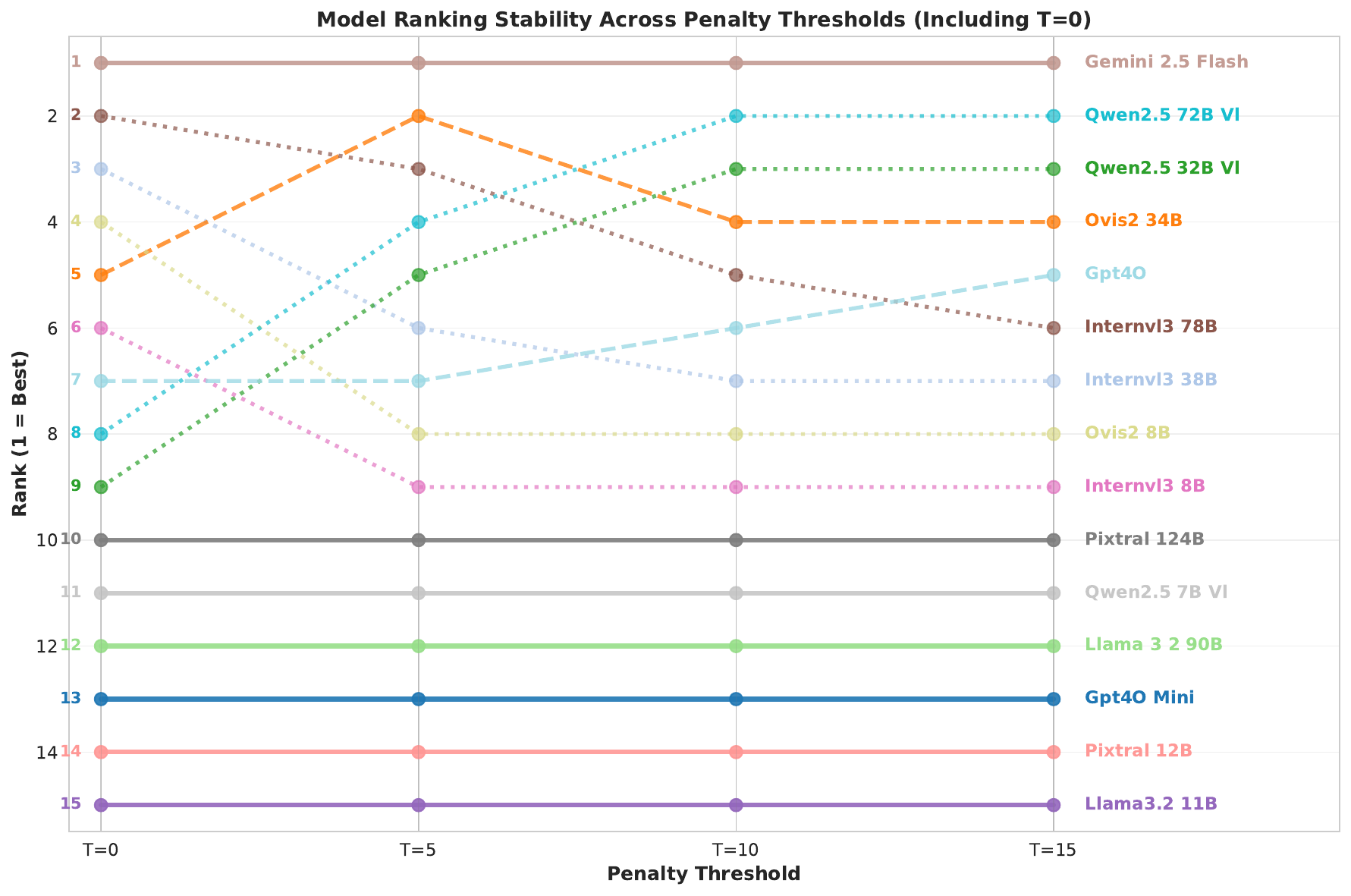}
\caption{Model ranking trajectories under varying penalty thresholds. 
Three patterns emerge: robust (e.g., Gemini 2.5 Flash), 
minor over-correctors (e.g., GPT-4o), and critical over-correctors (e.g., InternVL3).}
\label{fig:bump_chart}
\end{figure}

Figure~\ref{fig:bump_chart} reveals how individual model rankings evolve across threshold values. 
Three distinct behavioral patterns emerge:

\begin{itemize}
\item \textbf{Robust models} (e.g., Gemini 2.5 Flash): Maintain top rankings regardless of threshold, 
demonstrating consistent faithful transcription
\item \textbf{Minor over-correctors} (e.g., GPT-4o, Qwen2.5 VL family): Improve rankings at higher thresholds, 
indicating predominantly small-scale corrections ($\Delta S < 10$)  
\item \textbf{Critical over-correctors} (e.g., Ovis2 8B, InternVL3 family): Deteriorate under stricter thresholds, 
revealing frequent substantial alterations ($\Delta S > 10$)
\end{itemize}

Notably, Gemini 2.5 Flash's consistency may stem from its reasoning architecture, 
enabling better preservation of original content despite contextual pressure to correct errors.

\begin{figure}[h]
\centering
\includegraphics[width=\linewidth]{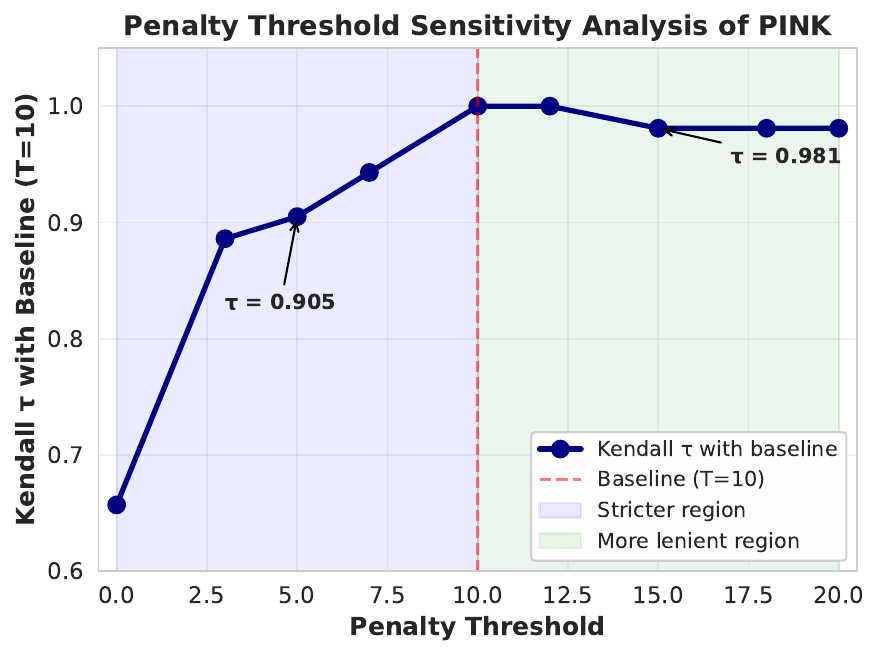}
\caption{Ranking stability under different penalty thresholds $T$, where models are penalized when $\Delta S (S_{\text{model}} - S_{\text{oracle}}) > T$ (i.e., VLM score exceeds student's true score by more than $T$ points). Kendall's $\tau$ measures ranking consistency relative to $T{=}10$ baseline. High stability ($\tau > 0.9$) in the range $T \in [5, 15]$ validates our threshold choice.}
\label{fig:kendall_tau}
\end{figure}

\section{Detailed Methodology for Metric Design and Evaluation}
\label{app_additional}

In this section, we provide supplementary visual materials to further detail our evaluation framework, including the pedagogical basis for our rubric design and the user interface protocol used for human evaluation.

\subsection{Pedagogical Basis and Rubric Specification}
\label{subsec:rubric_details}

\begin{figure}[ht]
    \centering
    \includegraphics[width=\linewidth]{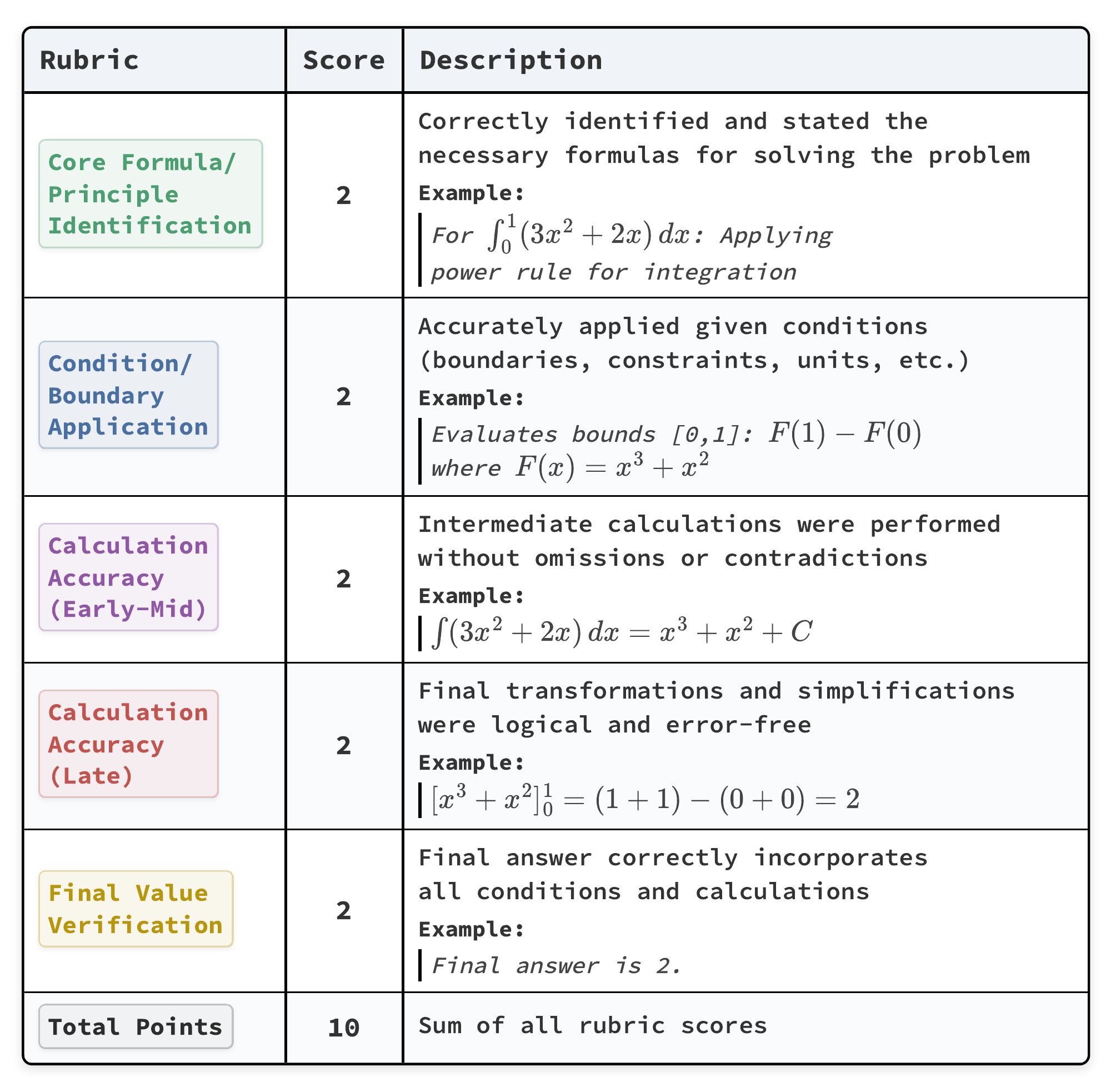}
    \caption{Overview of Auto-Grading System's Rubrics}
    \label{fig_app:figure_rubric}
\end{figure}

Our grading rubric is not arbitrarily defined but is rigorously grounded in process-oriented assessment principles from educational theory. Traditional binary scoring (correct/incorrect) has long been criticized for failing to reflect a learner's problem-solving ability, conceptual understanding, and strategic choices~\cite{polya1945solve, schoenfeld1985mathematical}. 

To address this, international standard assessments such as AP Calculus and Cambridge A-Level adopt a \textbf{Partial Credit} system. This system is structured around sequential problem-solving steps: ``Problem Understanding $\to$ Condition Application $\to$ Calculation Procedure $\to$ Result Verification''~\cite{collegeboard2023ap, cambridge2022alevel}. This structure is widely recognized for ensuring transparency in evaluation and aiding in the diagnosis of specific error types~\cite{nctm2014}.

Consistent with these standards, we designed a fixed 5-component rubric (Figure~\ref{fig_app:figure_rubric}), allocating 20 points to each stage (Total 100). By adopting this established educational framework, our metric can systematically capture mathematical thinking processes and error patterns, mirroring the standards of human expert grading.

\subsection{Human Evaluation Interface and Protocol}
\label{subsec:labeling_tool}

\begin{figure}[t]
    \centering
    \includegraphics[width=\linewidth]{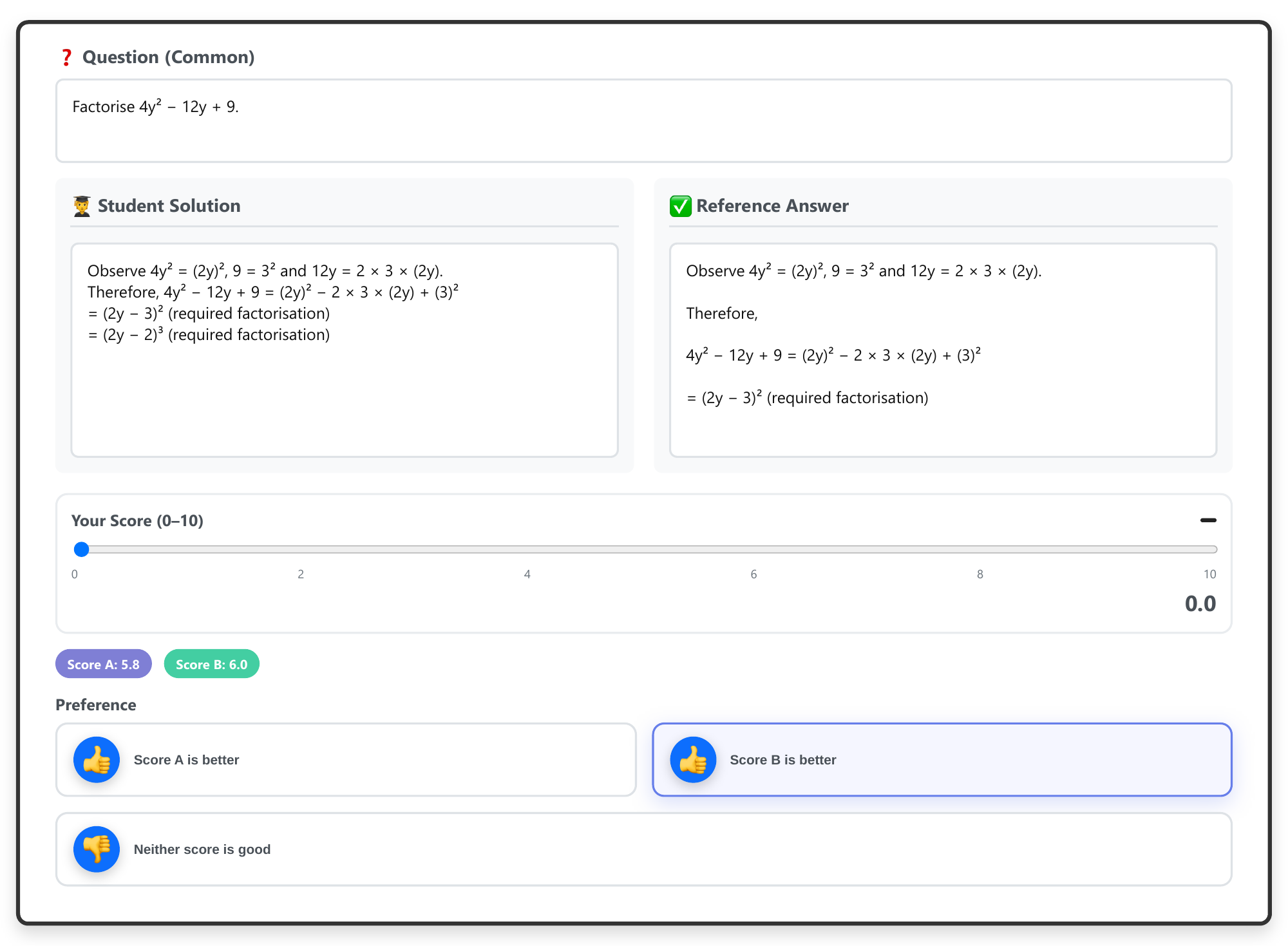}
    \caption{Human Evaluation: Labeling Tool}
    \label{fig_app:labeling_tool}
\end{figure}

Figure~\ref{fig_app:labeling_tool} illustrates the interface used for our human expert study. The evaluation protocol was designed to simulate a realistic grading scenario while ensuring blind comparison:

\begin{enumerate}
    \item \textbf{Simulated Grading Scenario:} Annotators are presented with a math problem and a ``Student Solution.'' Although this text is the VLM's OCR transcription, it is presented as a student's answer. Annotators are instructed to act as math teachers and grade the solution based on the provided Reference Answer.
    
    \item \textbf{Blind Metric Comparison:} After assigning their own score (0-10 scale), annotators are shown two automated scores labeled \textbf{``Score A''} and \textbf{``Score B.''} These correspond to the PINK and BLEU scores (normalized to the same scale). Crucially, the assignment of A/B is \textbf{randomized} for each sample to prevent positional bias and ensure that annotators remain unaware of which metric is which.
    
    \item \textbf{Preference Selection:} Finally, experts select which score (A or B) better reflects the true quality and faithfulness of the transcription compared to their own expert judgment.
\end{enumerate}

Also, annotators were recruited from our institution and compensated through regular employment. They were fully informed about the research objectives and how their evaluations would be used.

\section{Qualitative and Mechanistic Analysis of Over-Correction} 
\label{supp:oc_casestudy}
To provide a qualitative understanding of how Vision-Language Models (VLMs) fail to faithfully transcribe student errors, we present a detailed analysis from two perspectives. First, we classify Over-Correction into two distinct categories based on severity—\textbf{Simple-Level} and \textbf{Critical-Level}—and illustrate how these behaviors artificially inflate scores using real examples from the FERMAT dataset (\cref{fig_app:oc_case}). Second, we go beyond phenomenological categorization to empirically validate the underlying mechanism. By visualizing the model's attention maps, we demonstrate that Over-Correction is driven by the model's internal reasoning overriding visual perception (\cref{fig:attention_viz}).

\subsection{Type 1: Simple-Level (Local Fix)}
\label{subsec:simple_oc}

This type occurs when the model acts like a \textbf{mathematical spell-checker}, performing localized token substitutions to fix specific errors while preserving the overall structure.

\begin{itemize}
    \item \textbf{Scenario:} The student incorrectly wrote $\sin^{-1}$ instead of $\tan^{-1}$ (Top of Fig.~\ref{fig_app:oc_case}).
    \item \textbf{Model Action:} InternVL3-8B detects the inconsistency with the integral context ($\int \frac{dx}{1+x^2}$) and locally fixes the student's error by swapping ``sin'' for ``tan''.
    \item \textbf{Consequence:} The student's misconception is masked. A failing grade (Oracle: \textbf{34}) is inflated to a near-perfect score (\textbf{99}).
\end{itemize}

\subsection{Type 2: Critical-Level (Global Rewrite)}
\label{subsec:critical_oc}

A more severe failure mode where the model \textbf{discards the student's reasoning} and generates a correct summary based on its own internal calculation.

\begin{figure*}[t]
    \centering
    \includegraphics[width=0.95\linewidth]{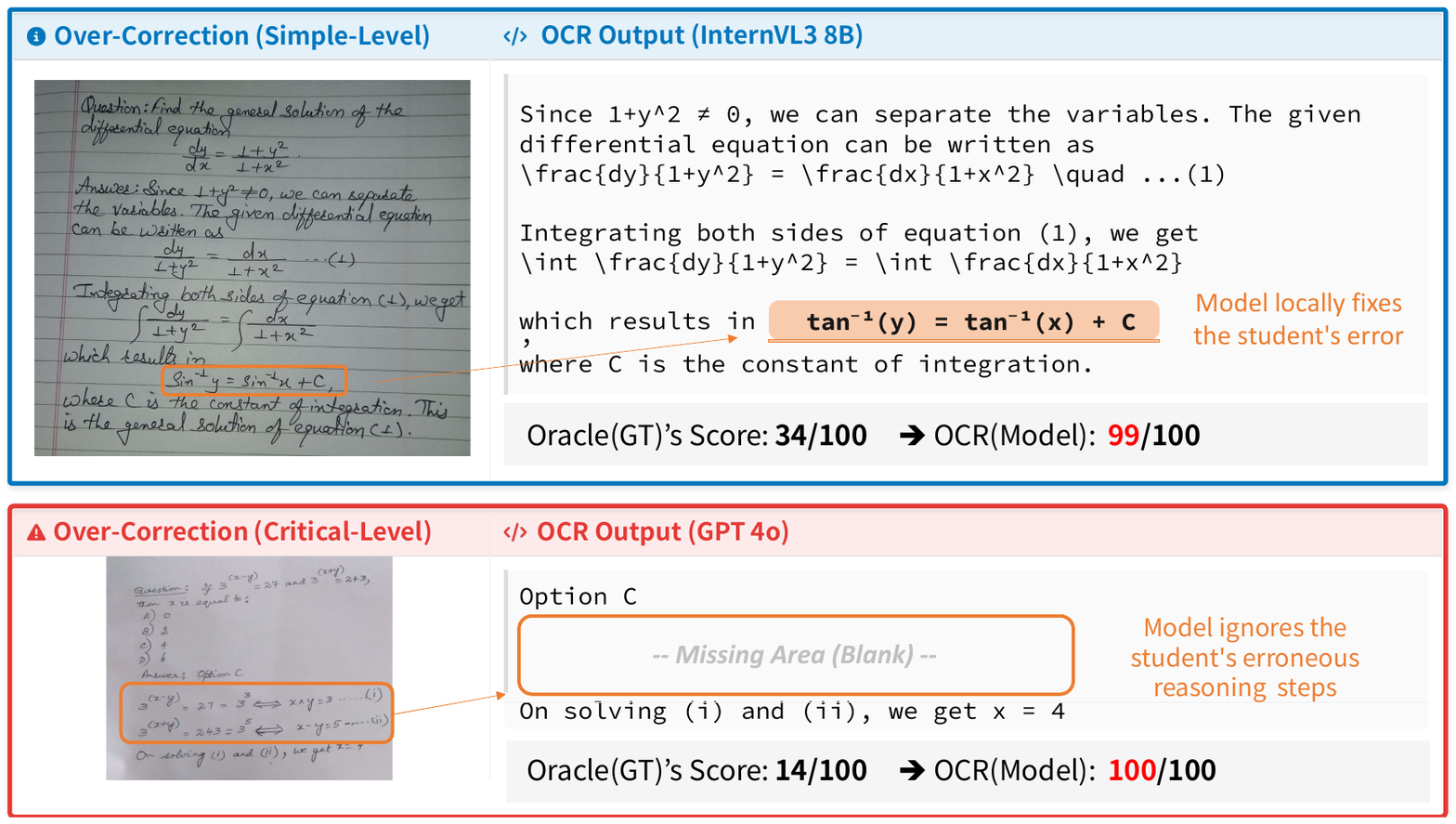}
    \caption{\textbf{Case Study of Over-Correction.} 
    \textbf{(Top) Simple-Level:} InternVL3-8B locally fixes the student's error ($\sin \to \tan$), inflating the score from 34 to 99. 
    \textbf{(Bottom) Critical-Level:} GPT-4o ignores the student's erroneous reasoning steps and provides a correct summary, inflating the score from 14 to 100.}
    \label{fig_app:oc_case}
\end{figure*}

\begin{itemize} 
    \item \textbf{Scenario:} The student failed to derive the correct solution for an exponential system (Bottom of Fig.~\ref{fig_app:oc_case}).
    \item \textbf{Model Action:} GPT-4o ignores the erroneous derivation steps entirely (visualized as the ``Missing Area'' in the figure) and simply outputs the correct final value: ``On solving... we get x=4''.
    \item \textbf{Consequence:} The evidence of the student's error is erased. A low score (Oracle: \textbf{14}) jumps to a perfect score (\textbf{100}).
\end{itemize}

\subsection{Detailed Analysis of Visual Grounding}
\label{subsec:attention_analysis}

To empirically validate our hypothesis that \textbf{Over-Correction} is an emergent behavior where a model's internal reasoning suppresses visual evidence, we visualized the attention maps of two distinct models: \textbf{InternVL3-8B} (prone to over-correction) and \textbf{Qwen2.5-VL 7B} (a faithful transcriber).

We hypothesize that as a VLM autoregressively generates the solution, the accumulation of mathematical logic creates \textbf{``contextual pressure.''} When the visual evidence (e.g., a student's error) conflicts with this internal context, the model treats the visual tokens as \textbf{noise} and suppresses them, prioritizing its internal prediction instead.

To prove this, we analyzed a case where a student incorrectly wrote $\sin^{-1}y = \sin^{-1}x + C$ following an integral $\int \frac{dx}{1+x^2}$. We extracted the attention weights at the exact moment the model generates the target token, given the full history of tokens the model had generated autoregressively up to that point.

As illustrated in Figure~\ref{fig:attention_viz}, the results confirm our hypothesis and justify the necessity of the PINK penalty mechanism:

\begin{itemize}
    \item \textbf{InternVL3-8B (Over-Correction):} Under the pressure of the \textbf{autoregressively accumulated context}, the model generates the mathematically correct token ``tan''. Crucially, the attention map reveals \textbf{negligible activation} on the visual region corresponding to the handwritten ``sin'', indicating that the model treated the visual evidence as noise. While this ``correction'' yields a high initial grading score of \textbf{99/100}, it falsely inflates the student's ability. Consequently, PINK identifies this unfaithful behavior and applies a penalty, drastically reducing the final score to \textbf{27/100}.
    
    \item \textbf{Qwen2.5-VL 7B (Faithful Transcription):} In contrast, this model faithfully transcribes the student's error (``sin''). The attention map demonstrates strong and precise activation on the handwritten strokes, showing that it successfully resisted contextual pressure to maintain visual grounding. Although the resulting score is lower (\textbf{34/100}), it accurately reflects the student's actual performance (matching the Oracle baseline), thus incurring \textbf{no penalty}.
\end{itemize}

These visualizations provide direct evidence that Over-Correction allows reasoning to override perception. This confirms that high raw scores in reasoning models can be misleading, validating PINK's penalty mechanism as essential for ensuring evaluation integrity.

\section{Over-Correction: Frequency and Ranking Impact}
\subsection{Per-Model Over-Correction Rates}
\label{supp:oc_rates}
\begin{figure*}[t]
    \centering
    \includegraphics[width=0.8\linewidth]{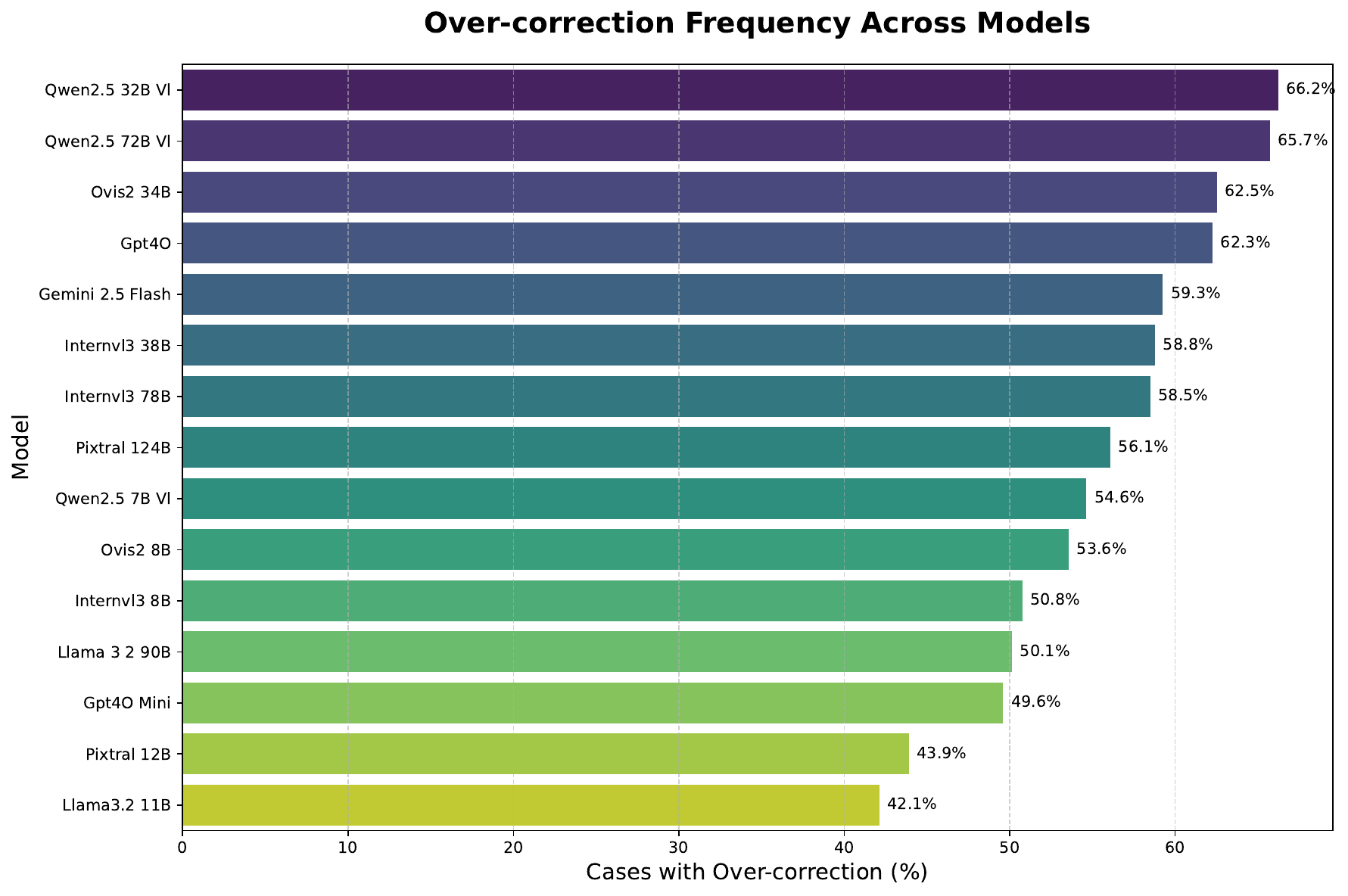} 
    \caption{Over-correction frequency across all 15 tested VLM models.}
    \label{fig_app:oc_frequency_bar}
\end{figure*}

\begin{figure}[h]
    \centering
    \includegraphics[width=\linewidth]{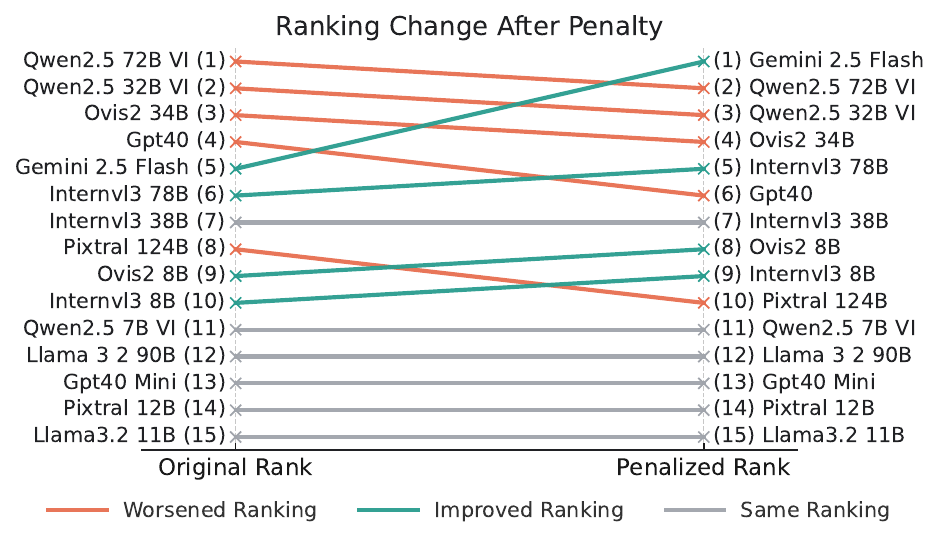}
    \caption{\textbf{Ranking changes after applying the over-correction penalty.} The PINK score leads to a significant reordering of models, highlighting the impact of penalizing unfaithful corrections.}
    \label{fig:ranking_change}
\end{figure}

Fig.\ref{fig_app:oc_frequency_bar} illustrates that over-correction is a pervasive issue, yet its prevalence varies significantly, ranging from \textbf{66.2\%} in Qwen2.5 32B VI down to \textbf{42.1\%} in Llama3.2 11B.

\subsection{Ranking Reshuffle via Penalty}
\label{app:ranking_reshuffle}

The application of our over-correction penalty dramatically reshapes the raw auto-graded leaderboard, as shown in \cref{fig:ranking_change}. This shift exposes a core limitation of relying solely on the initial auto-graded scores: they fail to distinguish between faithful transcription and cases where the model quietly ``fixes'' student errors.

A clear example is the contrast between \textbf{Gemini 2.5 Flash} and \textbf{GPT-4o}. Gemini rises from \textbf{5th} under the raw auto-graded ranking to \textbf{1st} after the penalty is applied, while GPT-4o drops from \textbf{4th} to \textbf{6th} due to frequent over-corrections. This reversal highlights that PINK surfaces fidelity failures that the raw scores overlook, yielding a model ranking that better reflects true OCR behavior.

\section{Robustness and Reproducibility}
\label{app:reproducibility}

\noindent We validate reproducibility by comparing \textbf{Qwen3-Next-80B-A3B-Thinking} (open-source) against GPT-5 across 15 VLMs and 33K+ samples.

\subsection{Cross-Grader Consistency Analysis} 
As summarized in Table~\ref{tab:pipeline_stats}, our analysis confirms that PINK is grader-agnostic.

\begin{table}[h]
  \centering
  \small
  \setlength{\tabcolsep}{4pt}
  \begin{tabular}{@{}llccc@{}}
    \toprule
    \textbf{Stage} & \textbf{Level} & \textbf{Pearson $r$} & \textbf{QWK} & \textbf{Kendall's $\tau$} \\
    \midrule
    Grading & Sample & 0.78 & 0.78 & -- \\
    OC Penalty & Sample & 0.74 & 0.73 & -- \\
    PINK Score & VLM & \textbf{0.99} & -- & -- \\
    Final Rank & VLM & -- & -- & \textbf{0.90} \\
    \bottomrule
  \end{tabular}
  \caption{\textbf{Reproducibility Pipeline.} Micro-variances at the sample level cancel out, yielding near-perfect macro agreement. N\,=\,33k (sample), N\,=\,15 (VLM).}
  \label{tab:pipeline_stats}
\end{table}

\begin{table}[h]
  \centering
  \scriptsize
  \setlength{\tabcolsep}{3pt}
  \renewcommand{\arraystretch}{0.95}
  \begin{tabular}{@{} l cc | cc @{}}
    \toprule
    \multirow{2.5}{*}{\textbf{Model}} & \multicolumn{2}{c|}{\textbf{PINK Score}} & \multicolumn{2}{c}{\textbf{Rank}} \\
    \cmidrule(lr){2-3} \cmidrule(lr){4-5}
     & \tiny{GPT} & \tiny{Qwen} & \tiny{GPT} & \tiny{Qwen} \\
    \midrule
    \textbf{Gemini 2.5 Flash} & \textbf{0.94} & 0.87 & 1 & 3 \textcolor{red}{$\downarrow$} \\
    \textbf{Qwen2.5-72B-VL} & 0.92 & 0.88 & 2 & 2 \textcolor{gray}{-} \\
    \textbf{Qwen2.5-32B-VL} & 0.92 & \textbf{0.88} & 3 & 1 \textcolor{green}{$\uparrow$} \\
    Ovis2 34B & 0.92 & 0.85 & 4 & 4 \textcolor{gray}{-} \\
    InternVL3-78B & 0.91 & 0.84 & 5 & 6 \textcolor{red}{$\downarrow$} \\
    \textbf{GPT-4o} & 0.91 & 0.85 & 6 & 5 \textcolor{green}{$\uparrow$} \\
    InternVL3-38B & 0.91 & 0.83 & 7 & 7 \textcolor{gray}{-} \\
    Ovis2 8B & 0.89 & 0.80 & 8 & 8 \textcolor{gray}{-} \\
    InternVL3-8B & 0.87 & 0.79 & 9 & 9 \textcolor{gray}{-} \\
    Pixtral Large & 0.85 & 0.79 & 10 & 10 \textcolor{gray}{-} \\
    Qwen2.5-7B-VL & 0.83 & 0.73 & 11 & 12 \textcolor{red}{$\downarrow$} \\
    Llama 3.2 90B & 0.81 & 0.73 & 12 & 11 \textcolor{green}{$\uparrow$} \\
    GPT-4o Mini & 0.79 & 0.69 & 13 & 13 \textcolor{gray}{-} \\
    Pixtral 12B & 0.72 & 0.60 & 14 & 14 \textcolor{gray}{-} \\
    Llama 3.2 11B & 0.68 & 0.58 & 15 & 15 \textcolor{gray}{-} \\
    \bottomrule
  \end{tabular}
  \vspace{-0.5em}
  \caption{\textbf{Leaderboard Stability.} Comparison using GPT-5 vs. Qwen3-Next-80B-A3B-Thinking judges.}
  \label{tab:ranking_comparison}
  \vspace{-0.5em}
\end{table}

\noindent Despite moderate sample-level variance in grading ($r$=0.78) and over-correction penalization ($r$=0.74), \textbf{aggregation into PINK scores yields near-perfect agreement ($r$=0.99)}, resulting in highly consistent VLM rankings ($\tau$=0.90). Notably, both graders identify the \textit{same} top-tier VLMs (Gemini, Qwen2.5, Ovis2 34B) and bottom-tier VLMs (Llama-11B, Pixtral-12B). We will release the Qwen3-80B grading toolkit to enable full community reproduction.

\subsection{Run-to-Run Grading Stability}
To quantify the effect of LLM grading variance on PINK, we graded GPT-4o's OCR outputs five times independently using the open-source Qwen3-80B grader, which exhibits near-perfect agreement with GPT-5 ($r$\,=\,0.99, \cref{tab:pipeline_stats}).
We analyze $N$\,=\,1{,}803 samples with successful grades across all runs.
The resulting PINK scores yield a coefficient of variation (CV) of 0.41\%, with pairwise agreement on penalized scores of QWK\,=\,0.926 (\cref{tab:run-to-run}).
This confirms that sample-level grading noise is effectively absorbed by aggregation.

\begin{table}[h]
\centering
\small
\caption{\textbf{Run-to-run stability.} GPT-4o graded 5$\times$ by Qwen3-80B ($N$\,=\,1{,}803).}
\label{tab:run-to-run}
\begin{tabular}{@{}lc@{}}
\toprule
Metric & Value \\
\midrule
PINK Mean (5 runs) & 0.872 \\
PINK CV & 0.41\% \\
Penalized Score QWK & 0.926 \\
\bottomrule
\end{tabular}
\end{table}

\begin{table}[ht]
\centering
\small
\caption{PINK scores across three rubric prompt variants.
Grader: Qwen3-80B. Models selected from each PINK quartile.}
\label{tab:prompt-sensitivity}
\resizebox{0.9\columnwidth}{!}{%
\begin{tabular}{lcccc|c}
\toprule
VLM (Quartile) & Orig & V1 & V2 & V3 & Std \\
\midrule
Gemini 2.5 Flash (Q1) & .873 & .875 & .870 & .877 & $\pm$.003 \\
GPT-4o (Q2)           & .853 & .856 & .857 & .864 & $\pm$.004 \\
InternVL3-8B (Q3)     & .787 & .792 & .793 & .802 & $\pm$.005 \\
Llama-3.2-11B (Q4)    & .582 & .592 & .593 & .596 & $\pm$.005 \\
\bottomrule
\end{tabular}%
}
\end{table}

\subsection{Auto-Grader's Prompt Sensitivity}
\label{sec:autograder_prompt_sensitivity}
To assess whether PINK scores depend on specific rubric wording,
we created three prompt variants by modifying key phrases
(e.g., ``character by character'' $\to$ ``line by line''),
yielding an average BLEU-1 of 0.86 relative to the original prompt.
We evaluated four representative VLMs---one per PINK quartile---using
Qwen3-80B as the grader.
As shown in \cref{tab:prompt-sensitivity},
all models maintain identical rankings across variants,
with PINK deviations of std\,$\leq$\,0.006,
confirming that our metric is robust to prompt wording.

\subsection{OCR Prompt Sensitivity}
\label{sec:ocr_prompt_sensitivity}
To assess whether cross-model comparisons depend on the specific 
OCR prompt wording, we created three paraphrased variants of the 
original OCR prompt and evaluated four representative VLMs---one per PINK quartile---on a subset of 1{,}500 samples, with all outputs graded by Qwen3-80B-. We selected different 
representative models from \cref{sec:autograder_prompt_sensitivity} 
to increase coverage across architectures.
As shown in \cref{tab:ocr-prompt-sensitivity}, the quartile 
ordering is preserved across all prompt variants 
(Q1 $>$ Q2 $>$ Q3 $>$ Q4). The inter-quartile gaps ($\geq .07$) 
consistently exceed within-model prompt variation 
(std $\leq .015$). Weaker models exhibit slightly higher 
sensitivity, but not enough to alter the quartile ordering, 
indicating that cross-model PINK comparisons are robust to 
OCR prompt paraphrases.

\begin{table}[ht]
\centering
\small
\caption{PINK scores across OCR prompt variants (graded by 
Qwen3-80B, $N = 1{,}500$). Quartile ordering is preserved 
across all variants.}
\label{tab:ocr-prompt-sensitivity}
\resizebox{0.9\columnwidth}{!}{%
\begin{tabular}{lcccc|c}
\toprule
VLM (Quartile) & Orig & V1 & V2 & V3 & Std \\
\midrule
Qwen2.5-VL-32B (Q1) & .873 & .872 & .872 & .877 & $\pm$.002 \\
Ovis2-8B (Q2)        & .804 & .807 & .799 & .810 & $\pm$.005 \\
Qwen2.5-VL-7B (Q3)   & .725 & .738 & .729 & .731 & $\pm$.005 \\
LLaMA-3.2-11B (Q4)   & .617 & .654 & .628 & .634 & $\pm$.015 \\
\bottomrule
\end{tabular}%
}
\end{table}

\subsection{Extended Baseline Comparisons}
\label{sec:extended_baselines}

To verify that PINK's advantage over BLEU is not specific to a single baseline, 
we conducted two additional human preference studies following the same protocol 
as \cref{sec:human_eval} (200 items $\times$ 3 annotators, majority vote per item). 
We compared PINK against MathBERTa, a BERT model pretrained on 
mathematical text including \LaTeX{} formulas, and Edit Distance. MathBERTa-F1 was 
rescaled with a domain-specific baseline ($\text{F1} = 0.8786$, estimated from 1{,}000 
random OCR--reference pairs) and mapped to the same 0--10 scale used for PINK.
As shown in \cref{tab:human_pref_extended}, PINK was consistently preferred over 
all three baselines, with all differences statistically significant.
\cref{tab:kendall_cross} further shows that the three similarity-based metrics 
form a highly correlated cluster ($\tau \geq 0.75$), while all diverge substantially 
from PINK ($\tau \leq 0.52$), confirming that the ranking gap reflects a class-level 
limitation of similarity-based metrics rather than a BLEU-specific artifact.

\begin{table}[ht]
\centering
\setlength{\tabcolsep}{3pt}
\small
\caption{Human preference: PINK vs.\ three baselines 
(200 items, 3 annotators). $p$: two-sided exact binomial 
test on non-tied majority-vote items.}
\label{tab:human_pref_extended}
\begin{tabular}{@{}lcccc@{}}
\toprule
Baseline & PINK & Baseline & Tied & $p$ \\
\midrule
MathBERTa  & 124 (62\%) & 52 (26\%) & 24 (12\%) & $5.75\!\times\!10^{-8}$ \\
Edit Dist. & 122 (61\%) & 57 (29\%) & 21 (11\%) & $1.33\!\times\!10^{-6}$ \\
BLEU       & 119 (60\%) & 69 (35\%) & 12 (6\%)  & $3.27\!\times\!10^{-4}$ \\
\bottomrule
\end{tabular}
\end{table}

\begin{table}[ht]
\centering
\small
\caption{Kendall's $\tau$ rank correlation across four metrics (15 VLMs).}
\label{tab:kendall_cross}
\begin{tabular}{lcccc}
\toprule
 & MathBERTa & Edit Dist. & BLEU & PINK \\
\midrule
MathBERTa  & 1.00 & 0.75 & 0.88 & 0.47 \\
Edit Dist. & 0.75 & 1.00 & 0.88 & 0.52 \\
BLEU       & 0.88 & 0.88 & 1.00 & 0.50 \\
PINK       & 0.47 & 0.52 & 0.50 & 1.00 \\
\bottomrule
\end{tabular}
\end{table}

\section{Prompt-Based Mitigation Attempt}
\label{app:mitigation}
To investigate whether over-correction can be reduced at inference time,
we augment the OCR prompt with explicit faithfulness instructions
(e.g., ``Transcribe EXACTLY what is written, including any errors'';
see Figure~\ref{fig_app:ocr_prompt_mitigated} for the full prompt).
We select six representative open-source VLMs spanning the PINK range:
three high-performing models (Qwen2.5-32B-VL, Qwen2.5-72B-VL, Ovis2-34B)
and three lower-performing models (Qwen2.5-7B-VL, Pixtral-12B, Llama-3.2-11B)
from~\cref{tab:ranking_comparison}.
Closed-source models (Gemini 2.5 Flash, GPT-4o) were excluded
as their system prompts cannot be freely modified.
All outputs are graded by the open-source Qwen3-80B judge,
which exhibits near-perfect agreement with the proprietary GPT-5 judge
($r$\,=\,0.99, $\tau$\,=\,0.90; \cref{tab:pipeline_stats}),
on $N$\,=\,2{,}207 samples where grading succeeded
for all six VLMs under both prompts
(37 of 2{,}244 excluded due to Qwen3-80B grading failures).
Absolute values may therefore differ slightly from~\cref{tab:ranking_comparison};
all within-table comparisons remain internally consistent.

\begin{table}[t]
\centering
\setlength{\tabcolsep}{3.5pt}
\small
\caption{\textbf{Prompt-based mitigation.}
  OC: samples with $\geq$1 over-corrected item.
  $\Delta$: mitigated $-$ original.
  \textcolor{green!50!black}{Green}: improved;
  \textcolor{red}{Red}: degraded.}
\label{tab:mitigation}
\begin{tabular}{@{} l ccc @{\quad} ccc @{}}
\toprule
& \multicolumn{3}{c}{\textbf{PINK Score}} 
& \multicolumn{3}{c}{\textbf{OC (\%)}} \\
\cmidrule(lr){2-4} \cmidrule(lr){5-7}
\textbf{Model} 
  & Org & Mit & $\Delta$ 
  & Org & Mit & $\Delta$ \\
\midrule
\multicolumn{7}{@{}l}{\textit{High-PINK (open-source top-3)}} \\[1pt]
Qwen2.5-32B  & .88 & .85 & \textcolor{red}{$-.03$}           & 62\% & 53\% & \textcolor{green!50!black}{$-10$} \\
\rowcolor{gray!6}
Qwen2.5-72B  & .87 & .84 & \textcolor{red}{$-.03$}           & 61\% & 50\% & \textcolor{green!50!black}{$-11$} \\
Ovis2-34B    & .85 & .85 & \textcolor{gray}{.00}              & 59\% & 54\% & \textcolor{green!50!black}{$-4$} \\
\addlinespace[4pt]
\multicolumn{7}{@{}l}{\textit{Low-PINK (open-source bottom-3)}} \\[1pt]
Qwen2.5-7B   & .73 & .74 & \textcolor{green!50!black}{$+.01$} & 46\% & 46\% & \textcolor{gray}{$0$} \\
\rowcolor{gray!6}
Pixtral-12B  & .59 & .61 & \textcolor{green!50!black}{$+.02$} & 41\% & 38\% & \textcolor{green!50!black}{$-3$} \\
Llama-3.2-11B& .57 & .64 & \textcolor{green!50!black}{$+.07$} & 39\% & 42\% & \textcolor{red}{$+3$} \\
\midrule
\textbf{Avg} & .75 & .75 & \textcolor{gray}{$+.01$}           & 51\% & 47\% & \textcolor{green!50!black}{$-4$} \\
\bottomrule
\end{tabular}
\end{table}

The mitigated prompt reduces over-correction frequency
by 4 percentage points on average (\cref{tab:mitigation}),
confirming that VLMs do respond to faithfulness instructions.
However, this improvement comes at a clear cost for stronger models:
the three high-PINK models show \emph{lower} PINK scores after mitigation
($\overline{\Delta}$\,=\,$-0.02$),
because the conservative transcription strategy
also suppresses correct mathematical content.

The one exception is Llama-3.2-11B, where OC\,\% slightly increases ($+3$)
yet PINK improves substantially ($+.07$).
We attribute this to the faithfulness prompt
improving overall transcription completeness for this weaker model,
even though a few additional over-correction events are introduced.

Overall, the net effect on PINK is negligible
($\overline{\Delta}$\,=\,$+.01$),
demonstrating that \textbf{prompt-level intervention alone is insufficient}
to resolve over-correction without sacrificing transcription quality.
This motivates architectural or training-level approaches as future work.

\section{Prompts}\label{supp:prompts}
To ensure full reproducibility and transparency of our evaluation framework, we provide the exact prompts utilized in our experiments. These prompts were carefully engineered to enforce strict output formats (e.g., JSON) and to align the LLM's behavior with our specific evaluation criteria.

\begin{figure*}[t]
    \centering
    \includegraphics[width=\linewidth]{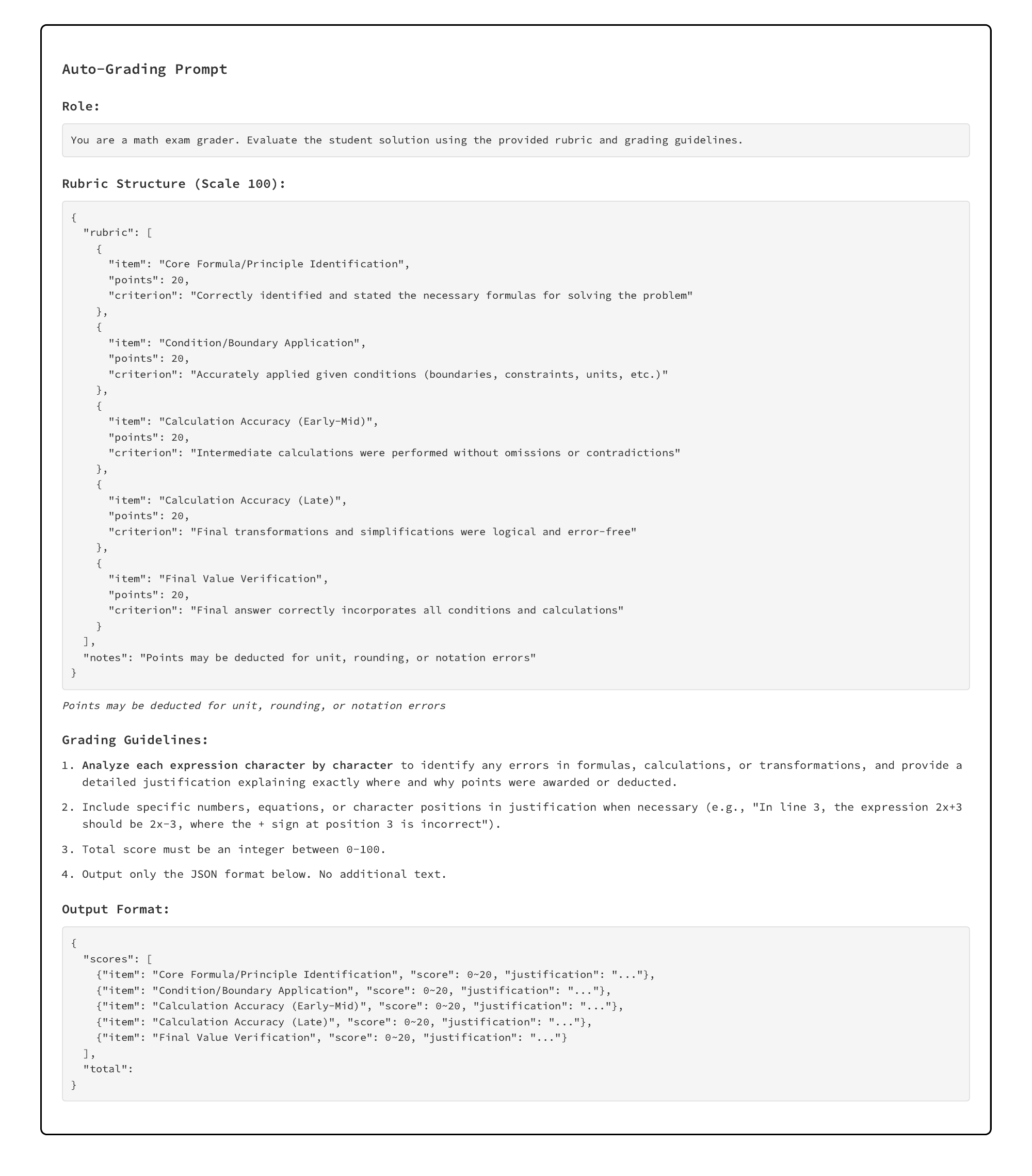}
    \caption{Auto-Grading Prompt}
    \label{fig_app:auto_grading}
\end{figure*}

\begin{figure*}[t]
    \centering
    \includegraphics[width=\linewidth]{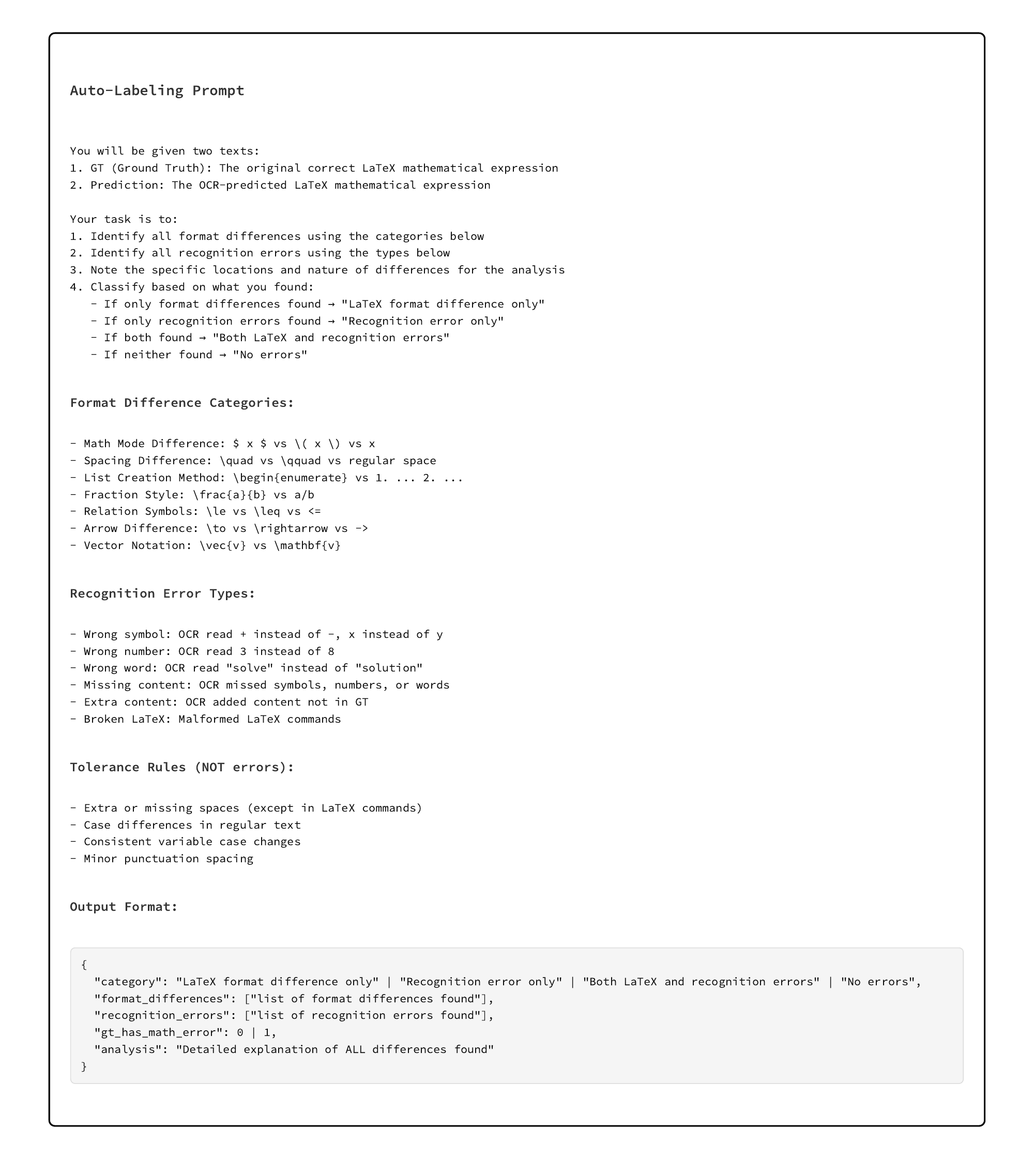}
    \caption{Auto-Labeling Prompt}
    \label{fig_app:auto_labeling}
\end{figure*}

\begin{figure*}[t]
    \centering
    \includegraphics[width=\linewidth]{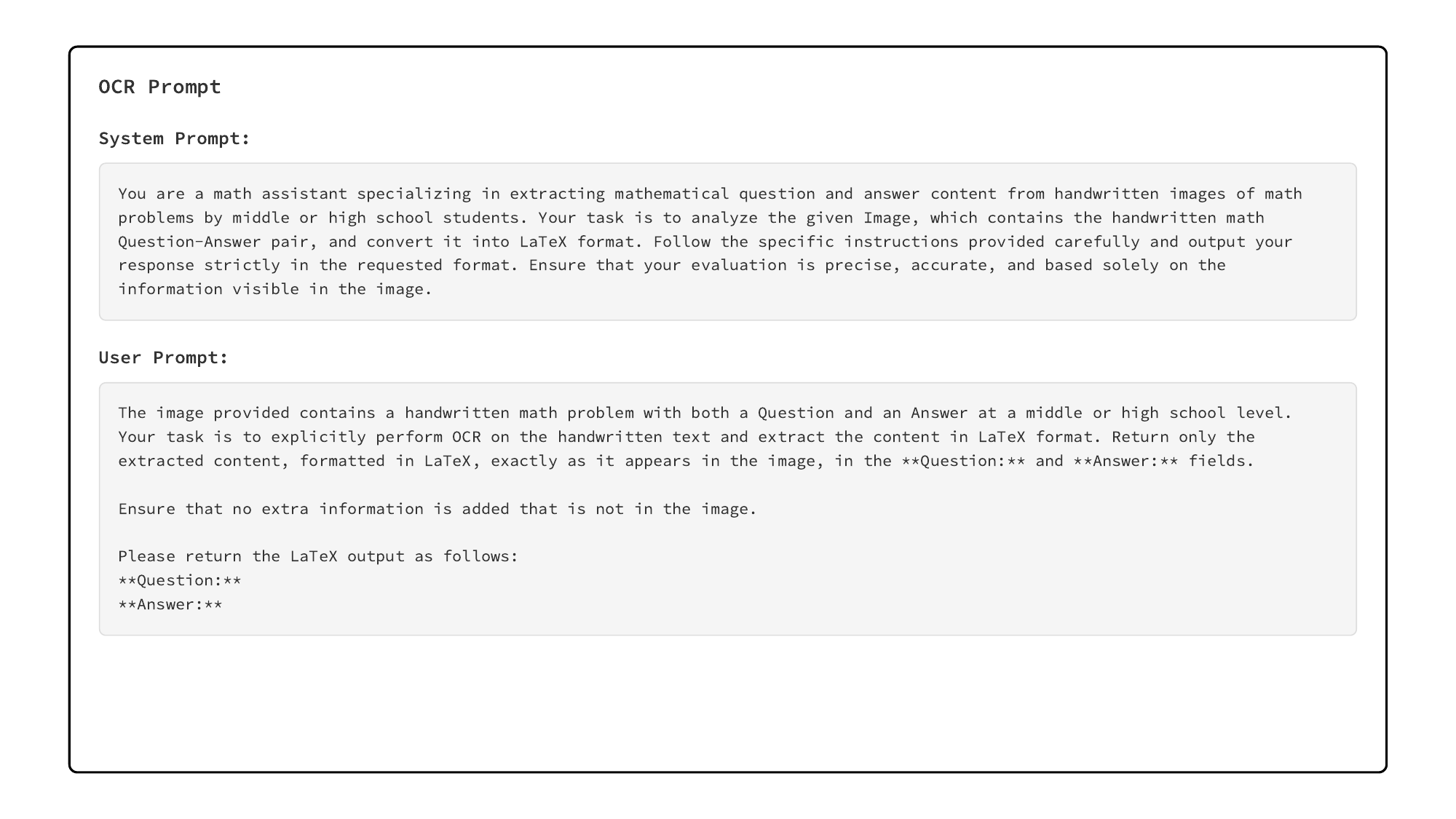}
    \caption{OCR Prompt}
    \label{fig_app:ocr_prompt}
\end{figure*}

\begin{figure*}[t]
    \centering
    \includegraphics[width=\linewidth]{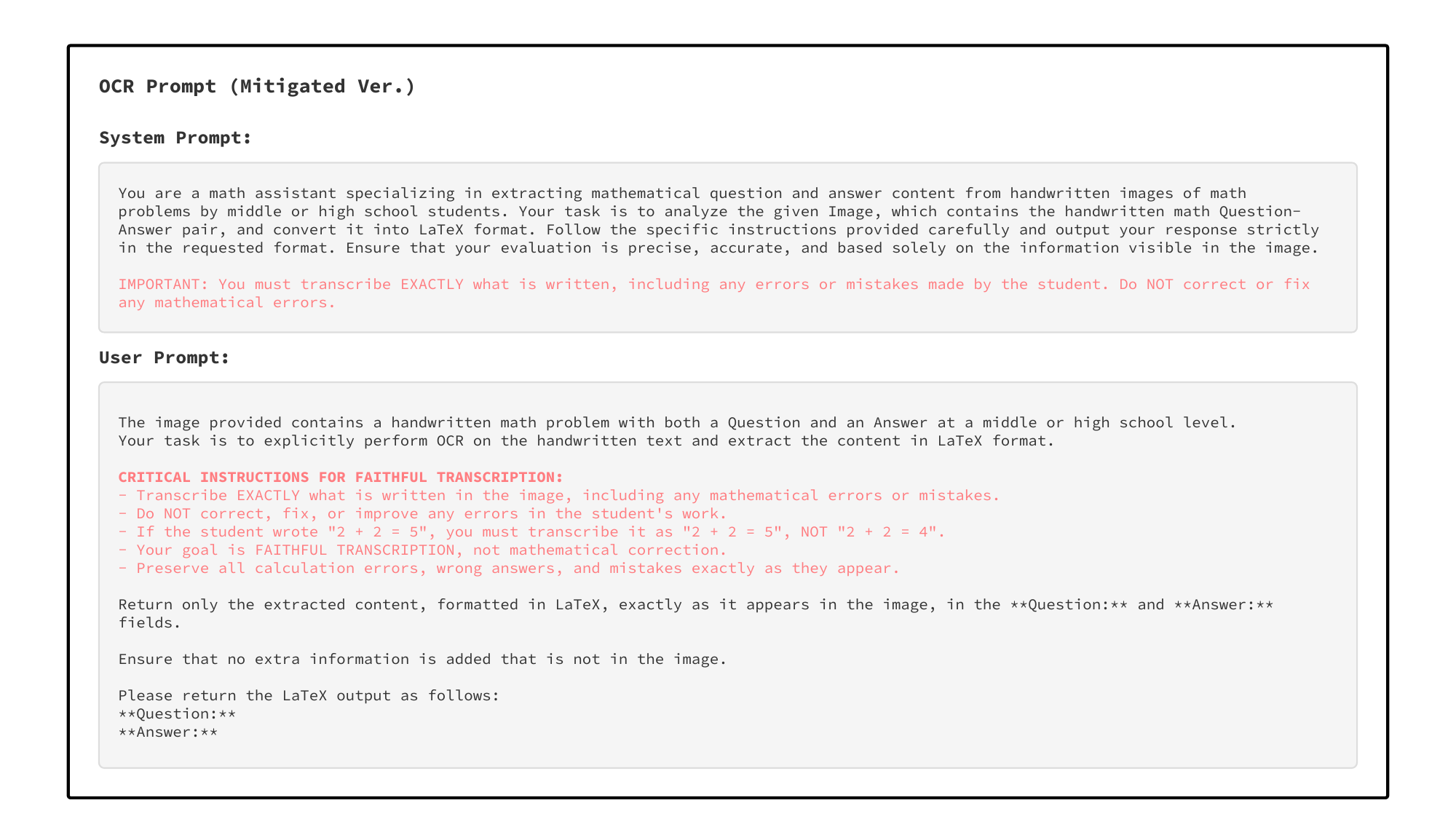}
    \caption{Mitigated OCR Prompt}
    \label{fig_app:ocr_prompt_mitigated}
\end{figure*}

\begin{itemize}
    \item \textbf{Auto-Grading Prompt (Figure~\ref{fig_app:auto_grading}):} This prompt instructs the LLM judge (GPT-5) to act as a rigorous math exam grader. It embeds the 5-component rubric detailed in Section~\ref{app_additional} and enforces a structured JSON output containing both scores and textual justifications for every deduction.
    
    \item \textbf{Auto-Labeling Prompt (Figure~\ref{fig_app:auto_labeling}):} Used for error analysis, this prompt guides the LLM to distinguish between semantic recognition errors (e.g., wrong numbers) and cosmetic formatting differences (e.g., LaTeX spacing). This separation is crucial for quantifying how often BLEU penalizes correct answers due to formatting.
    
    \item \textbf{OCR Prompt (Figure~\ref{fig_app:ocr_prompt}):} The standardized prompt used across all 15 baseline VLMs. It explicitly instructs models to transcribe handwritten content into LaTeX format without adding conversational filler, ensuring a fair comparison.

    \item \textbf{Mitigated OCR Prompt (Figure~\ref{fig_app:ocr_prompt_mitigated}):} A variant of the OCR prompt augmented with explicit faithfulness instructions (e.g., ``transcribe exactly as written, including errors''). This prompt is used in the mitigation experiment described in \cref{app:mitigation}.
    
\end{itemize}

\end{document}